\documentclass{ws-ijmpb}
\usepackage{subfigure}
\usepackage{graphicx,float}
\usepackage{slashed}
\begin{document}

\markboth{Thiago Prud\^{e}ncio}
{CONFINEMENT AND QUANTUM ANOMALY IN QUASI-1D SPINLESS FERMION CHAINS}

%
\catchline{}{}{}{}{}
%

\title{CONFINEMENT AND QUANTUM ANOMALY IN QUASI-1D SPINLESS FERMION CHAINS}

\author{THIAGO PRUD\^ENCIO\footnote{ERRATUM: Co-authorship \'Alvaro Ferraz removed under request. }}

\address{Instituto de F\'isica, Universidade de Bras\'ilia - UnB, CP: 04455, 70919-970, Bras\'ilia - DF, Brazil.}



\maketitle

\begin{history}
\received{FEV 2013}
\revised{2013}
\end{history}

\begin{abstract}
We consider the field renormalization group (RG) of two coupled one-spatial dimension (1D) spinless fermion chains under intraforward, 
interforward, interbackscattering and interumklapp interactions until two-loops order. Up to this order, 
we demonstrate the quantum confinement in the RG flow, where the interband chiral Fermi points reduce to single chiral Fermi points
and the renormalized interaction couplings have Luttinger liquid fixed points. We show that this quasi-1D system is equivalently 
described in terms of one and two-color interactions and address the problem of quantum anomaly, inherent 
to this system, as a direct consequence of coupling 1+1 free Dirac fields to one and two-color interactions.
\end{abstract}


\section{Introduction}

Luttinger liquids, characteristic of strongly correlated electron systems in one spatial dimension (1D), 
represent one of the most interesting examples 
of non-Fermi liquid behaviour, with intricate physical properties without counterpart in higher dimensions, 
as spin-charge separation \cite{alvaro1,alvaro4} 
and direct correspondence to bosonic excitations \cite{haldane}. It is a fundamental problem for quantum phase transitions 
in strongly correlated systems the crossover between a Luttinger liquid state and a Fermi liquid. 
When this occurs, the degrees of freedom of the system are reduced in 1D. Important systems as the edges of quantum Hall systems, 
quantum wires and high-Tc superconductors are related to the typical behaviour of 
Luttinger liquids. Intermediate between 1D and two spatial dimensions (2D), the coupling among 1D chains forms a 
quasi-1D system, that can serve as a prototype for the phenomenon of confinement \cite{ledowski2} and the study of 
CuO, CuO$_{2}$ compounds associated 
to high-Tc superconductors \cite{kondo}. As in 1D case, the Fermi surface of quasi-1D systems consist of discrete Fermi points, 
a set of Fermi points that form a discrete Fermi surface, in constrast to 2D and higher dimensions where 
the Fermi surface is continuous or a set of continuous \cite{correa}.  

Here, we consider a quasi-1D system of spinless fermions under intraforward, interforward, 
interbackscattering and interumklapp interactions. For a linearized dispersion relation, the intra and interband interactions 
terms act as an external field perturbation 
and the system as a whole behaves as free $1+1$ Dirac fields under an external field. As a consequence, 
the system presents an inherent chiral anomaly, as in the 
case of a $1+1$ Schwinger model \cite{schwinger1,schwinger2}, that is a result of the simultaneous 
requirement of vector current conservation (charge conservation) 
and axial current conservation (chiral symmetry) under vector and axial gauge transformations, respectively, appearing in 
the Ward-Takahashi identities in the one loop (1-loop) bubble polarizations with intra and interforward interactions 
\cite{alvaro2}. 

Applying a full field renormalization group (RG) until two loops (2-loops) order, taking into account a renormalization procedure that include the 
renormalized interband Fermi points in the physical prescription, 
we will derive the appropriate bare and renormalized quantities that 
lead to the set of RG flow equations including the renormalized interband Fermi points 
and the quasiparticle weight, considering the physical prescriptions of the one and two-particle irreducible functions until 
2-loops order at the Fermi surface. In 2-loops, the role of the quasiparticle weight $Z$ and the corresponding anomalous dimension $\gamma$ turns 
to be important in both self-energy and scattering channels, leading to physically important contributions to the physical properties of the system 
\cite{kochetov,alvaro4,freire1,freire2,freire3,alvaro5,fabrizio}, as a consequence we move forward in the previous results 
\cite{ledowski,ledowski3}, investigating the reduction of the
 renormalized interband chiral Fermi points to single renormalized chiral Fermi points. Taking into account the consistence 
with Ward-Takahashi relations, we will show that the quasi-1D system of two-coupled spinless fermions 
chains under intraforward, interforward, interbackscattering and interumklapp interactons can 
be equivalently described as $1+1$ free Dirac fields coupled to one and two-color interactions, deriving the corresponding
Ward-Takahashi identities \cite{ward,takahashi,kopietz1} and addressing the problem of quantum anomaly 
as a direct consequence of coupling 1+1 free Dirac fields to one and two-color interactions.

\section{Two 1D coupled chains}

\begin{figure}[h]
\centering
\includegraphics[scale=0.5]{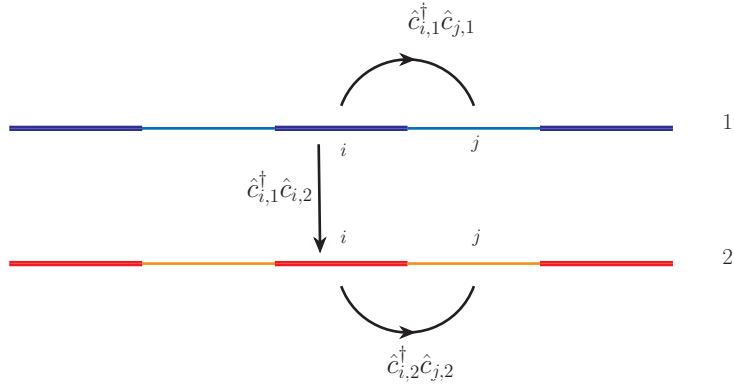}
\caption{(Color online) Scheme of two coupled 1D chains.}
\label{duascadeias}
\end{figure}

\begin{figure}[h]
\centering
\includegraphics[scale=0.7]{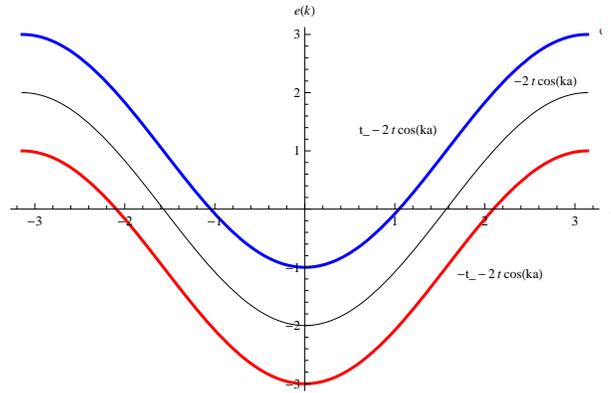}
\caption{(Color online) The blue line corresponds to a dispersion relation of the bonding band, $\varepsilon_{k}^{a}$,
the red line corresponds to the antibonding band, $\varepsilon_{k}^{b}$. 
The black line corresponds to a case without hopping term $t_{\perp}$.}
\label{duasbandas}
\end{figure}

The system of two coupled 1D spinless fermions chains is characterized by a transverse 
hopping $t_{\perp}$ that couples the two 1D chains\cite{fabrizio,ledermann} and the usual $1D$ chain intra-hopping term $t$ (figure \ref{duascadeias}) 
\begin{eqnarray}
\hat{H}_{0}&=& - t\sum_{<i,j>}\left(\hat{c}^{\dagger}_{i,1}\hat{c}_{j,1} + \hat{c}^{\dagger}_{i,2}\hat{c}_{j,2} + \texttt{h.c.} \right) 
\nonumber \\ &-&t_{\perp}\sum_{i}\left(\hat{c}^{\dagger}_{i,1}\hat{c}_{i,2} + \texttt{h.c.} \right),
\label{hamiltonianotcm1}
\end{eqnarray}
where $i$ and $j$ are neighbouring sites, $1$ and $2$ are the chain labels, 
$\hat{c}^{\dagger}$ and $\hat{c}$ are the criation and annihilation operators in the respective sites and chains.
We can write the momentum space representation of the 
operators $\hat{c}^{\dagger}$ and $\hat{c}$ by means of the following Fourier transform 
\begin{eqnarray}
\hat{c}_{i,1}&=&\frac{1}{\sqrt{N}}\sum_{k}e^{-ikx_{i}}\hat{c}_{k,1} \label{c1},\\
\hat{c}_{i,2}&=&\frac{1}{\sqrt{N}}\sum_{k}e^{-ikx_{i}}\hat{c}_{k,2} \label{c2}.
\end{eqnarray}
It follows that we can write
\begin{eqnarray}
\hat{c}_{i,1}^{\dagger}\hat{c}_{j,1}&=&\frac{1}{N}\sum_{k,k'}e^{ik'x_{i}}e^{-ikx_{j}}\hat{c}_{k',1}^{\dagger}\hat{c}_{k,1}.
\end{eqnarray}
Considering the sum in the first neighbours $x_{j}= x_{i} + \texttt{a}$ or $ x_{j}= x_{i} - \texttt{a}$, 
where $\texttt{a}$ is the constant lattice spacing,
\begin{eqnarray}
\sum_{<i,j>}\hat{c}_{i,1}^{\dagger}\hat{c}_{j,1} &=& \sum_{i}\left(\hat{c}_{i,1}^{\dagger}\hat{c}_{i+1,1} + \hat{c}_{i,1}^{\dagger}\hat{c}_{i-1,1} \right).
\label{soma1}
\end{eqnarray}
By summing the expressions
\begin{eqnarray}
\hat{c}_{i}^{\dagger}\hat{c}_{i+1}&=& \frac{1}{N}\sum_{k,k'} e^{-i(k-k')x_{i}}e^{-ik\texttt{a}}\hat{c}_{k'}^{\dagger}\hat{c}_{k}, \\
\hat{c}_{i}^{\dagger}\hat{c}_{i-1}&=& \frac{1}{N}\sum_{k,k'} e^{-i(k-k')x_{i}}e^{ik\texttt{a}}\hat{c}_{k'}^{\dagger}\hat{c}_{k}
\end{eqnarray}
and using the relation
\begin{eqnarray}
\frac{1}{N}\sum_{i}e^{-i(k-k')x_{i}}=\delta_{kk'},
\label{deltakk}
\end{eqnarray}
the equation (\ref{soma1}) can be rewritten as
\begin{eqnarray}
\sum_{<i,j>}\hat{c}_{i,1}^{\dagger}\hat{c}_{j,1} &=& 2\sum_{k}\cos(k\texttt{a})\hat{c}_{k,1}^{\dagger}\hat{c}_{k,1}.
\end{eqnarray}
The hamiltonian (\ref{hamiltonianotcm1}) is then written as
\begin{eqnarray}
\hat{H}_{0}&=& -2t\sum_{k}\cos(k\texttt{a})\left(\hat{c}_{k,1}^{\dagger}\hat{c}_{k,1} + \hat{c}_{k,2}^{\dagger}\hat{c}_{k,2} + \texttt{h.c.}\right) 
\nonumber \\ 
&-&t_{\perp}\sum_{i}\left(\hat{c}^{\dagger}_{i,1}\hat{c}_{i,2} + \texttt{h.c.} \right).
\end{eqnarray}
Now we can treat the term due to the coupling between the two chains. 
From the transforms (\ref{c1}) and (\ref{c2}), we arrive at 
\begin{eqnarray}
\hat{c}_{i,1}^{\dagger}\hat{c}_{i,2}&=&\frac{1}{N}\sum_{k,k'}e^{ik'x_{i}}e^{-ikx_{i}}\hat{c}_{k',1}^{\dagger}\hat{c}_{k,2}.
\end{eqnarray}
Then, from (\ref{deltakk}), we have, 
\begin{eqnarray}
\hat{H}_{0} &=& -2t\sum_{k}\cos(k\texttt{a})\left(\hat{c}_{k,1}^{\dagger}\hat{c}_{k,1} + \hat{c}_{k,2}^{\dagger}\hat{c}_{k,2} + \texttt{h.c.}\right) \nonumber \\ &-&t_{\perp}\left(\sum_{k}\hat{c}_{k,1}^{\dagger}\hat{c}_{k,2} + \texttt{h.c.} \right).
\label{h0nknd}
\end{eqnarray}
We see that the coupling term forbids the direct diagonalization of the hamiltonian. This problem can be solved
by means of the following canonical transformation
\begin{eqnarray}
\hat{b}_{k} &=&\frac{\hat{c}_{k,1} + \hat{c}_{k,2}}{\sqrt{2}}\\
\hat{a}_{k} &=&\frac{\hat{c}_{k,1} - \hat{c}_{k,2}}{\sqrt{2}}
\end{eqnarray}
which can also be written in the inverse form
\begin{eqnarray}
\hat{c}_{k,1} &=&\frac{\hat{b}_{k} + \hat{a}_{k}}{\sqrt{2}} \label{c1ab}\\
\hat{c}_{k,2} &=&\frac{\hat{b}_{k} - \hat{a}_{k}}{\sqrt{2}} \label{c2ab}
\end{eqnarray}
By using (\ref{c1ab}) and (\ref{c2ab}), we have
\small
\begin{eqnarray}
\hat{c}_{k,1}^{\dagger}\hat{c}_{k,1} &=& \frac{1}{2}\left(\hat{b}_{k}^{\dagger}\hat{b}_{k} + \hat{a}_{k}^{\dagger}\hat{a}_{k}  \right) + \frac{1}{2}\left(\hat{a}_{k}^{\dagger}\hat{b}_{k} + \hat{b}_{k}^{\dagger}\hat{a}_{k}  \right) \label{fs1} \\
\hat{c}_{k,2}^{\dagger}\hat{c}_{k,2} &=& \frac{1}{2}\left(\hat{b}_{k}^{\dagger}\hat{b}_{k} + \hat{a}_{k}^{\dagger}\hat{a}_{k}  \right) - \frac{1}{2}\left(\hat{a}_{k}^{\dagger}\hat{b}_{k} + \hat{b}_{k}^{\dagger}\hat{a}_{k}  \right) \label{fs2} \\
\hat{c}_{k,1}^{\dagger}\hat{c}_{k,2} &=& \frac{1}{2}\left(\hat{b}_{k'}^{\dagger}\hat{b}_{k} - \hat{a}_{k}^{\dagger}\hat{a}_{k}  \right) + \frac{1}{2}\left(\hat{a}_{k'}^{\dagger}\hat{b}_{k} - \hat{b}_{k}^{\dagger}\hat{a}_{k}  \right) \label{fs3} \\
\hat{c}_{k,2}^{\dagger}\hat{c}_{k,1} &=& \frac{1}{2}\left(\hat{b}_{k}^{\dagger}\hat{b}_{k} - \hat{a}_{k}^{\dagger}\hat{a}_{k}  \right) + \frac{1}{2}\left(-\hat{a}_{k}^{\dagger}\hat{b}_{k} + \hat{b}_{k}^{\dagger}\hat{a}_{k}  \right). \label{fs4}
\end{eqnarray}
\normalsize
From (\ref{fs1}), (\ref{fs2}), (\ref{fs3}) and (\ref{fs4}), we can rewrite (\ref{h0nknd}) in the diagonalized form
\begin{eqnarray}
\hat{H}_{0} &=& \sum_{k}\left(-2t\cos(k\texttt{a}) -t_{\perp}\right)\hat{b}_{k}^{\dagger}\hat{b}_{k} 
\nonumber \\
&+& \sum_{k} \left(-2t\cos(k\texttt{a}) +t_{\perp}\right)\hat{a}_{k}^{\dagger}\hat{a}_{k}.
\label{h0nknd}
\end{eqnarray}
By defining the following dispersion relations
\begin{eqnarray}
\varepsilon_{k}^{b} &=& -2t\cos(k\texttt{a}) - t_{\perp}, \\
\varepsilon_{k}^{a} &=& -2t\cos(k\texttt{a}) + t_{\perp},
\end{eqnarray}
we arrive finally at to the diagonalized form of the hamiltonian
\begin{eqnarray}
\hat{H}_{0} &=& \sum_{k}\left(\varepsilon_{k}^{a}\hat{a}_{k}^{\dagger}\hat{a}_{k} + \varepsilon_{k}^{b}\hat{b}_{k}^{\dagger}\hat{b}_{k} \right).
\label{h0nknd}
\end{eqnarray}
As a consequence, the presence of the coupling term divides the energy spectrum in two bands (figure \ref{duasbandas}). 
We call the dispersion relation $\varepsilon_{k}^{a}$ bonding band and $\varepsilon_{k}^{b}$ antibonding band. Note 
that the energy gap between the bonding and antibonding bands is constant
\begin{eqnarray}
\Delta \varepsilon = \varepsilon_{k}^{a}-\varepsilon_{k}^{b} = 2 t_{\perp}.
\end{eqnarray}

Expanding the Taylor series of the dispersion relations for the bonding and antibonding
bands around the respective Fermi points $k_{F}^{a}$, $-k_{F}^{a}$, $k_{F}^{b}$ e $-k_{F}^{b}$, 
at first order, we arrive at
linearized dispersion relations for the bond and antibonding bands
\begin{eqnarray}
\varepsilon_{+}^{a}(k)&=& \frac{d\varepsilon^{a}(k)}{dk}\vert_{k=k_{F}^{a}}(k-k_{F}^{a}),\label{epak}\\
\varepsilon_{-}^{a}(k)&=& \frac{d\varepsilon^{a}(k)}{dk}\vert_{k=-k_{F}^{a}}(k+k_{F}^{a}),\label{emak} \\
\varepsilon_{+}^{b}(k)&=& \frac{d\varepsilon^{b}(k)}{dk}\vert_{k=k_{F}^{b}}(k-k_{F}^{b}),\label{epbk} \\
\varepsilon_{-}^{b}(k)&=& \frac{d\varepsilon^{b}(k)}{dk}\vert_{k=-k_{F}^{b}}(k+k_{F}^{b})\label{embk}.
\end{eqnarray}
We can write the respective Fermi velocities
\begin{eqnarray}
v_{F}^{a} &=& \frac{d\varepsilon^{a}(k)}{dk}\vert_{k=k_{F}^{a}} = -\frac{d\varepsilon^{a}(k)}{dk}\vert_{k=-k_{F}^{a}} \label{vFaa}\\
v_{F}^{b}&=& \frac{d\varepsilon^{b}(k)}{dk}\vert_{k=k_{F}^{b}} = -\frac{d\varepsilon^{b}(k)}{dk}\vert_{k=-k_{F}^{b}}.
\label{vFbb}
\end{eqnarray}
Then from (\ref{vFaa}) and (\ref{vFbb}), we can write (\ref{emak}), (\ref{epak}), (\ref{embk}) and (\ref{epbk}) as follows
\begin{eqnarray}
\varepsilon_{+k}^{a}&=& v_{F}^{a}(k-k_{F}^{a}),\label{epak}\\
\varepsilon_{-k}^{a}&=& v_{F}^{a}(-k-k_{F}^{a}),\label{emak} \\
\varepsilon_{+k}^{b}&=& v_{F}^{b}(k-k_{F}^{b}),\label{epbk} \\
\varepsilon_{-k}^{b}&=& v_{F}^{b}(-k-k_{F}^{b}).\label{embk}
\end{eqnarray}
Now, we can write the hamiltonian for the linearized two coupled 1D chains in terms of fermionic fields 
$\hat{\psi}_{\alpha k}$, in the following form 
\begin{eqnarray}
\hat{H}_{0}=\sum_{k,\alpha,s}\varepsilon_{sk}^{\alpha}\hat{\psi}_{sk}^{\alpha\dagger}\hat{\psi}_{sk}^{\alpha},
\end{eqnarray}
where $\alpha=a,b$ is the color index, $s=\pm$ is the chirality index, corresponding tho the left and right-hand sides, 
$a$ is the bonding and $b$ is the antibonding band. 

This quasi-1D system is then characterized by two colors ($a$ and $b$) and chirality $s$ given by 
left ($-$) and right-hand ($+$) side of chiral spinless fermions. By means of a Legendre transform, we 
arrive at the following Lagrangean
\begin{eqnarray}
\mathcal{L}_{0}= \sum_{k,\alpha,s}\psi^{\dagger\alpha}_{sk}\left(k_{0} 
-\varepsilon_{sk}^{\alpha}\right)\psi_{sk}^{\alpha},
\label{lzero}
\end{eqnarray}
where the frequency $k_{0}$ comes from a Fourier transform on time derivative. 
Alternativelly, we can write this in the configuration space
\begin{eqnarray}
\mathcal{L}_{0}= \int d^{2}x \psi_{s}^{\alpha\dagger}(\vec{x})\left(i\partial_{t} -v_{F}\partial_{x} + v_{F}k_{F}^{\alpha}\right)\psi_{s}^{\alpha\dagger}(\vec{x})
\end{eqnarray}
\begin{figure}[h]
\centering
\includegraphics[scale=0.5]{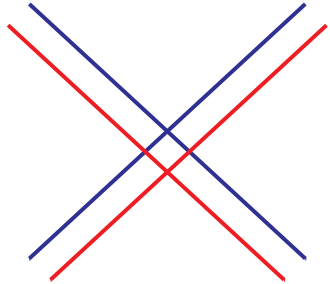}
\caption{(Color online) Linearized dispersion relations to the bonding band, $\varepsilon_{k}^{a}$, in blue,
 and the antibonding band, $\varepsilon_{k}^{b}$, in red.}
\label{duasbandas2}
\end{figure}
Taking into account the $1+1$ gamma matrices with upper-index 
$\gamma^{0}$, $\gamma^{1}$ and $\gamma^{5}=\gamma^{0}\gamma^{1}$,
\begin{eqnarray}
\gamma^{0}=\left( 
\begin{array}{cc}
0 & 1 \\ 
1 & 0%
\end{array}%
\right),
\end{eqnarray}
\begin{eqnarray}
\gamma^{1}=\left( 
\begin{array}{cc}
0 & -1 \\ 
1 & 0%
\end{array}%
\right),
\end{eqnarray}
\begin{eqnarray}
\gamma^{5}=\left( 
\begin{array}{cc}
1 & 0 \\ 
0 & -1
\end{array}%
\right),
\end{eqnarray}
and the low-index $\gamma_{0}=\gamma^{0}$, $\gamma_{5}=-\gamma^{5}$ and $\gamma_{1}=-\gamma^{1}$,
\begin{eqnarray}
\gamma_{1}=\left( 
\begin{array}{cc}
0 & 1 \\ 
-1 & 0%
\end{array}%
\right).
\end{eqnarray}
Defining the pseudo-spinors $\bar{\psi}^{\alpha}=\psi^{\alpha\dagger}\gamma^{0}$,
\begin{eqnarray}
\psi^{\alpha}&=&\left( 
\begin{array}{c}
\psi_{+}^{\alpha}   \\
\psi_{-}^{\alpha}    
\end{array}%
\right),
\psi^{\alpha\dagger}=\left( 
\begin{array}{cc}
\psi_{+}^{\alpha\dagger} & \psi_{-}^{\alpha\dagger}  
\end{array}
\right),
\end{eqnarray}
where $\alpha=a,b$, we can write the action as a $1+1$ Dirac field with an additional term.
This can be easily seen, taking $v_{F}=1$, $x^{0}=t$ e $x^{1}=x$, with $\slashed\partial=\gamma^{\mu}\partial_{\mu}$, 
$\mu=0,1$, such that the free action corresponding to the quasi-1D spinless fermions chains can be written in the following 
form
\begin{eqnarray}
\mathcal{S}_{0}= \sum_{\alpha} \int d^{2}x\bar{\psi}^{\alpha}\left(\slashed\partial + \gamma^{0}v_{F}k_{F}^{\alpha}\right)\psi^{\alpha},
\end{eqnarray}
that is particularly useful for the analysis of quantum anomaly when we couple the band interaction terms.

\section{Intraband and interband interactions}

\begin{figure}[h]
\centering
\includegraphics[scale=0.4]{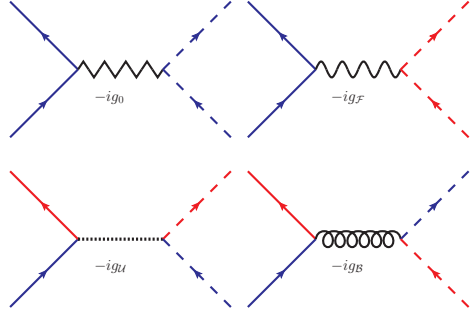}
\caption{(Color online) Intraband and interband interactions} 
\label{interacoesTCM}
\end{figure}

\begin{figure}[h]
\centering
\includegraphics[scale=0.4]{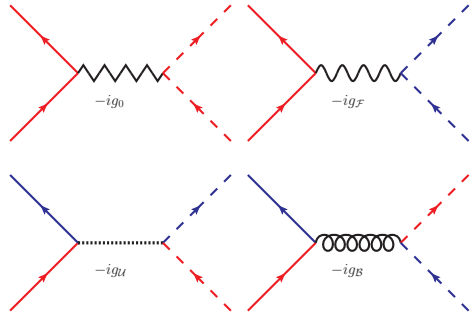}
\caption{(Color online) Intraband and interband interactions (color exchange $a\leftrightarrows b$)} 
\label{interacoesTCM2}
\end{figure}

We consider all possible interactions with particles of opposite chirality that preserve both chiral symmetry
 and realizes all possible color exchanges ($a\leftrightarrows b$ exchanges), reducing 
to spinless Luttinger liquids with interactions of type $-ig_{2}$ and without interactions of type $-ig_{4}$ in the $1D$ 
limit\cite{alvaro1}. 
Under this consideration, we 
have four possible band interactions: intraforward, interforward, interbackscattering 
and interumklapp (figures \ref{interacoesTCM} and \ref{interacoesTCM2}). The 
intraforward interaction is an intraband interaction of type 
$-ig_{2}$. The interforward $-ig_{\mathcal{F}}$ consists of a interband scattering of oposite chirality without color exchange. 
The interbackscattering $-ig_{\mathcal{B}}$ is a scattering of opposite chirality with color exchange, where 
the initial colors in opposite chiralities are different. 
The interumklapp interaction $-ig_{\mathcal{U}}$ is a scattering of opposite chirality with color exchange, 
where the initial colors in oposite chiralities are the same.
The interaction hamiltonian can be written as
\begin{eqnarray}
\hat{H}_{int} &=& \frac{1}{2}\sum_{k,k',q,s,\alpha}g_{0} 
\hat{\psi}^{\alpha\dagger}_{sk}\hat{\psi}^{\alpha\dagger}_{-sk'}
\hat{\psi}_{s(k+q)}^{\alpha}\hat{\psi}_{-s(k'-q)}^{\alpha} \nonumber \\
&+& \frac{1}{2}\sum_{k,k',q,s,\alpha\neq\beta}g_{\mathcal{F}} 
\hat{\psi}^{\alpha\dagger}_{sk}\hat{\psi}^{\beta\dagger}_{-sk'}
\hat{\psi}_{s(k+q)}^{\alpha}\hat{\psi}_{-s(k'-q)}^{\beta} \nonumber \\
&+& \frac{1}{2}\sum_{k,k',q,s,\alpha\neq\beta}g_{\mathcal{B}} 
\hat{\psi}^{\alpha\dagger}_{sk}\hat{\psi}^{\beta\dagger}_{-sk'}
\hat{\psi}_{s(k+q)}^{\beta}\hat{\psi}_{-s(k'-q)}^{\alpha} \nonumber \\
&+& \frac{1}{2}\sum_{k,k',q,s,\alpha\neq\beta}g_{\mathcal{U}} 
\hat{\psi}^{\beta\dagger}_{sk}\hat{\psi}^{\beta\dagger}_{-sk'}
\hat{\psi}_{s(k+q)}^{\alpha}\hat{\psi}_{-s(k'-q)}^{\alpha}, \nonumber \\
\end{eqnarray}
where $s=\pm$ is the chirality index, $\alpha,\beta=a,b$ are the color indexes, and $k,k',q$  are the corresponding moments
involved in the interactions. The term of $1/2$ it is due to the explicit indroduction fo the chirality term $s$.

We can now write the total lagrangean corresponding to the model in the renormalized form, with the indroduction of 
the couterterms $\delta Z$, $\delta g_{0,R}$, $\delta g_{\mathcal{F},R}$, $\delta g_{\mathcal{B},R}$ and 
$\delta g_{\mathcal{U},R}$,
\begin{eqnarray}
\mathcal{L}_{R}&=&\sum_{k,s,\alpha}(1+\delta Z)\hat{\psi}^{\alpha\dagger}_{sk,R}
\left[k_{0}-\varepsilon_{sk}^{\alpha}\right]\hat{\psi}_{sk,R}^{\alpha} \nonumber \\
&-& \frac{1}{2}\sum_{k,k',q,s,\alpha}g_{0,R} 
\hat{\psi}^{\alpha\dagger}_{sk,R}\hat{\psi}^{\alpha\dagger}_{-sk',R}
\hat{\psi}_{s(k+q),R}^{\alpha}\hat{\psi}_{-s(k'-q),R}^{\alpha} \nonumber \\
&-& \frac{1}{2}\sum_{k,k',q,s,\alpha\neq\beta}g_{\mathcal{F},R} 
\hat{\psi}^{\alpha\dagger}_{sk,R}\hat{\psi}^{\beta\dagger}_{-sk',R}
\hat{\psi}_{s(k+q),R}^{\alpha}\hat{\psi}_{-s(k'-q),R}^{\beta} \nonumber \\
&-& \frac{1}{2}\sum_{k,k',q,s,\alpha\neq\beta}g_{\mathcal{B},R} 
\hat{\psi}^{\alpha\dagger}_{sk,R}\hat{\psi}^{\beta\dagger}_{-sk',R}
\hat{\psi}_{s(k+q),R}^{\beta}\hat{\psi}_{-s(k'-q),R}^{\alpha} \nonumber \\
&-& \frac{1}{2}\sum_{k,k',q,s,\alpha\neq\beta}g_{\mathcal{U},R} 
\hat{\psi}^{\beta\dagger}_{sk,R}\hat{\psi}^{\beta\dagger}_{-sk',R}
\hat{\psi}_{s(k+q),R}^{\alpha}\hat{\psi}_{-s(k'-q),R}^{\alpha}, \nonumber\\
&-& \frac{1}{2}\sum_{k,k',q,s,\alpha}\delta g_{0,R} 
\hat{\psi}^{\alpha\dagger}_{sk,R}\hat{\psi}^{\alpha\dagger}_{-sk',R}
\hat{\psi}_{s(k+q),R}^{\alpha}\hat{\psi}_{-s(k'-q),R}^{\alpha} \nonumber \\
&-& \frac{1}{2}\sum_{k,k',q,s,\alpha\neq\beta}\delta g_{\mathcal{F},R} 
\hat{\psi}^{\alpha\dagger}_{sk,R}\hat{\psi}^{\beta\dagger}_{-sk',R}
\hat{\psi}_{s(k+q),R}^{\alpha}\hat{\psi}_{-s(k'-q),R}^{\beta} \nonumber \\
&-& \frac{1}{2}\sum_{k,k',q,s,\alpha\neq\beta}\delta g_{\mathcal{B},R} 
\hat{\psi}^{\alpha\dagger}_{sk,R}\hat{\psi}^{\beta\dagger}_{-sk',R}
\hat{\psi}_{s(k+q),R}^{\beta}\hat{\psi}_{-s(k'-q),R}^{\alpha} \nonumber \\
&-& \frac{1}{2}\sum_{k,k',q,s,\alpha\neq\beta}\delta g_{\mathcal{U},R} 
\hat{\psi}^{\beta\dagger}_{sk,R}\hat{\psi}^{\beta\dagger}_{-sk',R}
\hat{\psi}_{s(k+q),R}^{\alpha}\hat{\psi}_{-s(k'-q),R}^{\alpha}, \nonumber\\
\label{principallagrangeana} 
\end{eqnarray}
where the index $R$ means renormalized and the counterterms $\delta Z$, $\delta g_{0,R}$, 
$\delta g_{\mathcal{F},R}$, $\delta g_{\mathcal{B},R}$ and 
$\delta g_{\mathcal{U},R}$ correspond to the respective interactions, and the $\hat{\psi}$'s 
are given in terms of $\vec{k}=(k_{0},k)$.

The renormalized lagrangean (\ref{principallagrangeana}) involves a prescription of renormalization. By 
including into the prescription the Fermi moments and Fermi velocities, we have to reserve couterterms also for such terms, i. e.,
for the Fermi velocity and Fermi moment associated to each band, leading to a correction in in the 
free dispersion relation $\varepsilon_{sk}^{\alpha}$. The free term is then renormalized into the following form 

\begin{eqnarray}
\mathcal{L}_{0R}&=& \sum_{k,s,\alpha} \hat{\psi}^{\alpha\dagger}_{sk, R}\left[k_{0,R} - v_{F,R}\left(sk -k_{F,R}^{\alpha}\right)\right]\hat{\psi}^{\alpha}_{sk, R} \nonumber \\
&+& \sum_{k,s,\alpha}\hat{\psi}^{\alpha\dagger}_{sk, R}\left[ k_{0,R}\delta Z -sk\delta v_{FR} + v_{F,R}\delta k_{F}^{\alpha}\right]\hat{\psi}^{\alpha}_{sk, R}, \nonumber \\
\end{eqnarray}
where we have counterterms associated to the Fermi velocity $\delta v_{F,R}$ and to the Fermi moments $\delta k_{F}^{\alpha}$. Note that 
for simplicity we consider the same Fermi velocity $v_{F,R}$. We can describe the couterterm associated to the Fermi velocity in terms of 
$\delta Z$, by means of the relation
 $\delta v_{F,R}=v_{F,R}\delta Z$. We can then write the previous lagrangean in the following form
\begin{eqnarray}
\mathcal{L}_{0,R}&=& \sum_{k,s,\alpha} \hat{\psi}^{\alpha\dagger}_{sk, R}\left[k_{0,R} - v_{F,R}\left(sk -k_{F,R}^{\alpha}\right)\right]\hat{\psi}^{\alpha}_{sk, R} \nonumber \\
&+& \sum_{k,s,\alpha}\hat{\psi}^{\alpha\dagger}_{sk, R}\left[ k_{0,R}\delta Z -skv_{F,R}\delta Z + v_{F,R}\delta k_{F}^{\alpha}\right]\hat{\psi}^{\alpha}_{sk, R}. \nonumber \\
\end{eqnarray}
Now it involves a renormalized term $\varepsilon_{sk,R}^{\alpha}=v_{F,R}\left(sk -k_{F,R}^{\alpha}\right)$ and the counterterms 
associated.

The total renormalized lagrangean can be written with the explicit contributions due to the couterterms
\begin{eqnarray}
\mathcal{L}_{R}&=&\sum_{k,s,\alpha}\hat{\psi}^{\alpha\dagger}_{sk,R}
\left[k_{0,R}-\varepsilon_{sk,R}^{\alpha}\right]\hat{\psi}_{sk,R}^{\alpha} \nonumber \\
&-& \frac{1}{2}\sum_{k,k',q,s,\alpha}g_{0,R} 
\hat{\psi}^{\alpha\dagger}_{sk,R}\hat{\psi}^{\alpha\dagger}_{-sk',R}
\hat{\psi}_{s(k+q),R}^{\alpha}\hat{\psi}_{-s(k'-q),R}^{\alpha} \nonumber \\
&-& \frac{1}{2}\sum_{k,k',q,s,\alpha\neq\beta}g_{\mathcal{F},R} 
\hat{\psi}^{\alpha\dagger}_{sk,R}\hat{\psi}^{\beta\dagger}_{-sk',R}
\hat{\psi}_{s(k+q),R}^{\alpha}\hat{\psi}_{-s(k'-q),R}^{\beta} \nonumber \\
&-& \frac{1}{2}\sum_{k,k',q,s,\alpha\neq\beta}g_{\mathcal{B},R} 
\hat{\psi}^{\alpha\dagger}_{sk,R}\hat{\psi}^{\beta\dagger}_{-sk',R}
\hat{\psi}_{s(k+q),R}^{\beta}\hat{\psi}_{-s(k'-q),R}^{\alpha} \nonumber \\
&-& \frac{1}{2}\sum_{k,k',q,s,\alpha\neq\beta}g_{\mathcal{U},R}  
\hat{\psi}^{\beta\dagger}_{sk,R}\hat{\psi}^{\beta\dagger}_{-sk',R}
\hat{\psi}_{s(k+q),R}^{\alpha}\hat{\psi}_{-s(k'-q),R}^{\alpha}, \nonumber \\
&+& \sum_{k,s,\alpha}\hat{\psi}^{\alpha\dagger}_{sk, R}\left[ k_{0,R}\delta Z -skv_{F,R}\delta Z + v_{F,R}\delta k_{F,R}^{\alpha}\right]
\hat{\psi}^{\alpha}_{sk, R} \nonumber \\
&-& \frac{1}{2}\sum_{k,k',q,s,\alpha}\delta g_{0,R} 
\hat{\psi}^{\alpha\dagger}_{sk,R}\hat{\psi}^{\alpha\dagger}_{-sk',R}
\hat{\psi}_{s(k+q),R}^{\alpha}\hat{\psi}_{-s(k'-q),R}^{\alpha} \nonumber \\
&-& \frac{1}{2}\sum_{k,k',q,s,\alpha\neq\beta}\delta g_{\mathcal{F},R} 
\hat{\psi}^{\alpha\dagger}_{sk,R}\hat{\psi}^{\beta\dagger}_{-sk',R}
\hat{\psi}_{s(k+q),R}^{\alpha}\hat{\psi}_{-s(k'-q),R}^{\beta} \nonumber \\
&-& \frac{1}{2}\sum_{k,k',q,s,\alpha\neq\beta}\delta g_{\mathcal{B},R} 
\hat{\psi}^{\alpha\dagger}_{sk,R}\hat{\psi}^{\beta\dagger}_{-sk',R}
\hat{\psi}_{s(k+q),R}^{\beta}\hat{\psi}_{-s(k'-q),R}^{\alpha} \nonumber \\
&-& \frac{1}{2}\sum_{k,k',q,s,\alpha\neq\beta}\delta g_{\mathcal{U},R} 
\hat{\psi}^{\beta\dagger}_{sk,R}\hat{\psi}^{\beta\dagger}_{-sk',R}
\hat{\psi}_{s(k+q),R}^{\alpha}\hat{\psi}_{-s(k'-q),R}^{\alpha}. \nonumber \\
\label{renormalizadofinal}
\end{eqnarray}
The procedure of renormalization group realizes a modification in the fermionic fields, relating to the $\texttt{bare}$ fields 
by means of the following relation
\begin{eqnarray}
\hat{\psi}_{sk,R}^{\alpha}= Z^{-1/2}\hat{\psi}_{sk,\texttt{bare}}^{\alpha},
\label{barern}
\end{eqnarray}
where $Z = 1 + \delta{Z}$ is the weight of the quasiparticle.

\section{The $\texttt{bare}$ quantities}

The $\texttt{bare}$ fields are fields invariant under the renormalization group, such that, written in terms. In terms of such 
fields the lagrangean takes the following form
\begin{eqnarray}
\mathcal{L}_{\texttt{bare}}&=&\sum_{k,s,\alpha}\hat{\psi}^{\alpha\dagger}_{sk,\texttt{bare}}
\left[k_{0,\texttt{bare}}-\varepsilon_{sk,\texttt{bare}}^{\alpha}\right]\hat{\psi}_{sk,\texttt{bare}}^{\alpha}\nonumber \\
&-& \frac{1}{2}\sum_{k,k',q,s,\alpha}g_{0,\texttt{bare}} 
\hat{\psi}^{\alpha\dagger}_{sk,\texttt{bare}}\hat{\psi}^{\alpha\dagger}_{-sk',\texttt{bare}} 
 \nonumber \\
&\times&\hat{\psi}_{s(k+q),\texttt{bare}}^{\alpha}\hat{\psi}_{-s(k'-q),\texttt{bare}}^{\alpha} \nonumber \\
&-& \frac{1}{2}\sum_{k,k',q,s,\alpha\neq\beta}g_{\mathcal{F},\texttt{bare}}
\hat{\psi}^{\alpha\dagger}_{sk,\texttt{bare}}\hat{\psi}^{\beta\dagger}_{-sk',\texttt{bare}}\nonumber \\
&\times& 
\hat{\psi}_{s(k+q),\texttt{bare}}^{\alpha}\hat{\psi}_{-s(k'-q),\texttt{bare}}^{\beta} \nonumber \\
&-& \frac{1}{2}\sum_{k,k',q,s,\alpha\neq\beta}g_{\mathcal{B},\texttt{bare}} 
\hat{\psi}^{\alpha\dagger}_{sk,\texttt{bare}}\hat{\psi}^{\beta\dagger}_{-sk',\texttt{bare}}\nonumber \\
&\times& 
\hat{\psi}_{s(k+q),\texttt{bare}}^{\beta}\hat{\psi}_{-s(k'-q),\texttt{bare}}^{\alpha} \nonumber \\
&-& \frac{1}{2}\sum_{k,k',q,s,\alpha\neq\beta}g_{\mathcal{U},\texttt{bare}}  
\hat{\psi}^{\beta\dagger}_{sk,\texttt{bare}}\hat{\psi}^{\beta\dagger}_{-sk',\texttt{bare}}\nonumber \\
&\times&
\hat{\psi}_{s(k+q),\texttt{bare}}^{\alpha}\hat{\psi}_{-s(k'-q),\texttt{bare}}^{\alpha}, 
\label{barelagrange}
\end{eqnarray}
In this form, the lagrangean has the form of the non-renormalized model. It is in fact equivalent to the renormalized model, such that
using the relation (\ref{barern}) in (\ref{barelagrange}), we have
\begin{eqnarray}
\mathcal{L}_{R}&=&\sum_{k,s,\alpha}Z\hat{\psi}^{\alpha\dagger}_{sk,R}
\left[k_{0,\texttt{bare}}-\varepsilon_{sk,\texttt{bare}}^{\alpha}\right]\hat{\psi}_{sk,R}^{\alpha} \nonumber \\
&-& \frac{1}{2}\sum_{k,k',q,s,\alpha}Z^{2}g_{0,\texttt{bare}} 
\hat{\psi}^{\alpha\dagger}_{sk,R}\hat{\psi}^{\alpha\dagger}_{-sk',R} \nonumber \\
&\times& \hat{\psi}_{s(k+q),R}^{\alpha}\hat{\psi}_{-s(k'-q),R}^{\alpha} \nonumber \\
&-& \frac{1}{2}\sum_{k,k',q,s,\alpha\neq\beta}Z^{2}g_{\mathcal{F},\texttt{bare}} 
\hat{\psi}^{\alpha\dagger}_{sk,R}\hat{\psi}^{\beta\dagger}_{-sk',R}\nonumber \\
&\times&
\hat{\psi}_{s(k+q),R}^{\alpha}\hat{\psi}_{-s(k'-q),R}^{\beta} \nonumber \\
&-& \frac{1}{2}\sum_{k,k',q,s,\alpha\neq\beta}Z^{2}g_{\mathcal{B},\texttt{bare}} 
\hat{\psi}^{\alpha\dagger}_{sk,R}\hat{\psi}^{\beta\dagger}_{-sk',R}\nonumber \\
&\times&
\hat{\psi}_{s(k+q),R}^{\beta}\hat{\psi}_{-s(k'-q),R}^{\alpha} \nonumber \\
&-& \frac{1}{2}\sum_{k,k',q,s,\alpha\neq\beta}Z^{2}g_{\mathcal{U},\texttt{bare}}  
\hat{\psi}^{\beta\dagger}_{sk,R}\hat{\psi}^{\beta\dagger}_{-sk',R}\nonumber \\
&\times&
\hat{\psi}_{s(k+q),R}^{\alpha}\hat{\psi}_{-s(k'-q),R}^{\alpha}, 
\label{barelagrange2}
\end{eqnarray}
The equation (\ref{barelagrange2}) is identical to the equation (\ref{renormalizadofinal}) for the renormalized 
lagrangean, where we have the following relations between $\texttt{bare}$ and renormalized $R$ quantities
\begin{eqnarray}
Z k_{0,\texttt{bare}} &=& k_{0,R} +\delta Z k_{0,R} \\
Z\varepsilon_{sk,\texttt{bare}}^{\alpha} &=& \varepsilon_{sk,R}^{\alpha} + skv_{F,R}\delta Z - v_{F,R}\delta k_{F,R}^{\alpha} \label{ekR} \\
Z^{2}g_{0,\texttt{bare}} &=& g_{0,R} + \delta g_{0,R} \\
Z^{2}g_{\mathcal{F},\texttt{bare}} &=& g_{\mathcal{F},R} + \delta g_{\mathcal{F},R} \\ 
Z^{2}g_{\mathcal{B},\texttt{bare}} &=& g_{\mathcal{B},R} + \delta g_{\mathcal{B},R} \\
Z^{2}g_{\mathcal{U},\texttt{bare}} &=& g_{\mathcal{U},R} + \delta g_{\mathcal{U},R} 
\end{eqnarray}
From the equation (\ref{ekR}), we also arrive at the following relations
\begin{eqnarray}
Zv_{F,\texttt{bare}} &=& v_{F,R} + v_{F,R}\delta Z  = Z v_{F,R}.
\end{eqnarray}
It follows the relation $v_{F,\texttt{bare}}=v_{F,R}$ and
\begin{eqnarray}
Zk_{F,\texttt{bare}}^{\alpha} &=& k_{F,R}^{\alpha} + \delta k_{F,R}^{\alpha}. 
\label{thiskbare}
\end{eqnarray}
Defining as $\Delta k_{F}=k_{F}^{b}-k_{F}^{a}$ the difference between the Fermi points from the bands $b$ and $a$, 
we also can write such difference relating the $\texttt{bare}$ and renormalized quantities
\begin{eqnarray}
Z\Delta k_{F,\texttt{bare}} &=& \Delta k_{F,R} + \delta \Delta k_{F,R}, 
\label{thisdkbare}
\end{eqnarray}
where we have defined $\delta \Delta k_{F,R}= \delta k_{F,R}^{b}-\delta k_{F,R}^{a}$.

\section{Green's functions}

We now consider the free propagators written in the form $k_{0,R}-\varepsilon_{sk,R}^{\alpha}$ and the Green's functions 
associated to them
\begin{eqnarray}
G^{\alpha}_{s,R}(\vec{k})&=&\frac{\theta^{\alpha>}_{sk,R}}{k_{0,R} -\varepsilon_{sk,R}^{\alpha} + i\delta} + \frac{\theta^{\alpha<}_{sk,R}}{k_{0,R} -\varepsilon_{sk,R}^{\alpha} - i\delta},
 \label{pa16} \nonumber \\
\end{eqnarray}
where $s=\pm$ correspond to the chirality index, $\alpha=a,b$ correspond to the color index and 
\begin{eqnarray}
\theta^{\alpha >}_{sk,R}&=&\theta(sk-k^{\alpha}_{F,R}), \label{pa13}\\
\theta^{\alpha <}_{sk,R}&=&\theta(k^{\alpha}_{F,R}-sk), \label{pa14}
\end{eqnarray}
are the corresponding step functions.

\begin{figure}[h]
\centering
\includegraphics[scale=0.5]{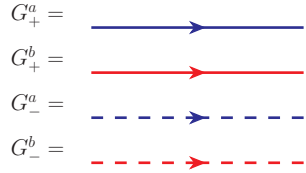}
\caption{(Color online) Feynman rules of the frees propagators.}
\label{greps}
\end{figure}

\begin{figure}[h]
\centering
\includegraphics[scale=0.5]{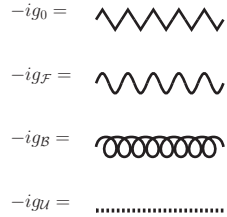}
\caption{(Color online) Feynman rules for the couplings.}
\label{interact}
\end{figure}

The figure \ref{greps} gives the Feynman rules for the free propagators. 
The Feynman rules for the interactions couplings can be written in the figure \ref{interact}.

\section{Energy-Momentum conservation, color exchanges and chiral symmetry}

As usual, we can represent the $\vec{p}=(p_{0},p)$ energy-momentum conservation 
in the Feynman diagrams by means of the ingoing $=$ outgoing relations
\begin{eqnarray}
\vec{p}_{in} = \vec{p}_{out}.
\end{eqnarray}
On the other hand, the quasi-1D system with intraforward, interforward, interbackscattering and interumklapp interactions
 presents some unusual interesting properties, corresponding to color and chiral indices.

\begin{figure}[h]
\centering
\includegraphics[scale=0.5]{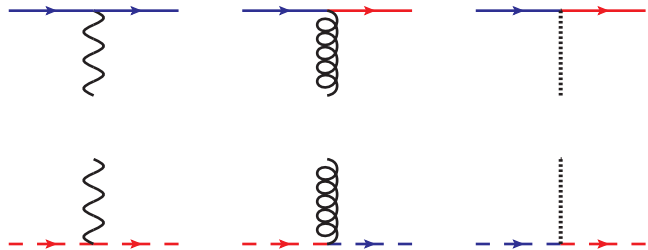}
\caption{(Color online) Color exchanges.}
\label{colors231w}
\end{figure}

\begin{figure}[h]
\centering
\includegraphics[scale=0.5]{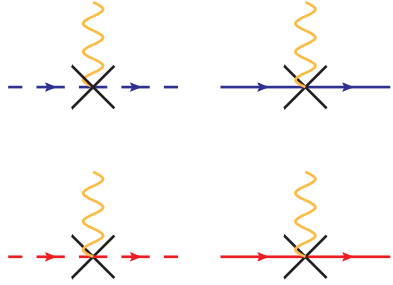}
\caption{(Color online) Vertex interactions.}
\label{colors234w}
\end{figure}

Color exchanges are mediated by interband interactions. In the interforward channel, there is 
no color exchange. In the interbackscattering and in the interumklapp channels there is always color exchange, where in 
the interumklapp the initial colors in oposite chiralities are the same and in the interbackscattering the initial colors 
in oposite chiralities are different (figure \ref{colors231w}).
As a consequence of this fact, the vertex interactions, where there is no color exchange, 
are associated to intraforward and interforward interactions (figure \ref{colors234w}).
The mediation of color exchanges by the band interactions also leads to a color number conservation, expressed 
by the fact of the number of ingoing colors is equal to the number of outgoing colors.

\begin{figure}[h]
\centering
\includegraphics[scale=0.5]{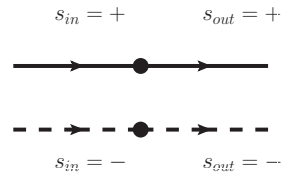}
\caption{(Color online) Chiral symmetry.}
\label{chiral}
\end{figure}

There is no chiral exchages, i.e., there is no interactions that meadiate change of chirality, implying (figure \ref{chiral})
\begin{eqnarray}
s_{in}= s_{out}.
\end{eqnarray}
Note that the chiral symmetry do not forbid $-ig_{4}$-type interactions, but excludes interactions of type $-ig_{1}$ 
and $-ig_{3}$ that are associated to intraband umklapp and backscattering interactions\cite{solyom}.

\section{One and two color polarization bubbles}

\begin{figure}[H]
\centering
\includegraphics[scale=0.7]{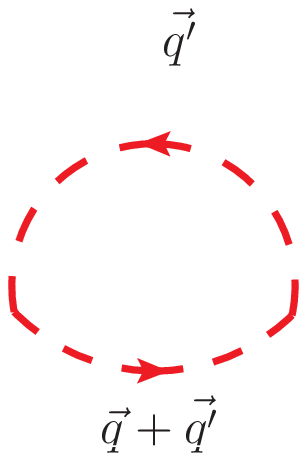}
\caption{(Color online) One-color polarization bubble.}
\label{polarizationFig1}
\end{figure}

One-color polarization bubbles are given by
\begin{eqnarray}
\chi_{s}^{\alpha}(\vec{q})= -\int_{\vec{q'}}iG_{s}^{\alpha}(\vec{q'})iG_{s}^{\alpha}(\vec{q'}+\vec{q}),
\end{eqnarray}
where the $-1$ sign is due to the fermionic loop. By computing the polarization in $b$
(figure \ref{polarizationFig1}) we arive at
\begin{eqnarray}
\chi_{-}^{b}(\vec{q})&=& -i\frac{q}{2\pi}G_{-}^{b}(q_{0},q-k_{F}^{b}). 
\end{eqnarray}
In the general case,
\begin{eqnarray}
\chi_{s}^{\alpha}(\vec{q})= i\frac{sq}{2\pi}G_{s}^{\alpha}(q_{0},q + sk_{F}^{\alpha}), \label{ak3456i}
\end{eqnarray}
where $s=\pm$, $\alpha=a,b$. 
\begin{figure}[H]
\centering
\includegraphics[scale=0.5]{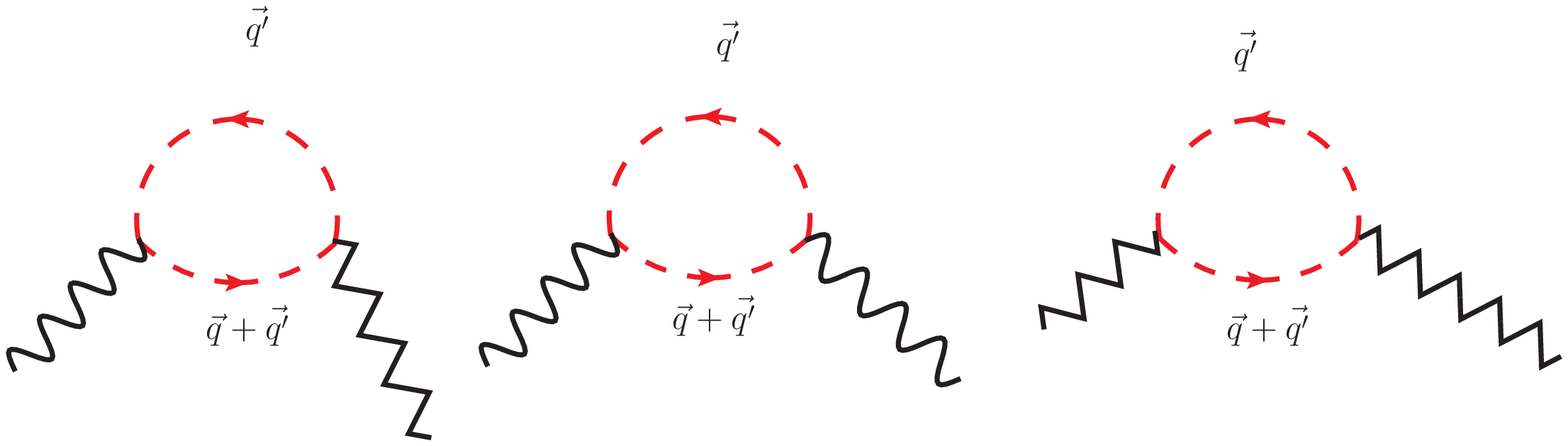}
\caption{(Color online) One-color polarization bubble in association to intraforward and interforward interactions.}
\label{polarizationFig1back}
\end{figure}
As there is no color exchange, the one-color polarization bubble is always 
associated to the intraforward and interforward interactions (figure \ref{polarizationFig1back}). 

\begin{figure}[H]
\centering
\includegraphics[scale=0.6]{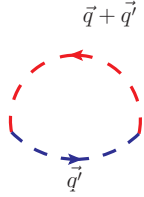}
\caption{(Color online) Two-color polarization bubble.}
\label{polarizationFig2}
\end{figure}

The two-color polarization bubbles are given by
\begin{eqnarray}
\chi_{s}^{\alpha\beta}(\vec{q})=-\int_{\vec{q'}}iG_{s}^{\alpha}(\vec{q'})iG_{s}^{\beta}(\vec{q'}+\vec{q}), \label{ak3456i1}
\end{eqnarray}
where $s=\pm$, $\alpha=a,b$ and $\alpha\neq\beta$. For instance, by computing the $ab$-polarization 
(figure \ref{polarizationFig2}), in the case where the Fermi velocities are equal, $\Delta v_{F,R} = 0$, we arrive at
\begin{eqnarray}
\chi_{-}^{ab}(\vec{q})=-\frac{i}{2\pi}(q+\Delta k_{F})G_{-}^{b}(q_{0},q-k_{F}^{a}). 
\end{eqnarray}
In the general case, we have
\begin{eqnarray}
\chi_{s}^{\alpha\beta}(\vec{q})= \frac{is}{2\pi}(q + k_{F}^{\alpha}-k_{F}^{\beta})G_{s}^{\beta}(q_{0},q + sk_{F}^{\alpha}), \label{ak3456i2}
\end{eqnarray}
where we have considered a momentum cutoff $\Lambda$ in the calculations.

\begin{figure}[H]
\centering
\includegraphics[scale=0.5]{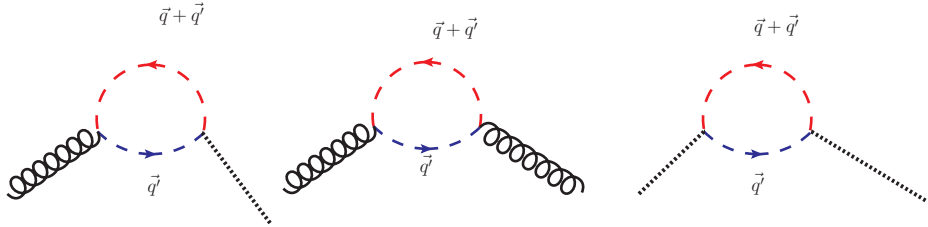}
\caption{(Color online) Two-color polarization bubble in association to interbackscattering and interumklapp interactions.}
\label{polarizationFig2back}
\end{figure}

\begin{figure}[h]
\centering
\includegraphics[scale=0.4]{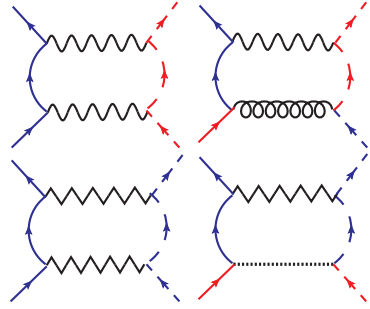}
\caption{(Color online) Polarization of different chiralities in association to two-particle interactions.}
\label{pppolback}
\end{figure}

As the two-color polarization bubbles involve color exchange, they always come in association to interbackscattering 
and interumklapp interactions (figure \ref{polarizationFig2back}).
\begin{figure}[h]
\centering
\includegraphics[scale=0.7]{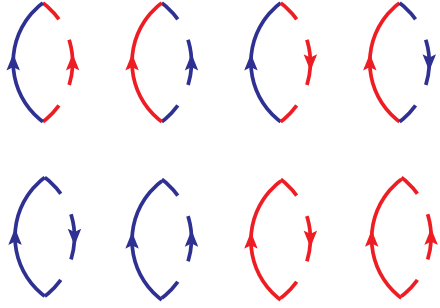}
\caption{(Color online) Polarizations of different chiralities}
\label{pppol}
\end{figure}
As there is no chirality exchanges, the diagrams of two-chiralities (figure \ref{pppol}) 
are always associated to two-particle interactions of opposite sides and depend on the external legs ( figure \ref{pppolback}). 
These type of polarization can be defined as follows
\begin{eqnarray}
\Pi_{ss'}^{\alpha\beta}(\vec{p})&=&\int_{\vec{q}}iG_{s}^{\alpha}(\vec{q})iG_{s'}^{\beta}(\vec{p}-\vec{q})\\
\chi_{ss'}^{\alpha\beta}(\vec{p})&=&\int_{\vec{q}}iG_{s}^{\alpha}(\vec{q})iG_{s'}^{\beta}(\vec{p}+\vec{q})
\end{eqnarray}
where $s\neq s'$, $s=\pm$, $s'=\pm$, $\alpha=a,b$ and $\beta=a,b$.

By calculating the scattering channels in 1-loops in 
the Fermi surface, we can obtain the explicitly the polarizations of different chirality at 
Fermi surface, $s=+$ and $s'=-$,
\begin{eqnarray}
\Pi_{+-}^{ab}(\vec{p}_{ab}) &=& -\frac{i}{\pi V_{F,R}}\ln\left(\frac{\Omega}{\omega}\right),
\\
\Pi_{+-}^{ba}(\vec{p}_{ab}) &=& -\frac{i}{\pi V_{F,R}}\ln\left(\frac{\Omega}{ V_{F,R}\Delta k_{F,R}}\right) \\
\Pi_{+-}^{a}(\vec{p}_{a}) &=& -\frac{i}{\pi V_{F,R}}\ln\left(\frac{\Omega}{\omega}\right),
\label{20f}
\\
\Pi_{+-}^{b}(\vec{p}_{a})&=& -\frac{i}{\pi V_{F,R}}\ln\left(\frac{\Omega}{\omega}\right),
\label{pb}
\\
\chi_{+-}^{ab}(\vec{p'}_{ab})&=& \frac{i}{\pi V_{F,R}}\ln\left(\frac{\Omega}{\omega}\right),
\\
\chi_{+-}^{ba}(\vec{p'}_{ab})&=& \frac{i}{\pi V_{F,R}}\ln\left(\frac{\Omega}{\omega}\right),
\\
\chi_{+-}^{a}(\vec{p'}_{a})&=& \frac{i}{\pi V_{F,R}}\ln\left(\frac{\Omega}{\omega}\right),
\\
\chi_{+-}^{b}(\vec{p'}_{a})&=& \frac{i}{\pi V_{F,R}}\ln\left(\frac{\Omega}{V_{F,R}\Delta k_{F,R}}\right)
\end{eqnarray}
where $\vec{p}_{a}$, $\vec{p}_{ab}$, $\vec{p'}_{ab}$ and 
$\vec{p'}_{a}$ are associated to the ingoing external legs 
\begin{eqnarray}
\vec{p}_{ab} &=& \vec{p}_{in,a} + \vec{p}_{in,b},\\
\vec{p'}_{ab} &=& \vec{p}_{in,a} - \vec{p}_{in,b}, \\
\vec{p}_{a} &=& \vec{p}_{in,a} + \vec{p}_{in,b}, \\
\vec{p'}_{a} &=& \vec{p}_{in,a} - \vec{p}_{in,b}, 
\end{eqnarray}
and $\Omega=V_{F,R}\Lambda$ is the energy cutoff associated to the momentum cutoff $\Lambda$.

\section{Self-energies and renormalized $\Gamma^{(2)}$ functions}

\begin{figure}[H]
\centering
\subfigure[]{
\includegraphics[scale=0.4]{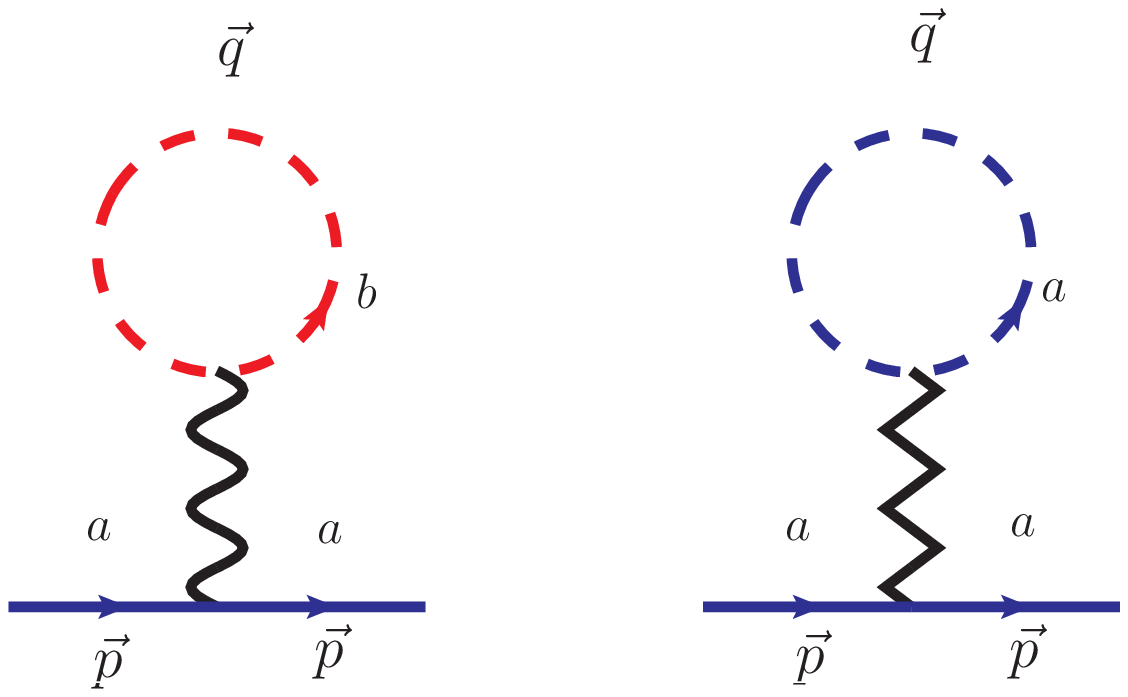}
\label{seap}
}
\subfigure[]{
\includegraphics[scale=0.4]{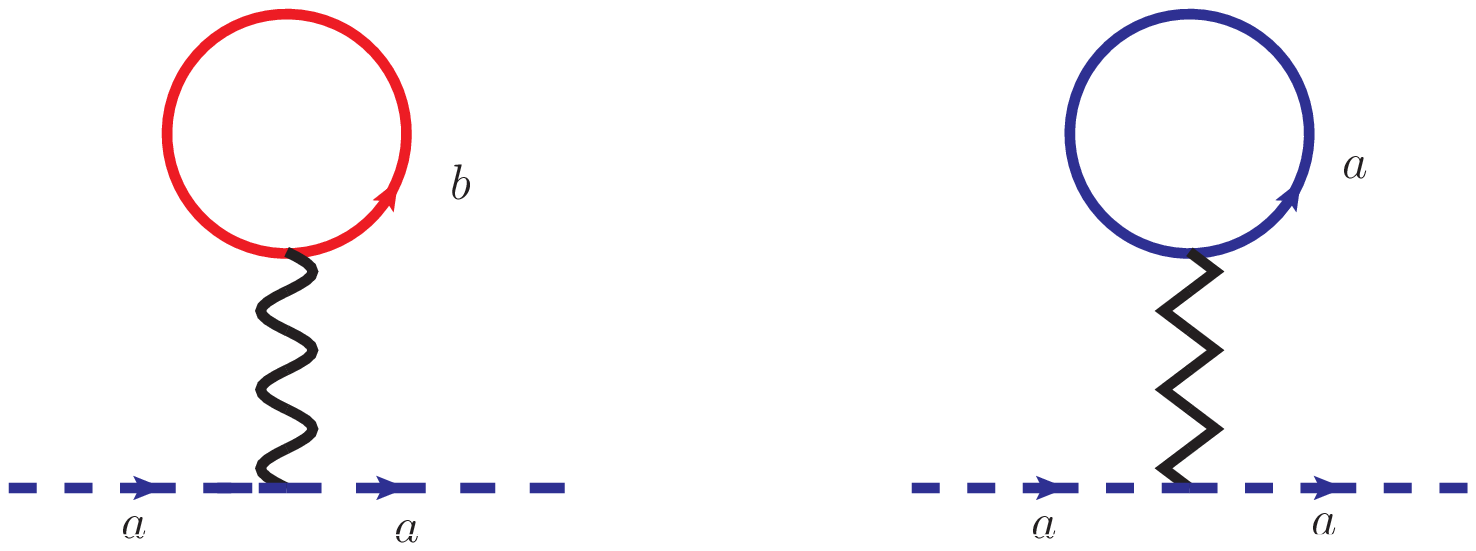}
\label{seam}
}
\subfigure[]{
\includegraphics[scale=0.4]{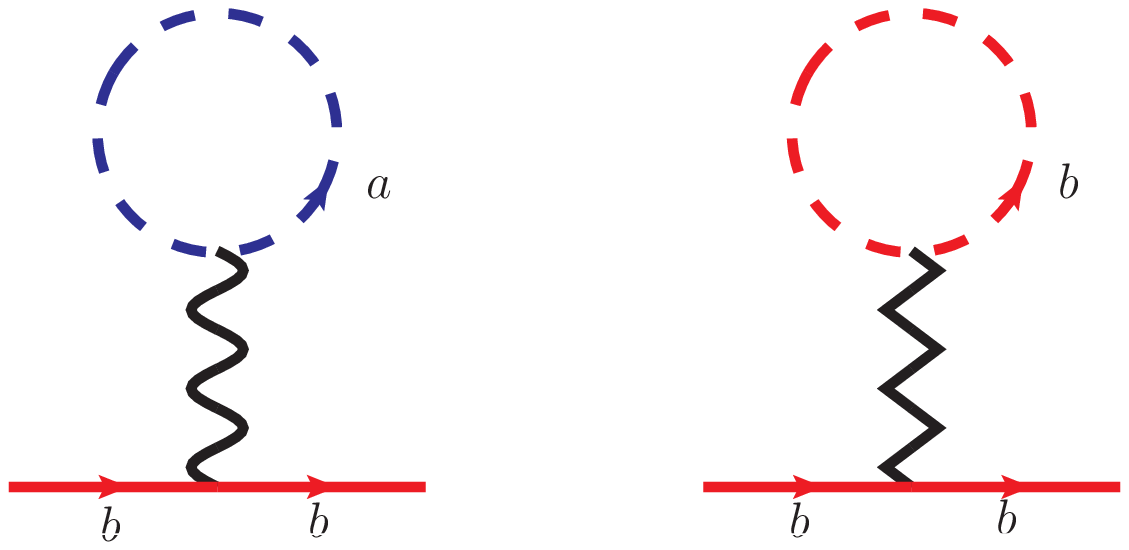}
\label{sebp}
}
\subfigure[]{
\includegraphics[scale=0.4]{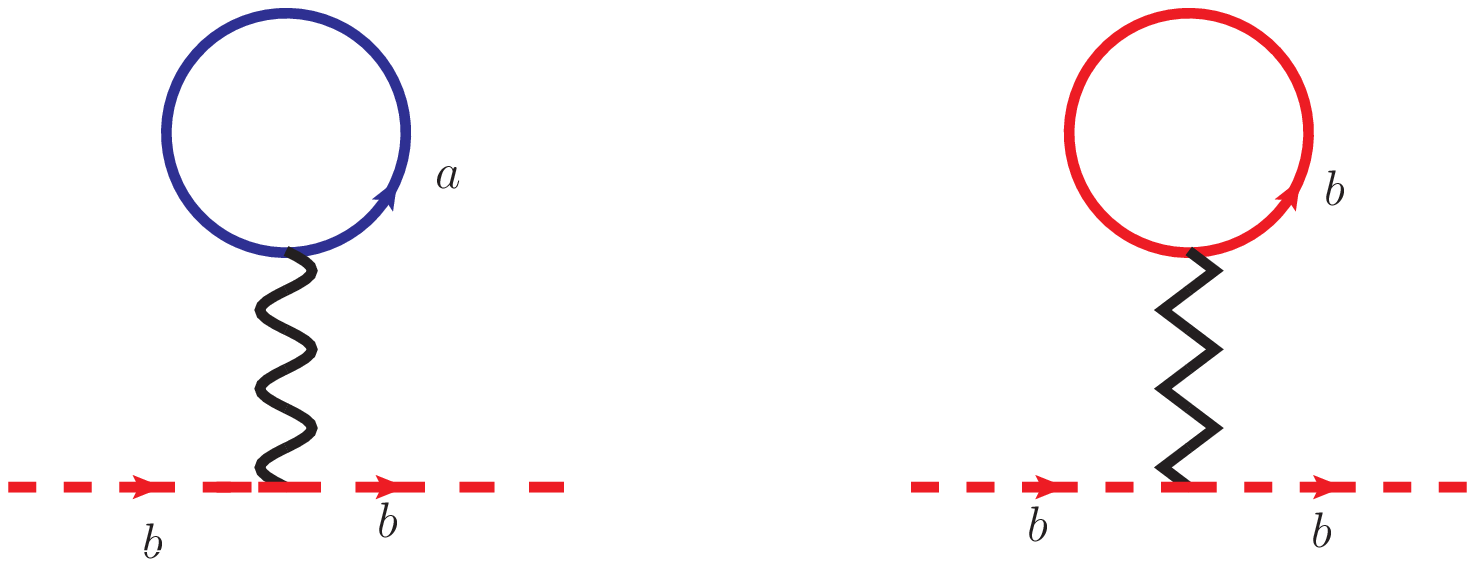}
\label{sebm}
}
\caption{(Color online) (a) Self-energy for the right hand side of band $a$.
(b) Self-energy for the left hand side of band $a$. (c) Self-energy for the right hand side of band $b$.
(d) Self-energy for the left hand side of band $b$}
\end{figure}
\begin{figure}[H]
\centering
\includegraphics[scale=0.5]{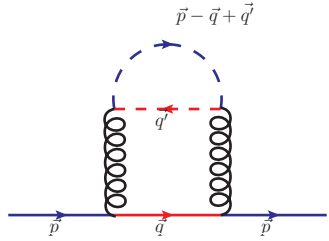}
\caption{(Color online) 2-loops self-energy diagram for interbackscattering interaction.}
\label{selfenergyFig3k}
\end{figure}

\begin{figure}[H]
\centering
\includegraphics[scale=0.5]{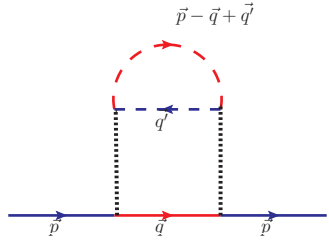}
\caption{2-loops self-energy diagram for interumklapp interaction.}
\label{selfenergyFig2k}
\end{figure}

\begin{figure}[H]
\centering
\includegraphics[scale=0.5]{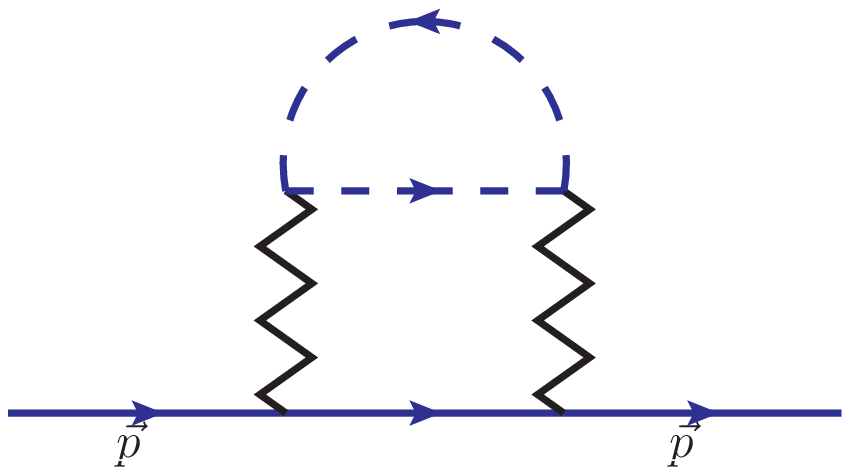}
\caption{(Color online) 2-loops self-energy diagram for intraforward interaction.}
\label{scatteringsx0k}
\end{figure}

\begin{figure}[H]
\centering
\includegraphics[scale=0.5]{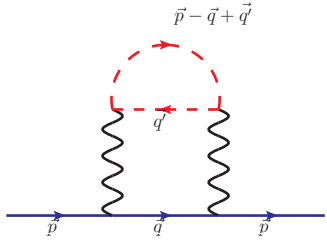}
\caption{(Color online) 2-loops self-energy diagram for interforward interaction.}
\label{selfenergyFigk}
\end{figure}

\begin{figure}[h]
\centering
\includegraphics[scale=0.5]{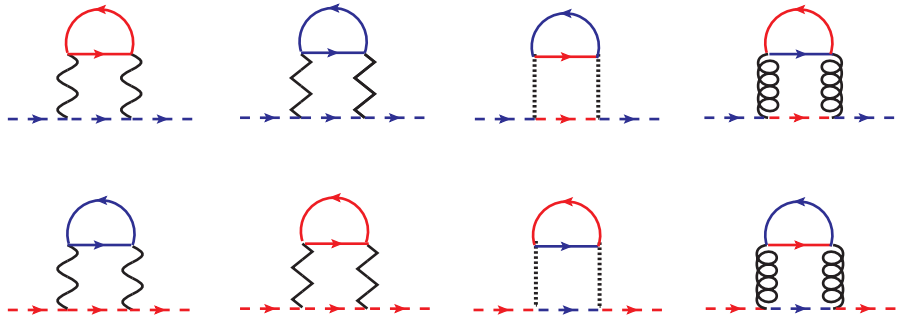}
\caption{(Color online) Self-energy diagrams in 2-loops, bands $a-$ and $b-$.}
\label{selfenergychiral}
\end{figure}

 
As the 1-loop self-energy diagrams involves only one-color polarization bubbles
 (see figures \ref{seap}, \ref{seam}, \ref{sebp} and \ref{sebm}), they are given by the following formula
\begin{eqnarray}
-i\Sigma_{s,R}^{\alpha}&=& g_{\mathcal{F},R}\int_{\vec{q}}G^{\beta}_{s,R}(\vec{q}) + g_{0,R}\int_{\vec{q}}G^{\alpha}_{s,R}(\vec{q}), \label{s1} 
\end{eqnarray}
where $\alpha=a,b$, $\beta=a,b$, $\alpha\neq\beta$.
By integrating in $\vec{q}$ the self-energy equations (\ref{s1}), including the momentum cuttoff $\Lambda$, in the Fermi surface, 
we arrive explicitly at 
\begin{eqnarray}
-i\Sigma_{-,R}^{a}&=& i\left( \bar{g}_{\mathcal{F},R} + \bar{g}_{0,R}\right)\Omega, \\
-i\Sigma_{+,R}^{a}&=& i\left( \bar{g}_{\mathcal{F},R} + \bar{g}_{0,R}\right)\Omega, \\
-i\Sigma_{-,R}^{b}&=& i\left( \bar{g}_{\mathcal{F},R} + \bar{g}_{0,R}\right)\Omega, \\
-i\Sigma_{+,R}^{b}&=& i\left( \bar{g}_{\mathcal{F},R} + \bar{g}_{0,R}\right)\Omega, 
\end{eqnarray}
where $\bar{g}_{\mathcal{F},R}=\frac{g_{\mathcal{F},R}}{\pi V_{F,R}}$, $\bar{g}_{0,R}=\frac{g_{0,R}}{\pi V_{F,R}}$ 
and $\Omega=\Lambda V_{F}$. Then, in 1-loop the self-energy contributions are given by intraforward and interforward interactions.
If we consider the 2-loops self-energy contributions, we have additional contributions including interbackscattering and interumklapp
 interactions, with two-color polarization bubbles and color exchanges (figures \ref{selfenergyFig3k} and \ref{selfenergyFig2k}).
The contribution of the intraforward interaction to 2-loops self-energy is given by (figure \ref{scatteringsx0k}) 
\begin{eqnarray}
-i\Sigma_{+,R}^{aa}(\vec{p})=(-ig_{0,R})^{2}\int_{\vec{q}}iG_{+,R}^{a}(\vec{q})\chi_{-,R}^{a}(\vec{p}-\vec{q}).
\end{eqnarray}
At the Fermi surface, $p_{0}=\omega$ and $p=k_{F,R}^{a}$,
\begin{eqnarray}
-i\Sigma_{+,R}^{aa}(k_{F,R}^{a},\omega)&=&\frac{ig_{0,R}^{2}\omega}{2(\pi V_{F,R})^{2}}
\ln\left(\frac{\Omega}{\omega}\right).
\end{eqnarray}
The self-energy contribution of the interforward interaction in 2-loops (\ref{selfenergyFigk}) is given by
\begin{eqnarray}
-i\Sigma_{+,R}^{ab}(\vec{p})=(-ig_{\mathcal{F},R})^{2}\int_{\vec{q}}iG_{+,R}^{a}(\vec{q})\chi_{-,R}^{b}(\vec{p}-\vec{q}),
\end{eqnarray}
that reduces, at the Fermi surface, $p_{0}=\omega$ and $p=k_{F}^{a}$, for equal Fermi velocities $v_{F,R}$, to the following
\begin{eqnarray}
-i\Sigma_{+,R}^{ab}(k_{F}^{a},\omega)&=&\frac{ig_{\mathcal{F},R}^{2} \omega}{2(\pi V_{F,R})^{2}}\ln\left(\frac{\Omega}{\omega}\right).
\end{eqnarray}
Both interforward and intraforward selfenergy contributions are one-color polarization bubbles contributions to the selfenergy, 
with no color exchange.
Now, considering the calculation of the two color contributions at 2-loops to the selfenergy, we have the
 interbackscattering interaction contribution (figure \ref{selfenergyFig3k}),
\begin{eqnarray}
\Sigma_{+-}^{bba}=(-ig_{B})^{2}\int_{\vec{q}}iG_{+}^{b}(\vec{q})\chi_{-}^{ba}(\vec{p}-\vec{q})
\end{eqnarray}
that at the Fermi surface $\vec{p}=(\omega,k_{F,R}^{a})$, in the case of same
 Fermi velocities, is given by
\begin{eqnarray}
-i\Sigma_{+,R}^{bba}(k_{F,R}^{a},\omega)&=&\frac{ig_{\mathcal{B},R}^{2}V_{F,R}\Delta k_{F,R}}{2(\pi V_{F,R})^{2}}\ln\left(\frac{\Omega}{V_{F,R}\Delta k_{F,R}}\right)\nonumber \\
&+&\frac{ig_{\mathcal{B},R}^{2}\omega}{2(\pi V_{F,R})^{2}}\ln\left(\frac{\Omega}{V_{F,R}\Delta k_{F,R}}\right),
\end{eqnarray}
and interumklapp interaction contribution (figure \ref{selfenergyFig2k}) 
\begin{eqnarray}
-i\Sigma_{+,R}^{bab}=(-ig_{\mathcal{U},R})^{2}\int_{\vec{q}}iG_{+,R}^{b}(\vec{q})\chi_{-,R}^{ab}(\vec{p}-\vec{q}),
\end{eqnarray}
that at the Fermi surface $\vec{p}=(k_{F,R}^{a},\omega)$, , in the case of same
 Fermi velocities, is
\begin{eqnarray}
-i\Sigma_{+,R}^{bab}(k_{F}^{a},\omega)= \frac{ig_{\mathcal{U},R}^{2}\omega}{2(\pi V_{F,R})^{2}}\ln\left(\frac{\Omega}{\omega}\right).
\end{eqnarray}
\begin{figure}[h]
\centering
\includegraphics[scale=0.5]{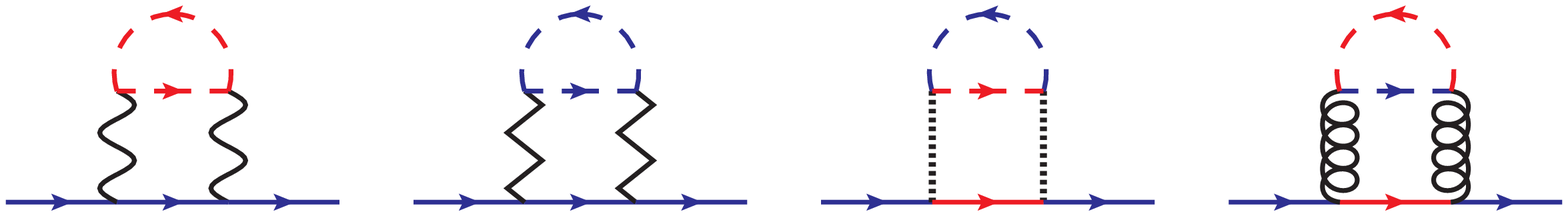}
\caption{(Color online) Self-energy diagrams in 2-loops, band $a+$.}
\label{selfenergy}
\end{figure}
Now, we can write the total contribution to the selfenergy in 2-loops for the right-hand side of the band $a$, 
at the Fermi surface (FS),
\begin{eqnarray}
-i\Sigma_{+,R}^{a}|_{FS} &=& \frac{ig_{\mathcal{F},R}^{2}\omega}{2(\pi V_{F,R})^{2}}
\ln\left(\frac{\Omega}{\omega}\right) + \frac{ig_{0,R}^{2}\omega}{2(\pi V_{F,R})^{2}}\ln\left(\frac{\Omega}{\omega}\right) \nonumber \\
&+& \frac{ig_{\mathcal{B},R}^{2}(V_{F,R}\Delta k_{F,R} + \omega)}{2(\pi V_{F,R})^{2}}\ln\left(\frac{\Omega}{V_{F,R}\Delta k_{F,R}}\right)
\nonumber \\ &+& \frac{ig_{\mathcal{U},R}^{2}\omega}{2(\pi V_{F,R})^{2}}\ln\left(\frac{\Omega}{\omega}\right)
\end{eqnarray} 
and the corresponding one-particle irreducible function $\Gamma^{(2)}$ at Fermi surface
\begin{eqnarray}
\Gamma^{(2)}_{a+R}(\omega,k_{F}^{a})&=& \omega - \left(\bar{g}_{0R}+\bar{g}_{\mathcal{F},R}\right)\Omega 
\nonumber \\ 
&+& \omega\left[\frac{\bar{g}_{\mathcal{F},R}^{2}}{2}
+\frac{\bar{g}_{0,R}^{2}}{2} + \frac{\bar{g}_{\mathcal{U},R}^{2}}{2}\right]\ln\left(\frac{\Omega}{\omega}\right) \nonumber \\
&+& \omega\frac{\bar{g}_{\mathcal{B},R}^{2}}{2}\ln\left(\frac{\Omega}{V_{F}\Delta k_{F}}\right)\nonumber \\
&+& (V_{F}\Delta k_{F})\frac{\bar{g}_{\mathcal{B},R}^{2}}{2}\ln\left(\frac{\Omega}{V_{F}\Delta k_{F}}\right)  \nonumber\\
&+& \omega\delta Z - k_{F,R}^{a} v_{F,R}\delta Z + v_{F,R}\delta k_{F,R}^{a}, \nonumber \\
\label{gamma21}
\end{eqnarray}
where $\delta Z$, $\delta k_{F,R}^{a}$ are the corresponding counterterms, introduced in the corresponding 
renormalized lagrangean.
\begin{figure}[h]
\centering
\includegraphics[scale=0.5]{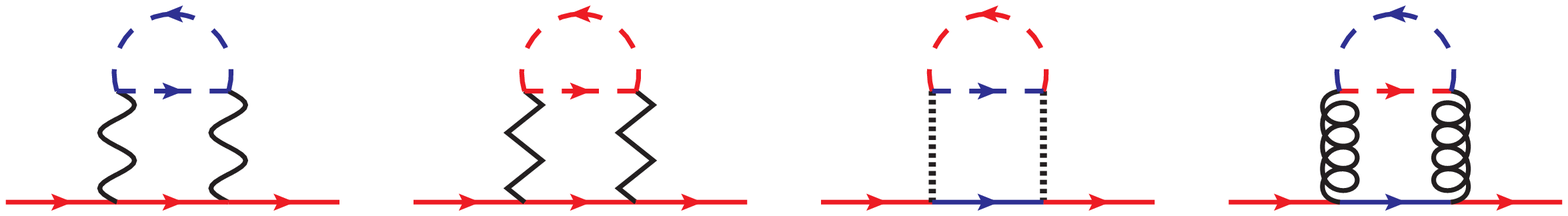}
\caption{(Color online) Self-energy diagrams in 2-loops, band $b+$.}
\label{selfenergyb}
\end{figure}
In the case of the band $b$ (figure \ref{selfenergyb}), 
\begin{eqnarray}
\Gamma^{(2)}_{b+R}(\omega,k_{F}^{b})&=& \omega - \left(\bar{g}_{0R}+\bar{g}_{\mathcal{F},R}\right)\Omega 
\nonumber \\ 
&+& \omega\left[\frac{\bar{g}_{\mathcal{F},R}^{2}}{2}
+\frac{\bar{g}_{0,R}^{2}}{2} + \frac{\bar{g}_{\mathcal{U},R}^{2}}{2}\right]\ln\left(\frac{\Omega}{\omega}\right) \nonumber \\
&+& \omega\frac{\bar{g}_{\mathcal{B},R}^{2}}{2}\ln\left(\frac{\Omega}{V_{F,R}\Delta k_{F,R}}\right)\nonumber \\
&-& (V_{F,R}\Delta k_{F,R})\frac{\bar{g}_{\mathcal{B},R}^{2}}{2}\ln\left(\frac{\Omega}{V_{F,R}\Delta k_{F,R}}\right)  \nonumber\\
&+& \omega\delta Z - k_{F,R}^{b} v_{F,R}\delta Z + v_{F,R}\delta k_{F,R}^{b}, \nonumber \\
\label{gamma22}
\end{eqnarray}
where now we have the counterterm $\delta k_{F,R}^{b}$ and 
we have introduced the notation $\bar{g}_{R}=g_{R}/\pi V_{F,R}$ for each corresponding coupling,
\begin{eqnarray}
\bar{g}_{0,R} &=& g_{0,R}(\omega)/\pi V_{F,R},  \\ 
\bar{g}_{\mathcal{F},R} &=& g_{\mathcal{F},R}(\omega)/\pi V_{F,R},  \\
\bar{g}_{\mathcal{B},R} &=&  g_{\mathcal{B},R}(\omega)/\pi V_{F,R}, \\
\bar{g}_{\mathcal{U},R} &=& g_{\mathcal{U},R}(\omega)/\pi V_{F,R}.  
\end{eqnarray}
Similar calculations are applied for the opposite chiralities $\Gamma^{(2)}_{a-R}$ and $\Gamma^{(2)}_{b-R}$ 
(figure \ref{selfenergychiral}).

\section{Physical prescription for $\Gamma^{(2)}_{R}$ and RG flow equation for $\Delta k_{F,R}$}

Now we consider the renormalization group precription for $\Gamma^{(2)}_{R}$ at the corresponding Fermi surfaces. 
At each band in the corresponding Fermi surfaces it takes the value $\Gamma^{(2)}_{R}=\omega$, such that 
from $\Gamma^{(2)}_{a+R}(\omega,k_{F}^{a})$, eq. (\ref{gamma21}) and $\Gamma^{(2)}_{b+R}(\omega,k_{F}^{b})$, eq. (\ref{gamma22}), 
we arrive at the following counterterms corresponding to the quasiparticle weight $Z$
\begin{eqnarray}
\delta Z &=& -\left[\frac{\bar{g}_{\mathcal{F},R}^{2}}{2}
+\frac{\bar{g}_{0,R}^{2}}{2} + \frac{\bar{g}_{\mathcal{U},R}^{2}}{2} + \frac{\bar{g}_{\mathcal{B},R}^{2}}{2} \right]\ln\left(\frac{\Omega}{\omega}\right), \nonumber \\
\end{eqnarray}
and the respective Fermi momentum counterterms
\begin{eqnarray}
\delta k_{F,R}^{a} &=& k_{F,R}^{a}\delta Z + \left(\bar{g}_{0,R}+\bar{g}_{\mathcal{F},R}\right)\frac{\Omega}{v_{F,R}} 
\nonumber \\
&-&\Delta k_{F,R}\bar{g}_{\mathcal{B},R}^{2}\ln\left(\frac{\Omega}{V_{F,R}\Delta k_{F,R}}\right),
\end{eqnarray}
\begin{eqnarray}
\delta k_{F,R}^{b} &=& k_{F,R}^{b}\delta Z + \left(\bar{g}_{0,R}+\bar{g}_{\mathcal{F},R}\right)\frac{\Omega}{v_{F,R}} \nonumber \\
&+& \Delta k_{F,R}\bar{g}_{\mathcal{B},R}^{2}\ln\left(\frac{\Omega}{V_{F,R}\Delta k_{F,R}}\right).
\end{eqnarray}
By adding $k_{F,R}^{\alpha}$, $\alpha=a,b$, in each corresponding counterterm for the 
Fermi momentum and taking into account the relation $\delta Z = Z -1$, we can 
write 
\begin{eqnarray}
\delta k_{F,R}^{a} + k_{F,R}^{a} &=& k_{F,R}^{a} Z + \left(\bar{g}_{0,R}+\bar{g}_{\mathcal{F},R}\right)\frac{\Omega}{v_{F,R}} \nonumber \\
&-&\Delta k_{F}\bar{g}_{bR}^{2}\ln\left(\frac{\Omega}{V_{F,R}\Delta k_{F,R}}\right),
\end{eqnarray}
\begin{eqnarray}
\delta k_{F,R}^{b} + k_{F,R}^{b} &=& k_{F,R}^{b} Z + \left(\bar{g}_{0,R}+\bar{g}_{\mathcal{F},R}\right)\frac{\Omega}{v_{F,R}} 
\nonumber \\ &+&\Delta k_{F,R}\bar{g}_{bR}^{2}\ln\left(\frac{\Omega}{V_{F,R}\Delta k_{F,R}}\right).
\end{eqnarray}
Taking into account the relation between renormalized and bare Fermi moments, eq. (\ref{thiskbare}),
\begin{eqnarray}
k_{F,\texttt{bare}}^{a}&=&\frac{\delta k_{F,R}^{a} + k_{F,R}^{a}}{Z},  \\
k_{F,\texttt{bare}}^{b}&=&\frac{\delta k_{F,R}^{b} + k_{F,R}^{b}}{Z},  
\end{eqnarray}
we can write the relation between $\Delta k_{F,\texttt{bare}}$ and $\Delta k_{F,R}$, eq. (\ref{thisdkbare}), 
explicitly 
\begin{eqnarray}
\Delta k_{F,\texttt{bare}} &=& k_{F,\texttt{bare}}^{b}-k_{F,\texttt{bare}}^{a} \nonumber \\
&=& \Delta k_{F,R}\left[ 1 + 2\frac{\bar{g}_{\mathcal{B},R}^{2}}{Z}\ln\left(\frac{\Omega}{V_{F,R}\Delta k_{F,R}}\right)\right]. \nonumber \\
\end{eqnarray}
As the bare quantities do not flow under RG, we arrive at the corresponding flow equation 
for $\Delta k_{F,R}$. We have 
\begin{eqnarray}
\omega\frac{d\Delta k_{F,\texttt{bare}}}{d\omega}&=&\omega\frac{d\Delta k_{F,R}}{d\omega}\left[ 1 + 2\frac{\bar{g}_{\mathcal{B},R}^{2}}{Z}\ln\left(\frac{\Omega}{V_{F,R}\Delta k_{F,R}}\right)\right]\nonumber \\
&+& \Delta k_{F,R}[ -2\frac{\bar{g}_{\mathcal{B},R}^{2}}{Z^{2}}\omega\frac{dZ}{d\omega}\ln\left(\frac{\Omega}{V_{F,R}\Delta k_{F,R}}\right) \nonumber \\
&-&2\frac{\bar{g}_{\mathcal{B},R}^{2}}{Z}\frac{1}{\Delta k_{F,R}}\omega\frac{d\Delta k_{F,R}}{d\omega}]\nonumber \\
&=& 0.
\end{eqnarray}
Including the anomalous dimension
\begin{eqnarray}
\gamma =\frac{\omega}{Z}\frac{dZ}{d\omega}
\end{eqnarray}
and multiplying both sides by $Z$, we arrive at
\begin{eqnarray}
\omega\frac{d\Delta k_{F,R}}{d\omega}
&=& \frac{2\Delta k_{F,R}\bar{g}_{\mathcal{B},R}^{2}\gamma\ln\left(\frac{\Omega}{V_{F,R}\Delta k_{F,R}}\right)}{Z + 2\bar{g}_{\mathcal{B},R}^{2}\left[\ln\left(\frac{\Omega}{V_{F,R}\Delta k_{F,R}}\right) -1\right]}.\nonumber \\
\end{eqnarray}
This can be also written in the following simplified form
\begin{eqnarray}
\omega\frac{d \ln\Delta k_{F,R}}{d\omega}
&=& \frac{\gamma}{1 -\frac{1}{\ln\left(\Lambda/\Delta k_{F,R}\right)}\left[ 1-\frac{Z}{2\bar{g}_{\mathcal{B},R}^{2}} \right]},
\label{deltakfrg}
\end{eqnarray}
where we take into account the cutoff $\Lambda=\Omega/V_{F,R}$.

In order to simplify the above equation and achieve to a cutoff independent equation, we can consider the cutoff sufficiently large
($\Lambda \rightarrow \infty$) such that the above equation is simplified to
\begin{eqnarray}
\omega\frac{d\ln\Delta k_{F,R}}{d\omega}=\gamma,
\end{eqnarray}
or, equivalently,
\begin{eqnarray}
\omega\frac{d\Delta k_{F,R}}{d\omega}=\gamma\Delta k_{F,R}.
\end{eqnarray}

\section{Two-particle irreducible function $\Gamma_{R}^{(4)}$ and the counterterms for the scattering channels}

Now, we consider the scattering channels corresponding to the interactions of intraforward, interforward, interbackscattering
and interumklapp, corresponding to the two-particle irreducible functions $\Gamma_{R}^{(4)}$ for each scattering channel.  

\begin{figure}[h]
\centering
\includegraphics[scale=0.4]{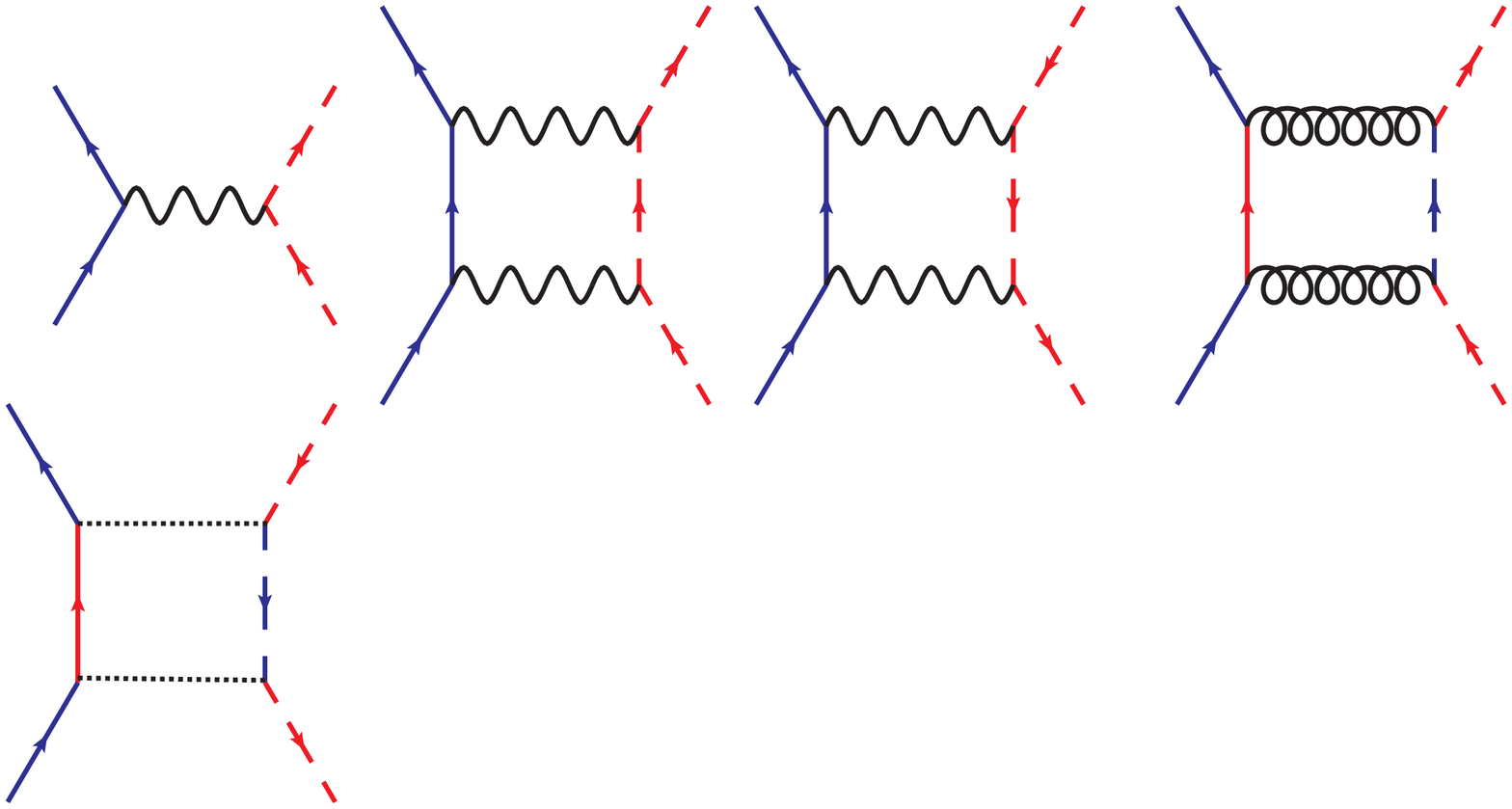}
\caption{(Color online) Interforward channel diagrams in 1-loop.}
\label{1fchk}
\end{figure}
\begin{figure}[h]
\centering
\includegraphics[scale=0.4]{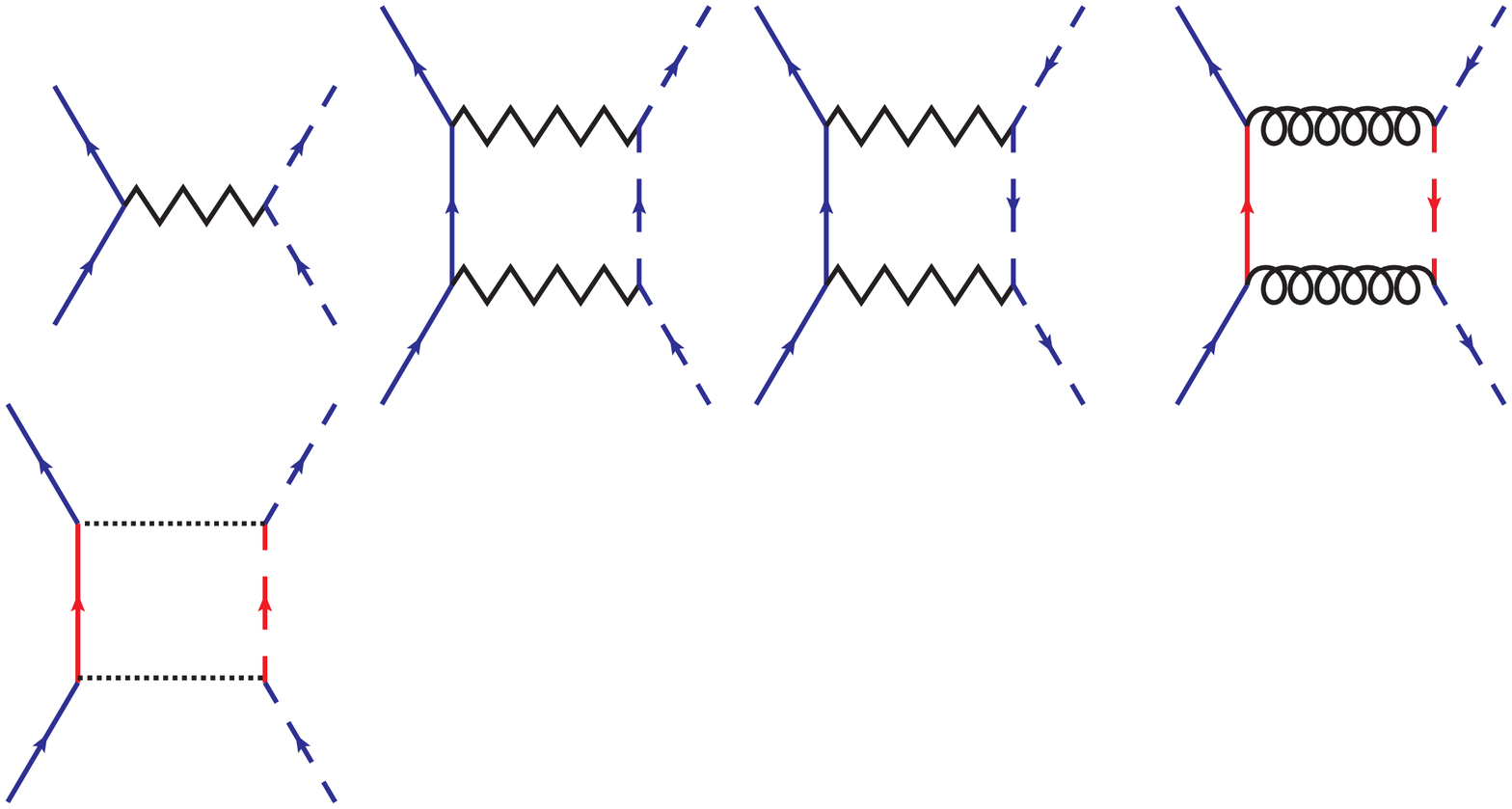}
\caption{(Color online) Intraforward channel diagrams in 1-loop.}
\label{12chk}
\end{figure}
\begin{figure}[h]
\centering
\includegraphics[scale=0.4]{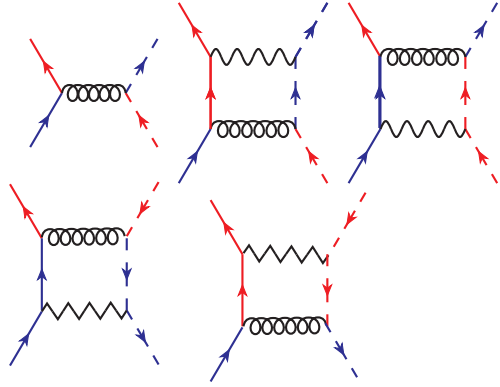}
\caption{(Color online) Interbackscatterig channel diagrams in 1-loop.}
\label{1bchk}
\end{figure}
\begin{figure}[h]
\centering
\includegraphics[scale=0.4]{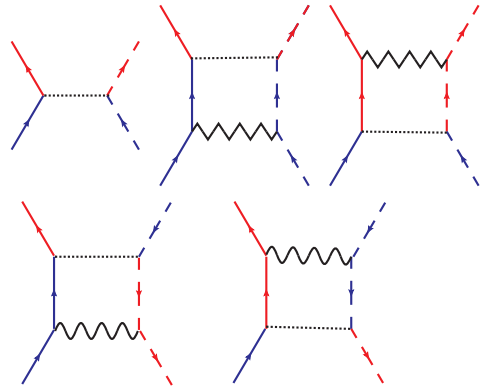}
\caption{(Color online) Interumklapp channel diagrams in 1-loop.}
\label{1uchk}
\end{figure}
In the intrafrontal channel (figure \ref{12chk}), the two-particle irreducible function in 1-loop has the following 
contribution  at the Fermi surface
\begin{eqnarray}
\Gamma_{0,R}^{(4)} &=& -ig_{0,R} -\frac{ig_{\mathcal{B},R}^{2}}{\pi V_{F}}\ln\left(\frac{\Omega}{V_{F,R}\Delta k_{F,R}}\right) \nonumber \\
&+& \frac{ig_{\mathcal{U},R}^{2}}{\pi V_{F,R}}\ln\left(\frac{\Omega}{\omega}\right) -i\delta g_{0,R},
\label{g0ch1}
\end{eqnarray}
where $-i\delta g_{0,R}$ is the corresponding counterterm for the intraforward channel in 1-loop. 
In the interfrontal channel (figure \ref{1fchk}) the two-particle irreducible function in 1-loop has the following 
contribution at the Fermi surface
\begin{eqnarray}
\Gamma_{\mathcal{F},R}^{(4)} &=& -ig_{\mathcal{F},R} + \frac{ig_{\mathcal{B},R}^{2}}{\pi V_{F,R}}\ln\left(\frac{\Omega}{V_{F,R}\Delta k_{F}}\right) \nonumber \\ 
&-&\frac{ig_{\mathcal{U},R}^{2}}{\pi V_{F,R}}\ln\left(\frac{\Omega}{\omega}\right) -i\delta g_{\mathcal{F},R},
\label{fch2}
\end{eqnarray}
where $-i\delta g_{\mathcal{F},R}$ is the corresponding counterterm for the interforward channel in 1-loop.
In the interbackscattering channel (figure \ref{1bchk}), we have the following 
contribution at the Fermi surface
\begin{eqnarray}
\Gamma_{\mathcal{B},R}^{(4)} &=& -ig_{\mathcal{B},R} + \frac{ig_{\mathcal{B},R}g_{\mathcal{F},R}}{\pi V_{F,R}}\ln\left(\frac{\Omega}{V_{F,R}\Delta k_{F,R}}\right) \nonumber \\
&+& \frac{ig_{\mathcal{F},R}g_{\mathcal{B},R}}{\pi V_{F,R}}\ln\left(\frac{\Omega}{\omega}\right) -\frac{ig_{\mathcal{B},R}g_{0,R}}{\pi V_{F,R}}\ln\left(\frac{\Omega}{V_{F}\Delta k_{F,R}}\right) \nonumber \\
&-& \frac{ig_{0,R}g_{\mathcal{B},R}}{\pi V_{F,R}}\ln\left(\frac{\Omega}{\omega}\right) -i\delta g_{\mathcal{B},R}, 
\label{Bch1}
\end{eqnarray}
where $-i\delta g_{\mathcal{B},R}$ is the corresponding counterterm for the interbackscattering channel in 1-loop.
In the interumklapp channel (figure \ref{1uchk}), we have the following 
contribution at the Fermi surface
\begin{eqnarray}
\Gamma_{\mathcal{U},R}^{(4)} &=& -ig_{\mathcal{U},R} + 2i\frac{g_{\mathcal{U},R}g_{0.R}}{\pi V_{F,R}}\ln\left(\frac{\Omega}{\omega}\right) \nonumber \\
&-& 2i\frac{g_{\mathcal{U},R}g_{\mathcal{F},R}}{\pi V_{F,R}}\ln\left(\frac{\Omega}{\omega}\right) -i\delta g_{\mathcal{U},R}, 
\label{uch1}
\end{eqnarray}
where $-i\delta g_{\mathcal{U},R}$ is the corresponding counterterm for the interumklapp channel in 1-loop.

\begin{figure}[!htb]
\centering
\includegraphics[scale=0.4]{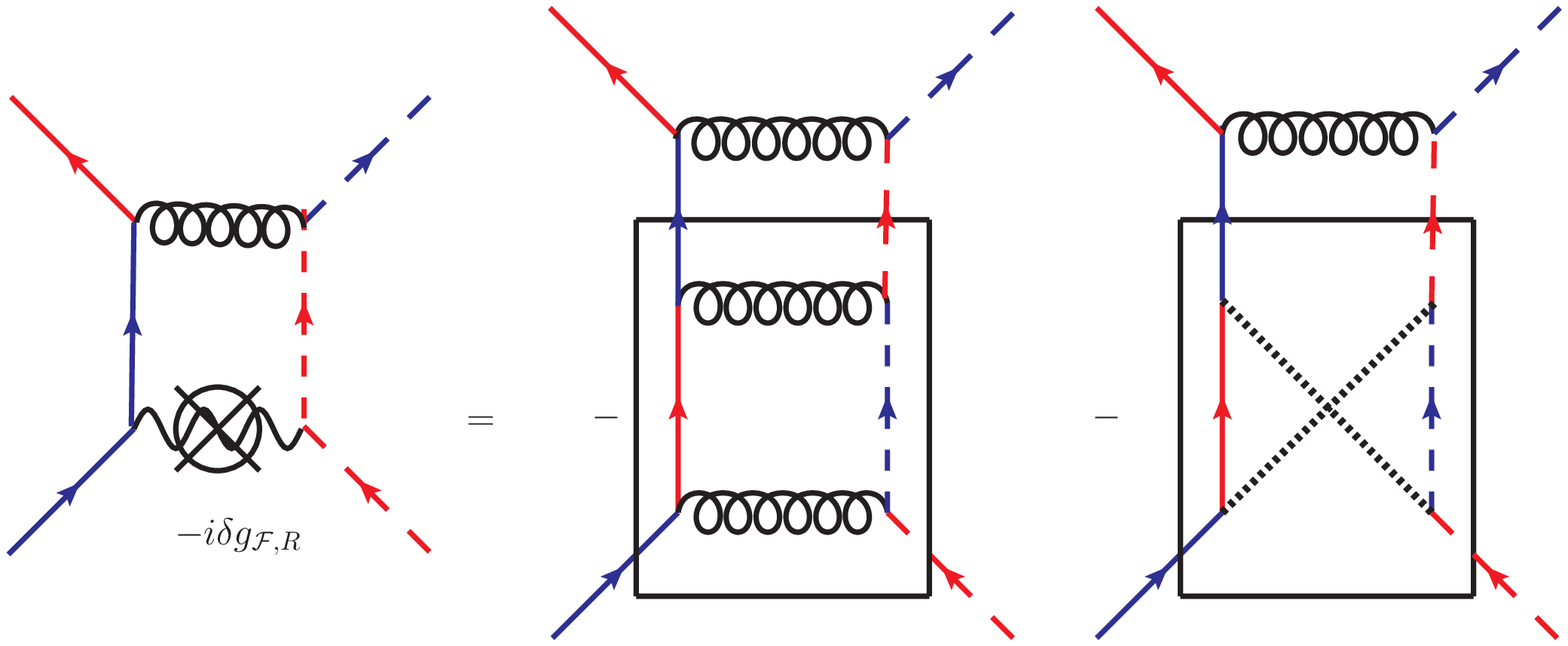}
\caption{(Color online) Counterterm diagram in 2-loops.}
\label{cGfDiagram2L1}
\end{figure}

\begin{figure}[!htb]
\centering
\includegraphics[scale=0.3]{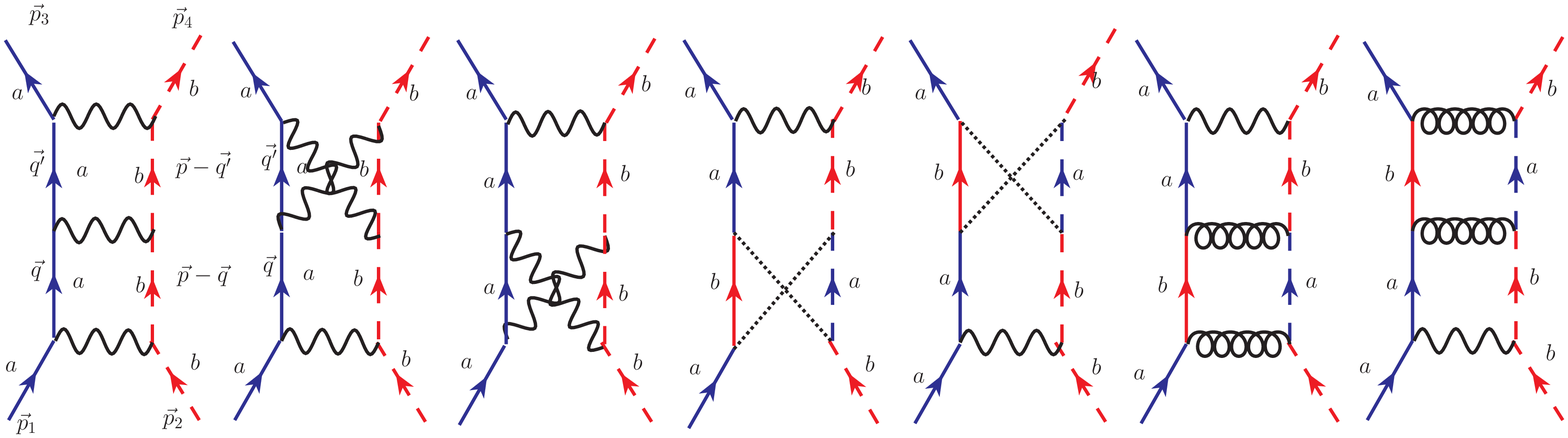}
\includegraphics[scale=0.3]{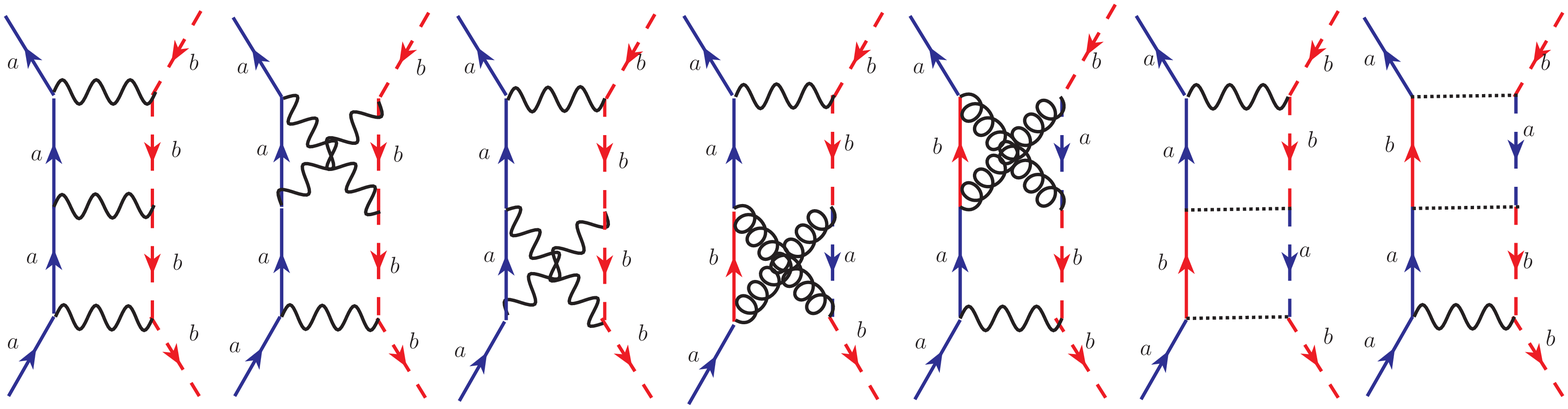}
\includegraphics[scale=0.3]{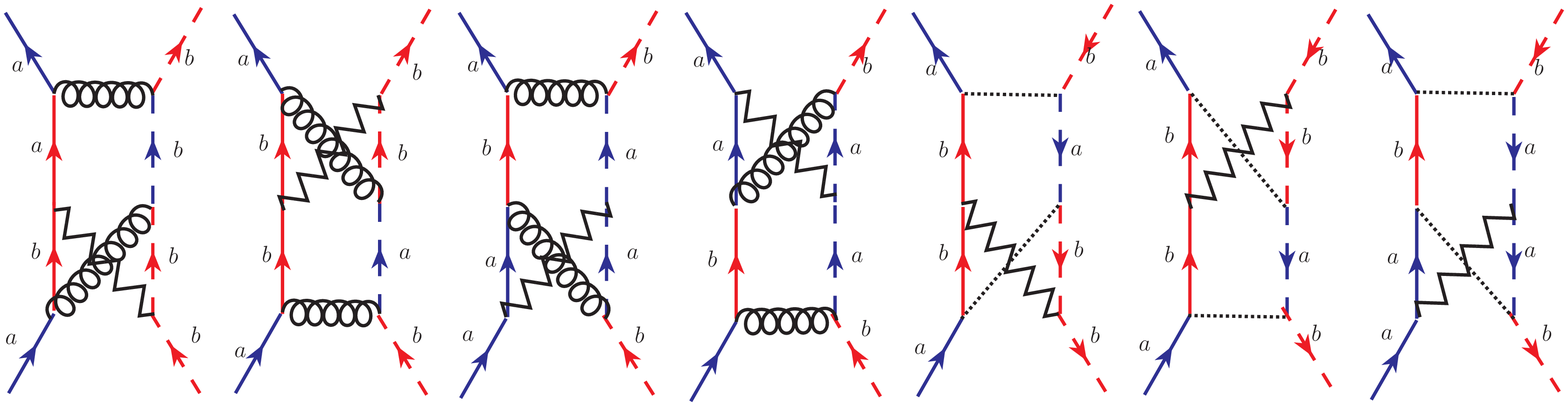}
\includegraphics[scale=0.3]{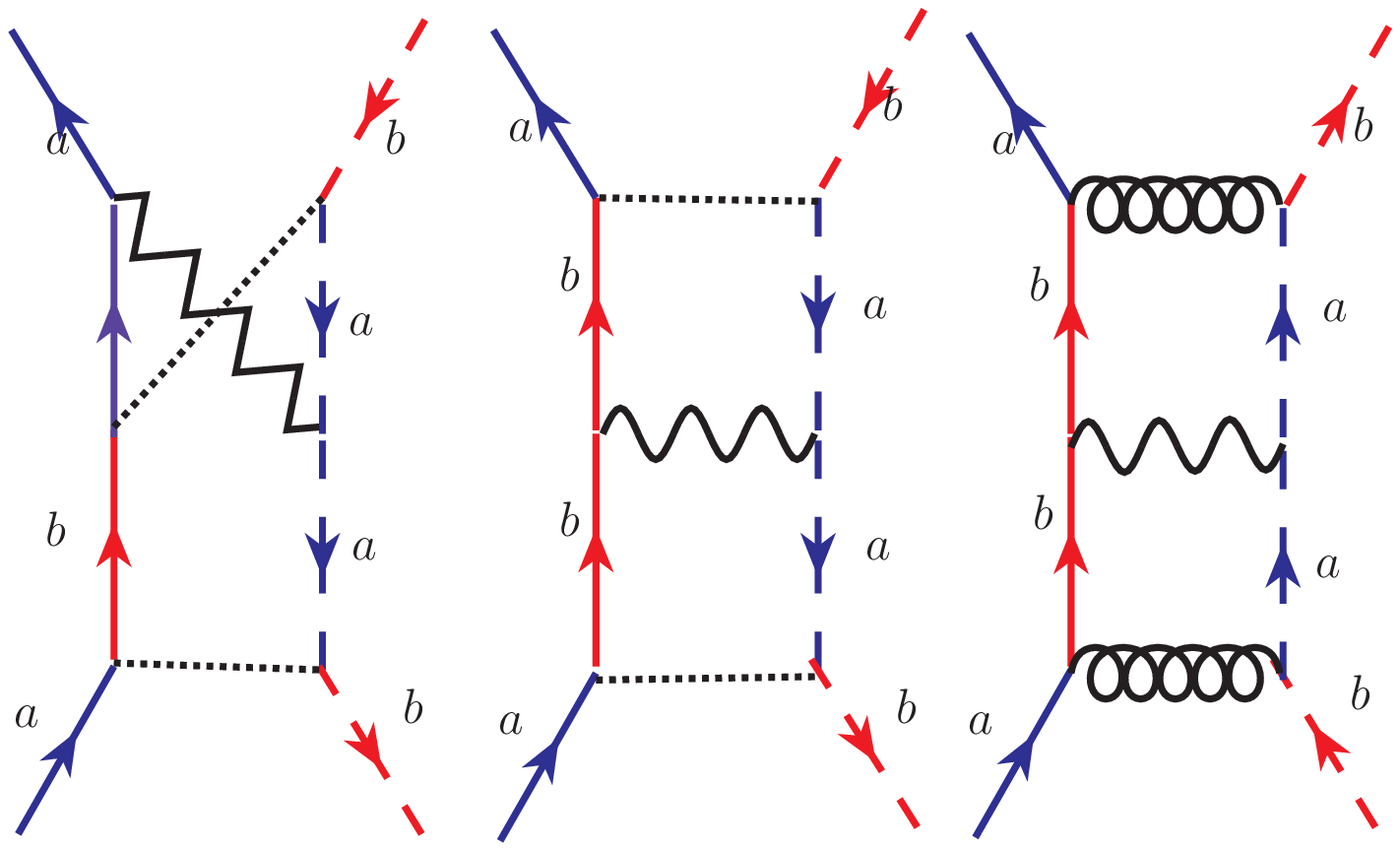}
\caption{(Color online) 2-loop parquet diagrams in the interforward scattering channel. Where we have $\vec{p} = \vec{p}_{1}+\vec{p}_{2}$.}
\label{2fc1}
\end{figure}

\begin{figure}[!htb]
\centering
\includegraphics[scale=0.3]{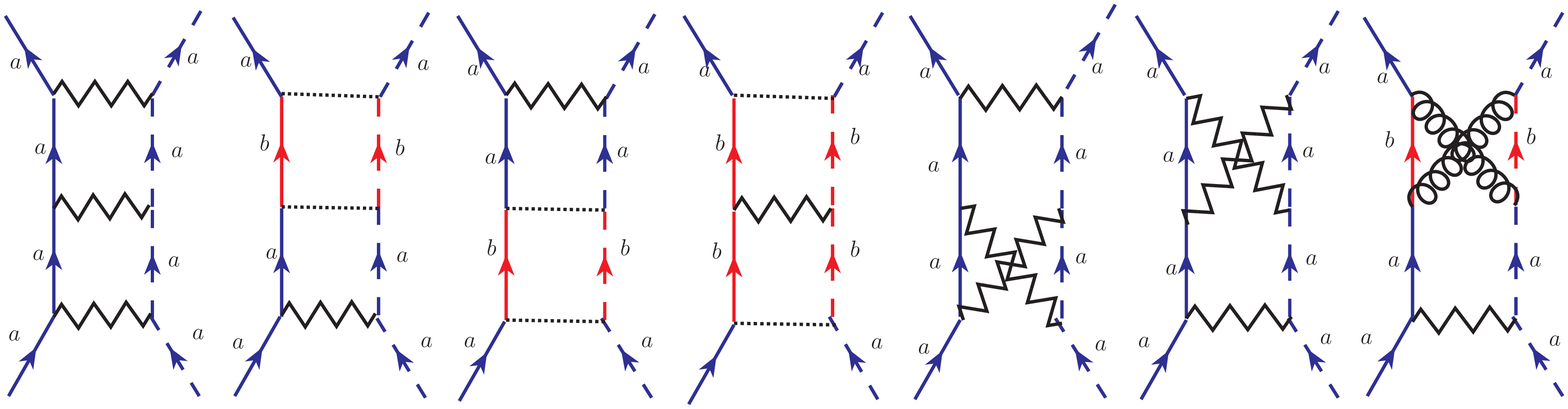}
\includegraphics[scale=0.3]{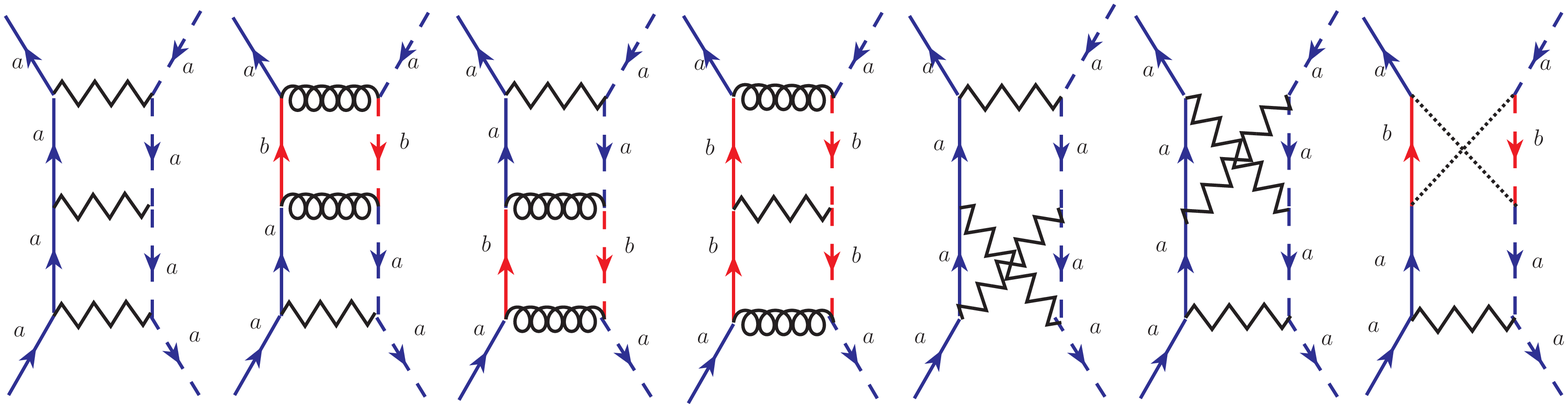}
\includegraphics[scale=0.3]{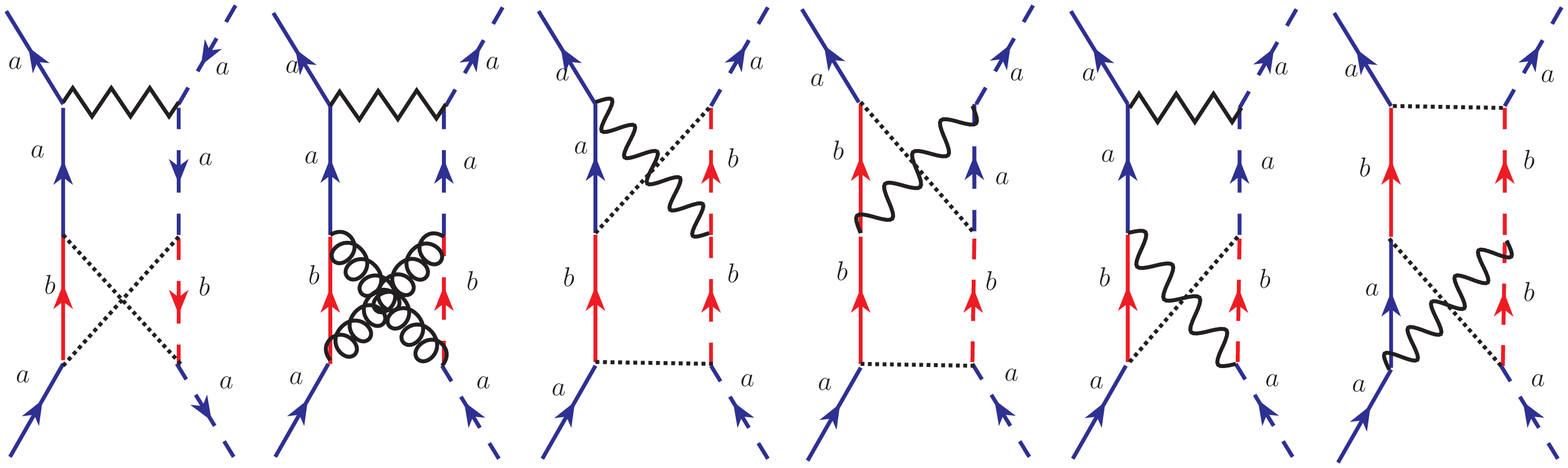}
\includegraphics[scale=0.3]{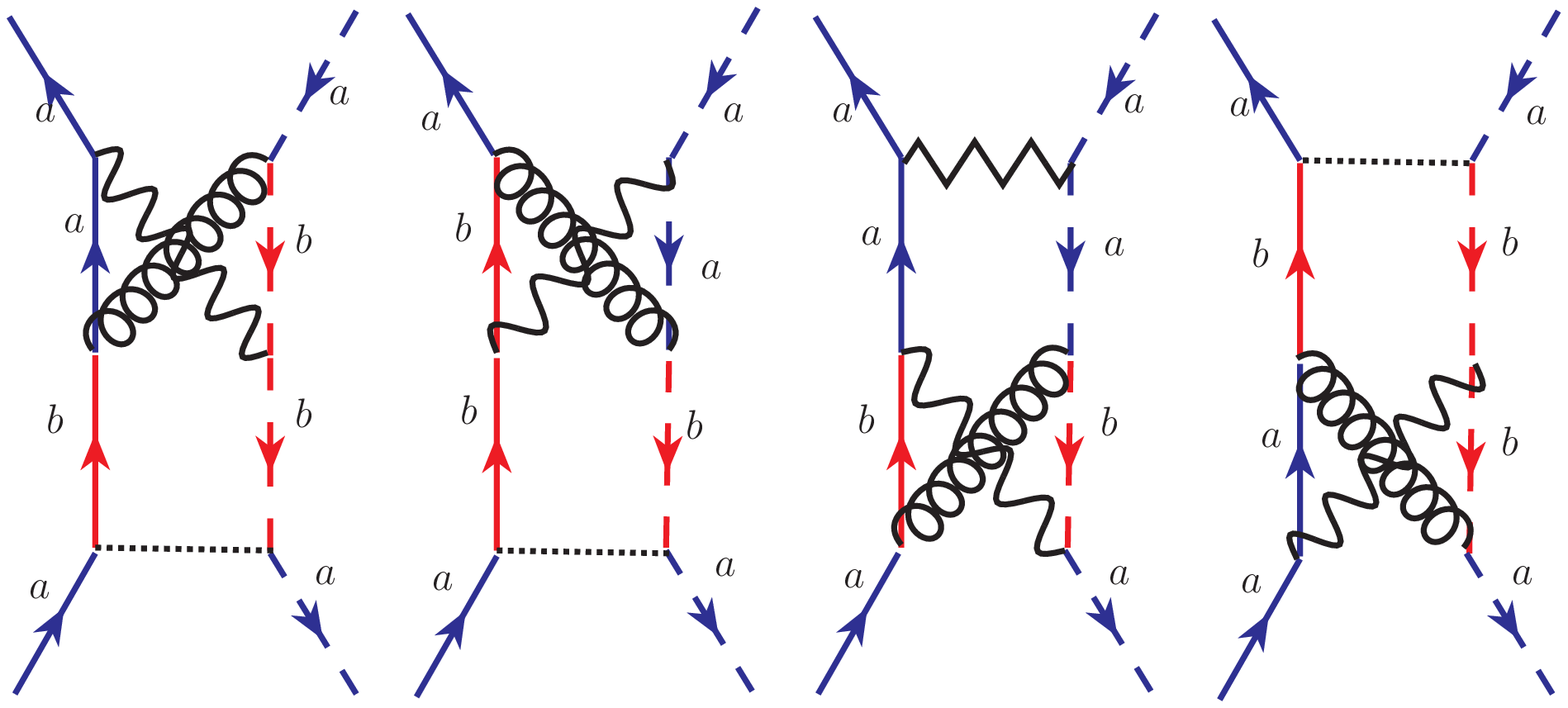}
\caption{(Color online) 2-loop parquet diagrams in the intraforward scattering channel.}
\label{22c}
\end{figure}

\begin{figure}[!htb]
\centering
\includegraphics[scale=0.3]{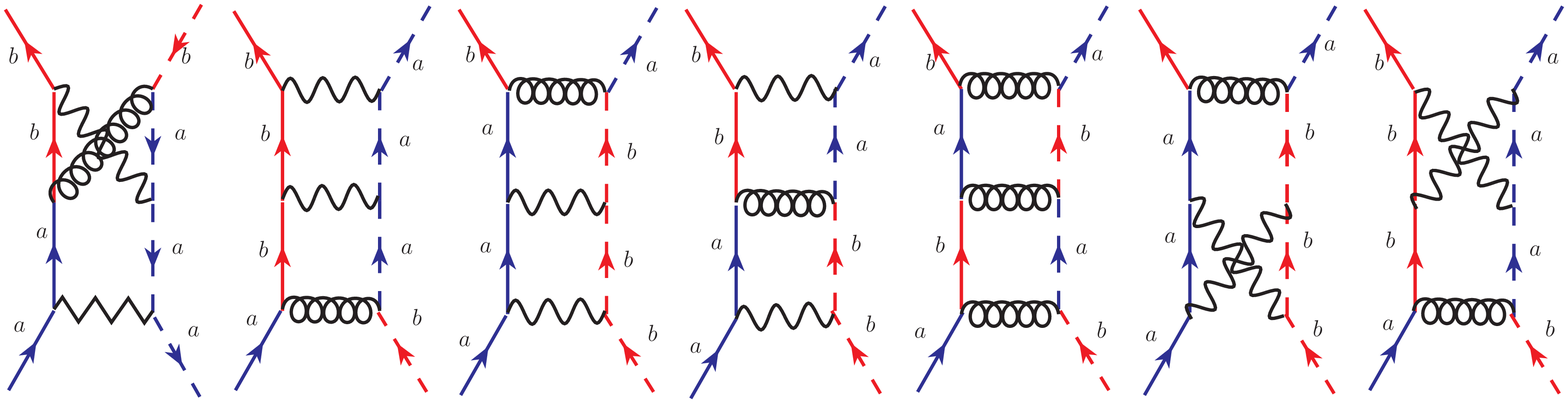}
\includegraphics[scale=0.3]{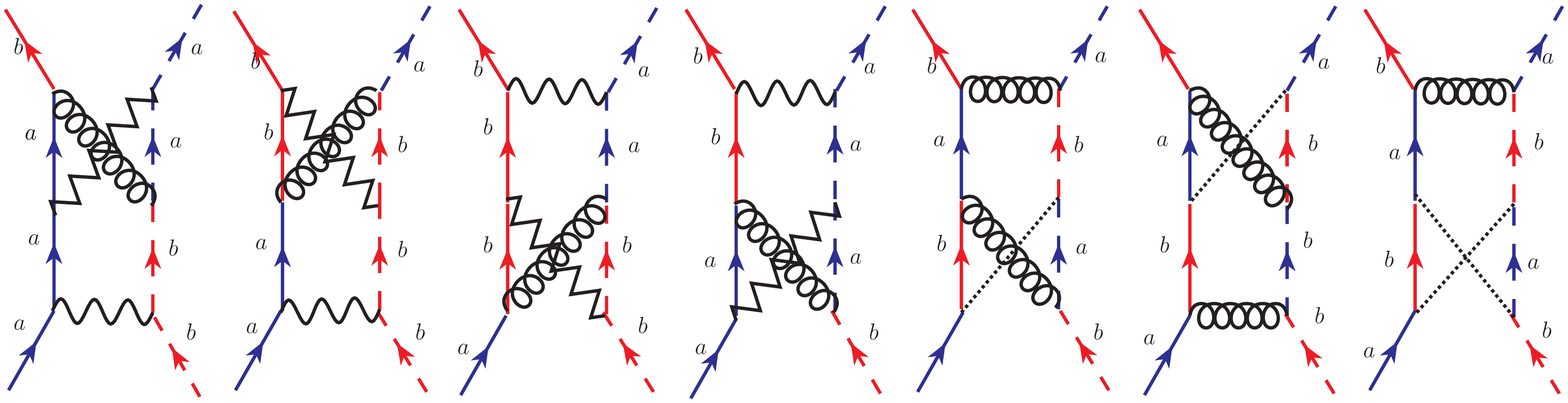}
\includegraphics[scale=0.3]{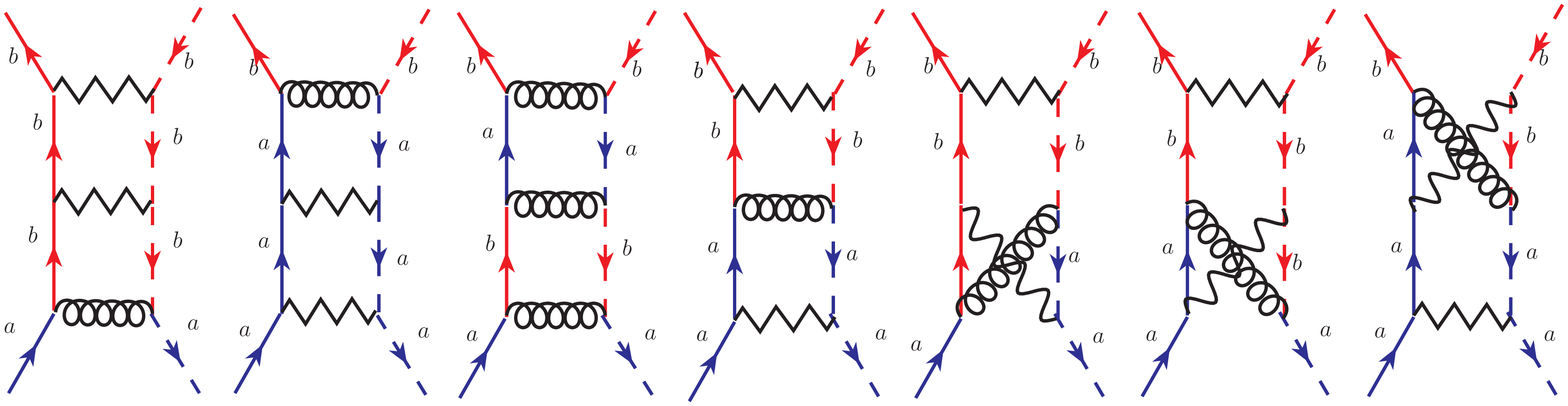}
\includegraphics[scale=0.3]{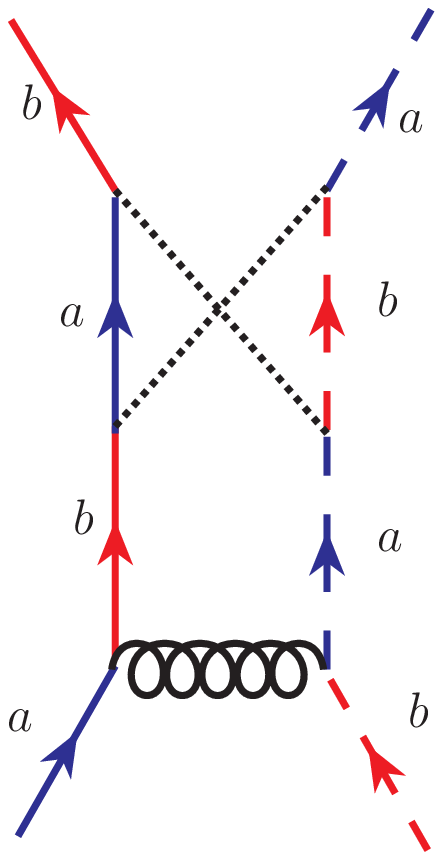}
\caption{(Color online) 2-loop parquet diagrams in the interbackscattering channel.}
\label{2bc}
\end{figure}

\begin{figure}[!htb]
\centering
\includegraphics[scale=0.3]{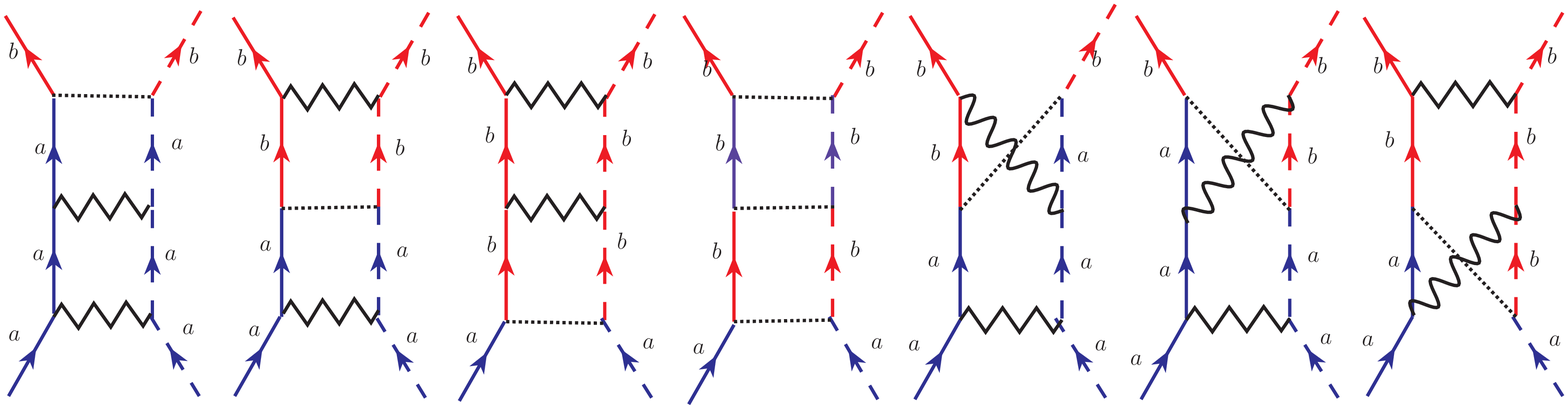}
\includegraphics[scale=0.3]{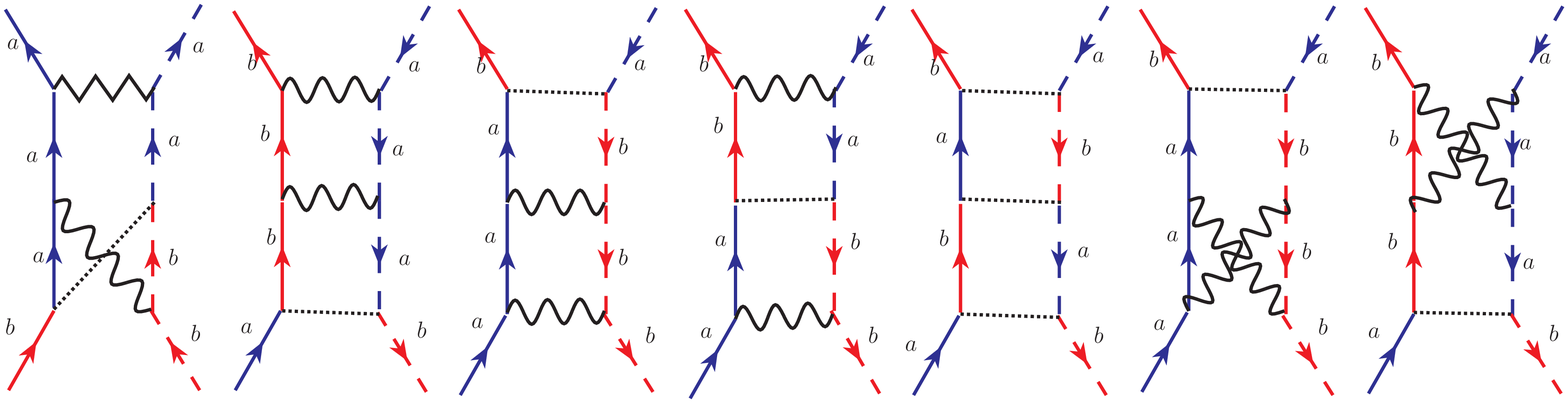}
\includegraphics[scale=0.3]{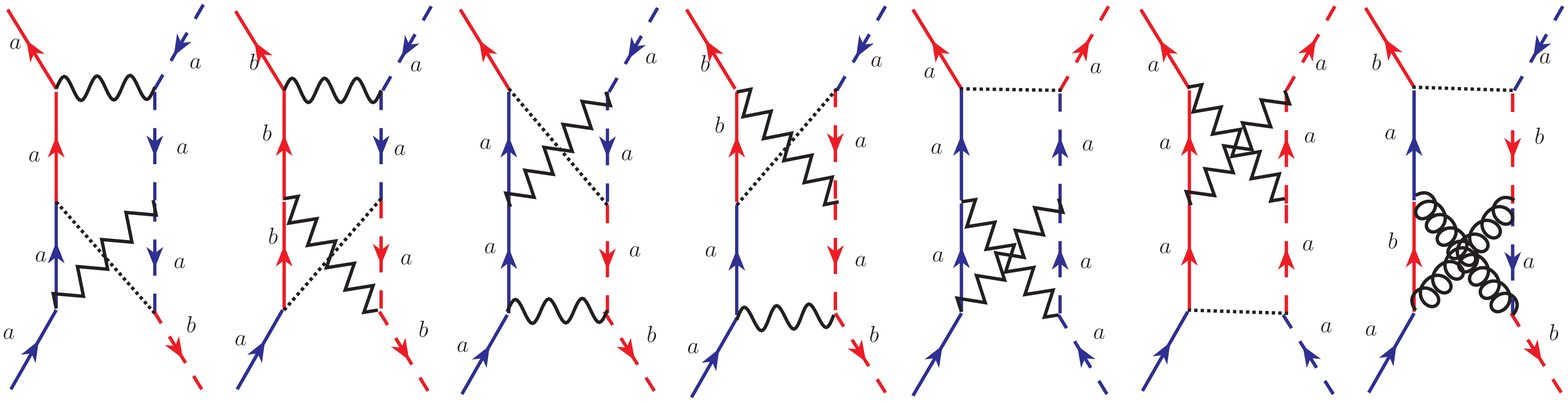}
\includegraphics[scale=0.3]{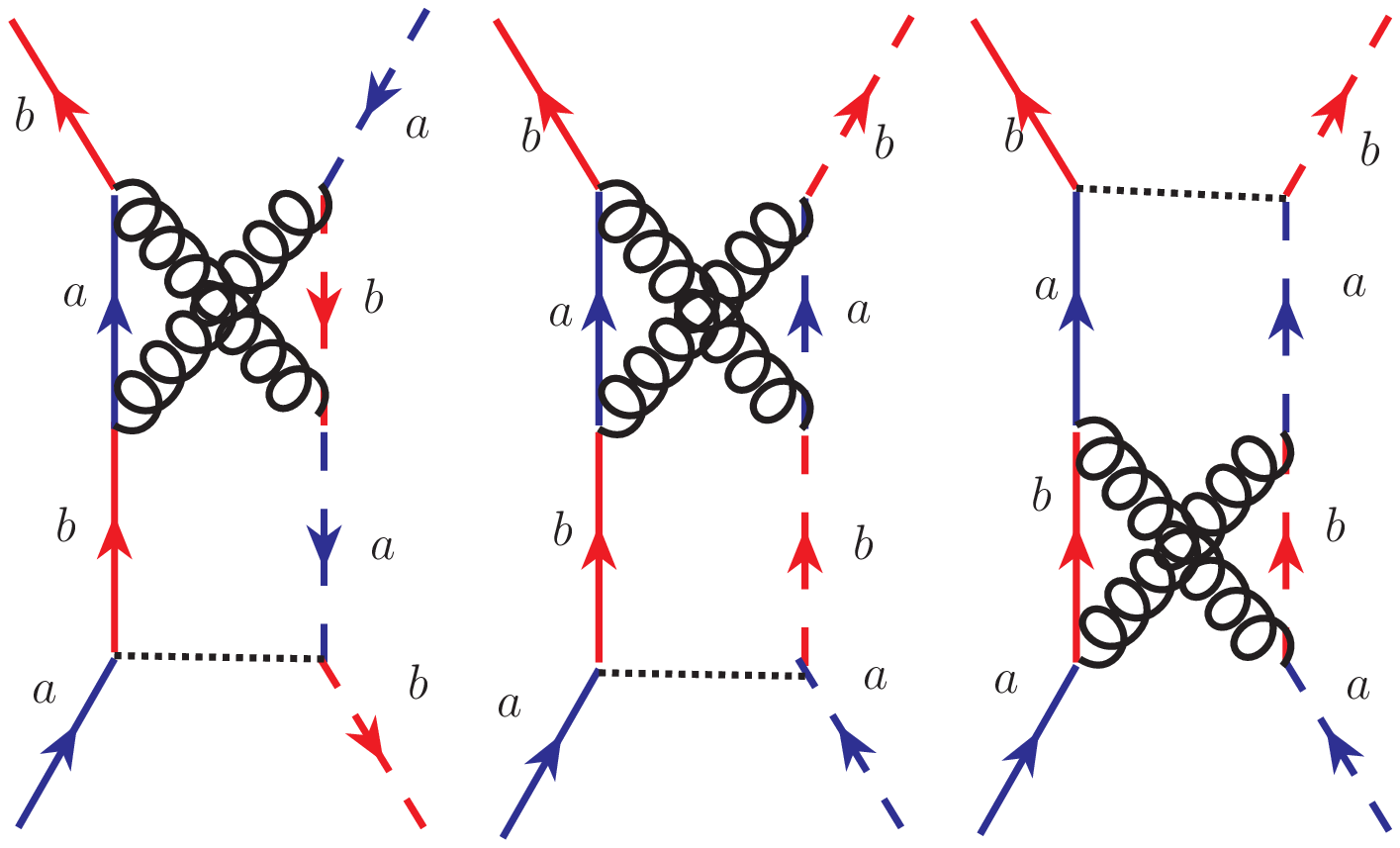}
\caption{(Color online) 2-loop parquet diagrams in the interumklapp scattering channel.}
\label{2uc}
\end{figure}

The renormalization prescription for the two-particle irreducible functions at the Fermi surface are the corresponding
 renormalized coupling
\begin{eqnarray}
\Gamma_{0,R}^{(4)} &=& -ig_{0,R}, \label{g0}\\
\Gamma_{\mathcal{F},R}^{(4)} &=& -ig_{\mathcal{F},R}, \label{gf}\\
\Gamma_{\mathcal{B},R}^{(4)} &=& -ig_{\mathcal{B},R}, \label{gB}\\
\Gamma_{\mathcal{U},R}^{(4)} &=& -ig_{\mathcal{U},R}. \label{gu} 
\end{eqnarray}
\begin{figure}[!htb]
\centering
\includegraphics[scale=0.4]{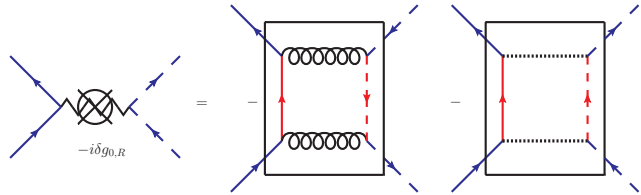}
\caption{(Color online) Counterterm diagram for intraforward.}
\label{cG0}
\end{figure}
\begin{figure}[!htb]
\centering
\includegraphics[scale=0.4]{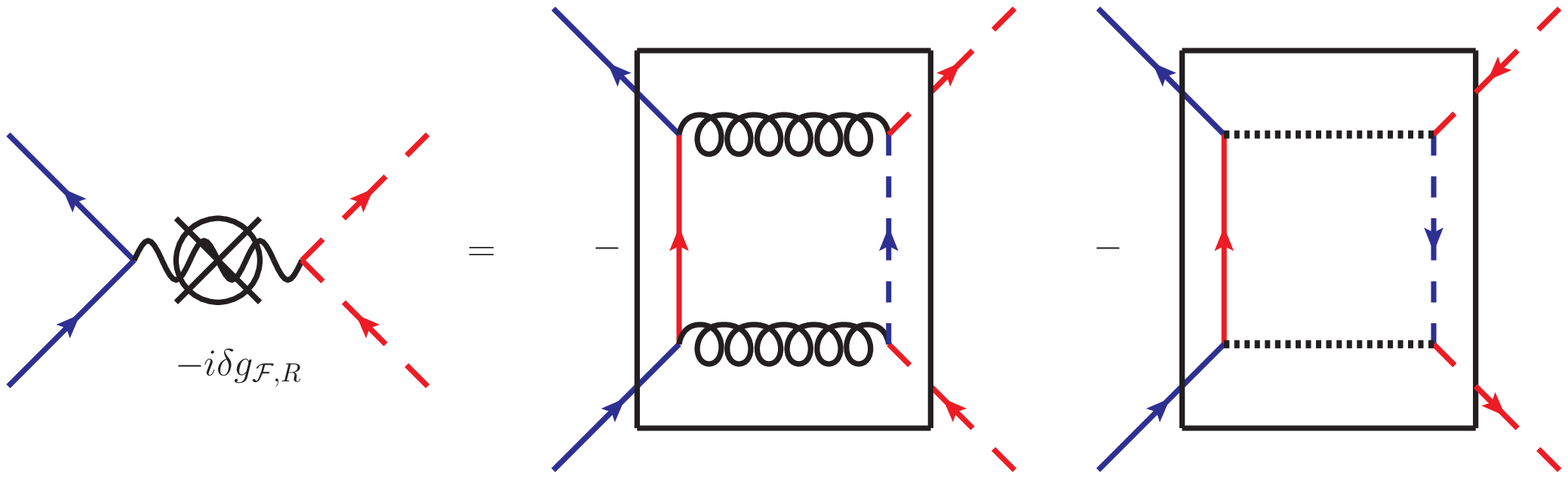}
\caption{(Color online) Counterterm diagram for interforward.}
\label{cGf}
\end{figure}
\begin{figure}[!htb]
\centering
\includegraphics[scale=0.4]{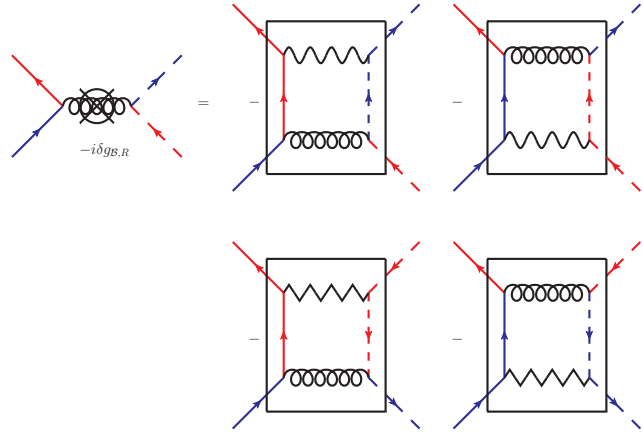}
\caption{(Color online) Counterterm diagram for interbackscattering.}
\label{cGb}
\end{figure}
\begin{figure}[!htb]
\centering
\includegraphics[scale=0.4]{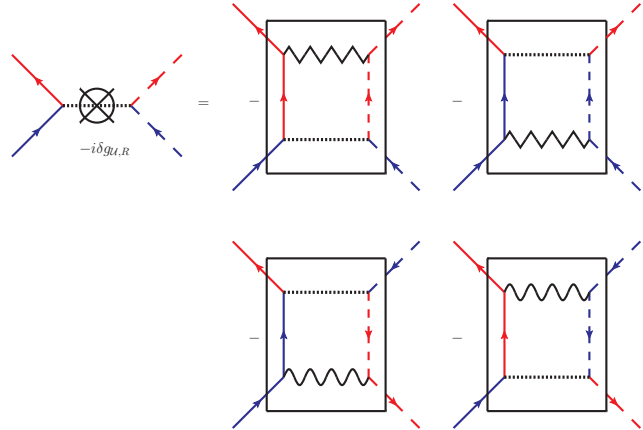}
\caption{(Color online) Counterterm diagram for interumklapp.}
\label{cGu}
\end{figure}
We can then calculate explicitly the counterterms in 1-loop for the intraforward channel (figure \ref{cG0})
\small
\begin{eqnarray}
\delta g_{0,R} &=& -\frac{g_{\mathcal{B}, R}^{2}}{\pi V_{F,R}}\ln\left(\frac{\Omega}{V_{F,R}\Delta k_{F,R}}\right)
+ \frac{g_{\mathcal{U},R}^{2}}{\pi V_{F,R}}\ln\left(\frac{\Omega}{\omega}\right), \nonumber \\
\label{d0}
\end{eqnarray}
\normalsize
for the interforward channel (figure \ref{cGf})
\small  
\begin{eqnarray}
\delta g_{\mathcal{F},R} &=& \frac{g_{\mathcal{B},R}^{2}}{\pi V_{F,R}}\ln\left(\frac{\Omega}{ V_{F,R}\Delta k_{F,R}}\right) \nonumber \\ 
&-&\frac{g_{\mathcal{U},R}^{2}}{\pi V_{F,R}}\ln\left(\frac{\Omega}{\omega}\right), 
\label{df}
\end{eqnarray}
\normalsize
for the interbackscattering channel (figure \ref{cGb})
\small
\begin{eqnarray}
\delta g_{\mathcal{B},R} &=& \frac{g_{\mathcal{B},R}g_{\mathcal{F},R}}{\pi V_{F}}\ln\left(\frac{\Omega}{ V_{F,R}\Delta k_{F,R}}\right) + \frac{g_{\mathcal{F},R}g_{\mathcal{B},R}}{\pi V_{F,R}}\ln\left(\frac{\Omega}{\omega}\right)\nonumber \\ 
&-&\frac{g_{\mathcal{B},R}g_{0,R}}{\pi V_{F,R}}\ln\left(\frac{\Omega}{V_{F,R}\Delta k_{F,R}}\right)  \nonumber \\
&-& \frac{g_{0,R}g_{\mathcal{B},R}}{\pi V_{F}}\ln\left(\frac{\Omega}{\omega}\right),
\label{dB}
\end{eqnarray} 
\normalsize 
and for the interumklapp channel (figure \ref{cGu})
\small
\begin{eqnarray}
\delta g_{\mathcal{U},R} &=& 2\frac{g_{0,R}g_{\mathcal{U},R}}{\pi V_{F,R}}\ln\left(\frac{\Omega}{\omega}\right) \nonumber \\
&-& 2\frac{g_{\mathcal{U},R}g_{\mathcal{F},R}}{\pi V_{F,R}}\ln\left(\frac{\Omega}{\omega}\right).
\label{du}
\end{eqnarray} 
\normalsize
The 2-loops diagrams from the 1-loop counterterms (figure \ref{cGfDiagram2L1}) cancel all the parquet diagrams in 2-loops 
(figures \ref{2fc1}, \ref{22c}, \ref{2bc}, \ref{2uc}). 

\begin{figure}[!htb]
\centering
\includegraphics[scale=0.3]{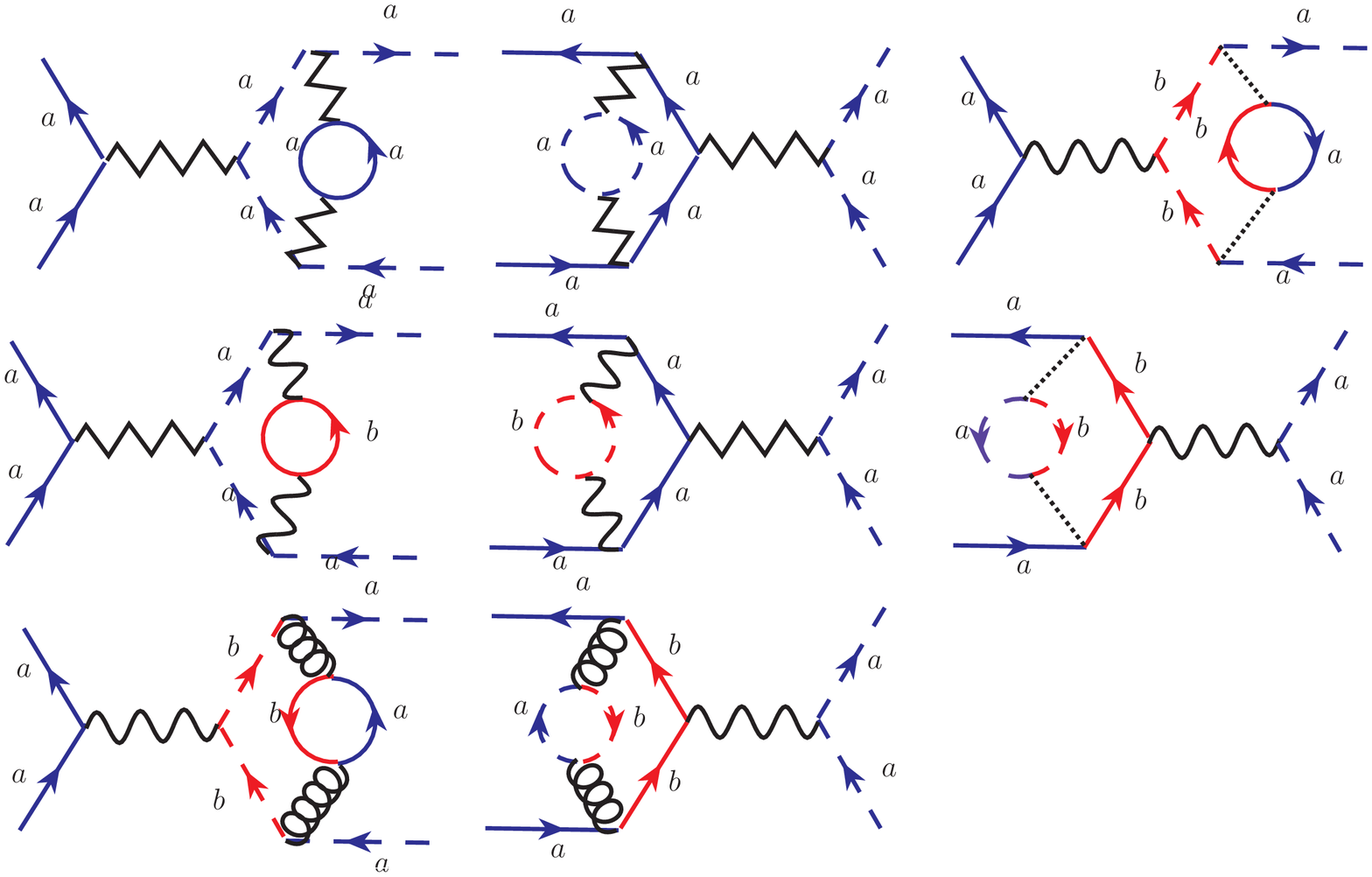}
\includegraphics[scale=0.3]{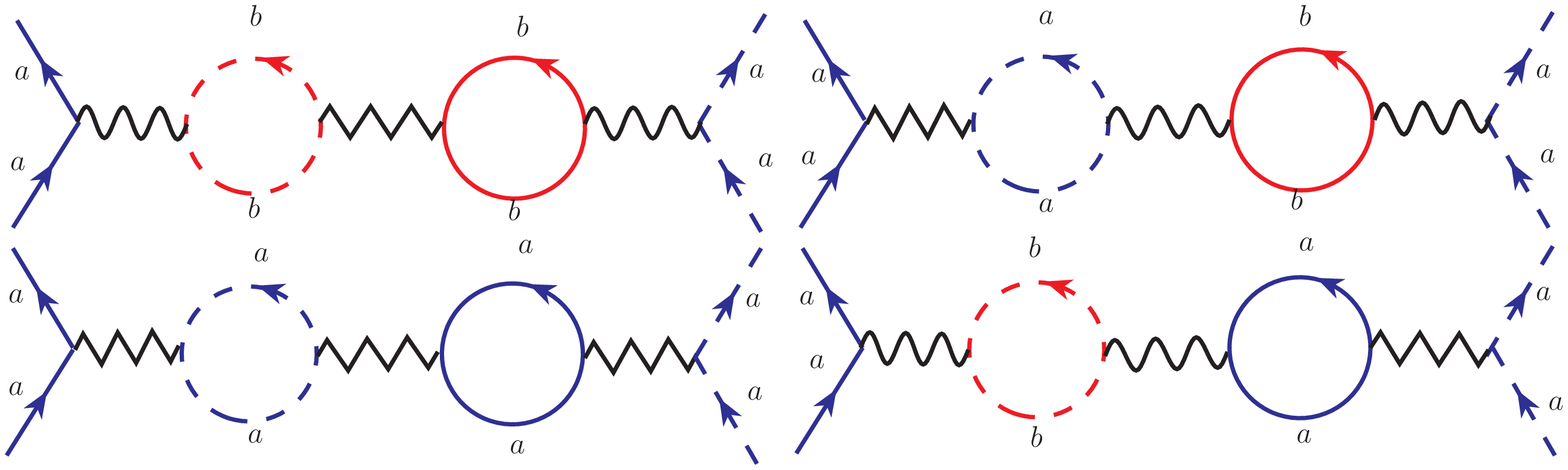}
\caption{(Color online) 2-loop non-parquet diagrams in the intraforward scattering channel.}
\label{22c2intraforward}
\end{figure}

By calculating the contributions of non-parquet diagrams in two-loops, including the 2-loops non-parquet contributions, we have in 
the intraforward channel (figure \ref{22c2intraforward}), the two-particle 
irreducible function in 2-loops, at the Fermi surface,

\begin{eqnarray}
\bar{\Gamma}_{0,R}^{(4)} &=& -i\bar{g}_{0,R} + i\bar{g}_{\mathcal{U},R}^{2}\ln\left(\frac{\Omega}{\omega}\right) 
-i\bar{g}_{\mathcal{B},R}^{2}\ln\left(\frac{\Omega}{V_{F,R}\Delta k_{F,R}}\right) \nonumber \\
&-& i\bar{g}_{0,R}^{3}\ln\left(\frac{\Omega}{\omega}\right) 
-i\bar{g}_{\mathcal{F},R}^{2}\bar{g}_{0,R}\ln\left(\frac{\Omega}{\omega}\right) \nonumber \\
&-& i\bar{g}_{\mathcal{U},R}^{2}\bar{g}_{\mathcal{F},R}\ln\left(\frac{\Omega}{\omega}\right) - i\bar{g}_{\mathcal{B},R}^{2}\bar{g}_{\mathcal{F},R}\ln\left(\frac{\Omega}{V_{F,R}\Delta k_{F,R}}\right) -i\delta \bar{g}_{0,R},
\label{g0ch2}
\end{eqnarray}
where $-i\delta g_{0,R}$ is the corresponding counterterm for the intraforward channel until 2-loops, where
$\bar{\Gamma}_{0,R}^{(4)}= \Gamma_{0,R}^{(4)}/\pi V_{F,R}$. 

\begin{figure}[!htb]
\centering
\includegraphics[scale=0.3]{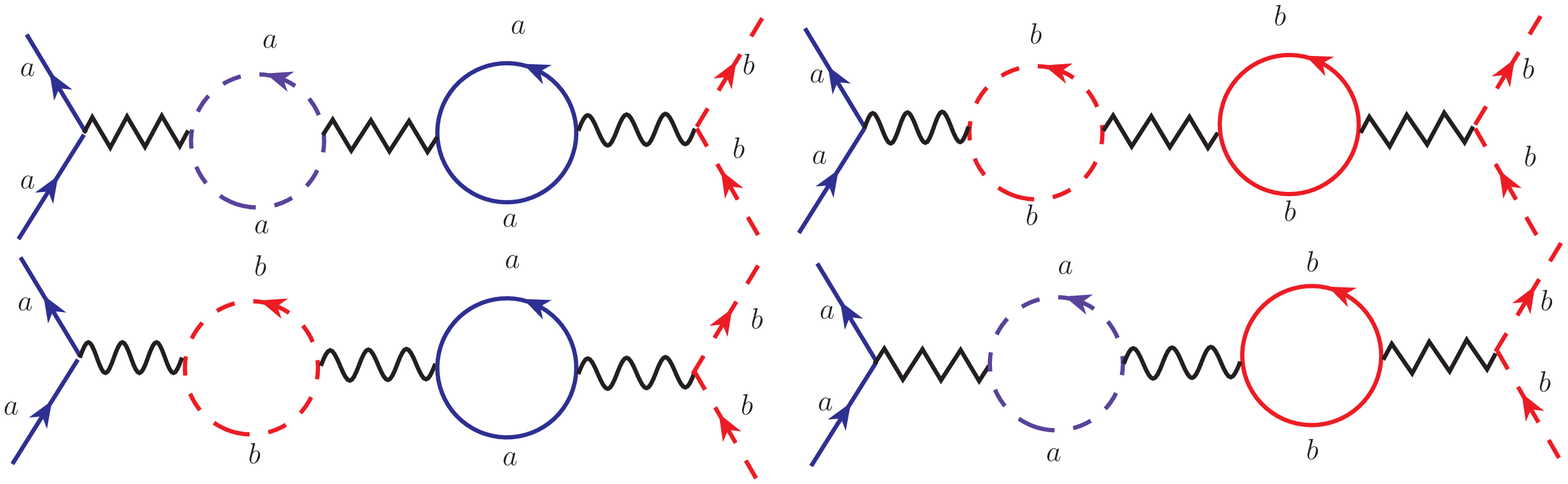}
\includegraphics[scale=0.3]{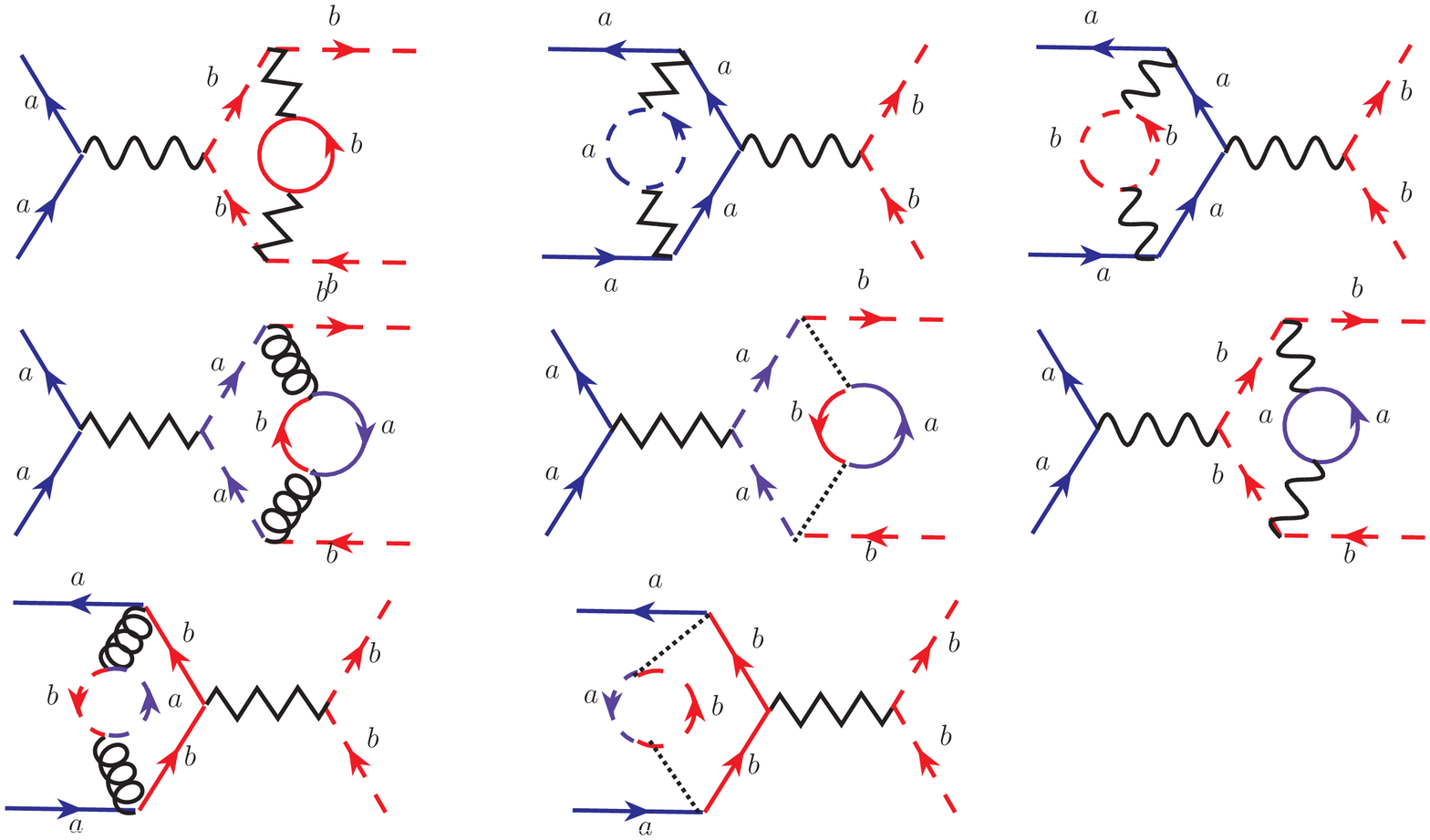}
\caption{(Color online) 2-loop non-parquet diagrams in the interforward scattering channel.}
\label{2fc2interforward}
\end{figure}

Including the non-parquet contribution to the interforward channel (figure \ref{2fc2interforward}) 
the two-particle irreducible function in 2-loops has the following contribution at the Fermi surface
\begin{eqnarray}
\bar{\Gamma}_{\mathcal{F},R}^{(4)} &=& -i\bar{g}_{\mathcal{F},R} -i\bar{g}_{\mathcal{U},R}^{2}\ln\left(\frac{\Omega}{\omega}\right) 
+ i\bar{g}_{\mathcal{B},R}^{2}\ln\left(\frac{\Omega}{V_{F,R}\Delta k_{F,R}}\right) \nonumber \\
&-& i\bar{g}_{0,R}^{2}\bar{g}_{\mathcal{F},R}\ln\left(\frac{\Omega}{\omega}\right) 
-i\bar{g}_{\mathcal{F},R}^{3}\ln\left(\frac{\Omega}{\omega}\right) \nonumber \\
&-& i\bar{g}_{\mathcal{U},R}^{2}\bar{g}_{0,R}\ln\left(\frac{\Omega}{\omega}\right) -i\bar{g}_{\mathcal{B},R}^{2}\bar{g}_{0,R}\ln\left(\frac{\Omega}{V_{F,R}\Delta k_{F,R}}\right) \nonumber \\
&-& i\delta \bar{g}_{\mathcal{F},R},
\label{fch2}
\end{eqnarray}
where $-i\delta \bar{g}_{\mathcal{F},R}=-i\delta g_{\mathcal{F},R}/\pi V_{F,R}$ is the corresponding 
counterterm for the interforward channel until 2-loops, where
$\bar{\Gamma}_{\mathcal{F},R}^{(4)}= \Gamma_{\mathcal{F},R}^{(4)}/\pi V_{F,R}$.

\begin{figure}[!htb]
\centering
\includegraphics[scale=0.3]{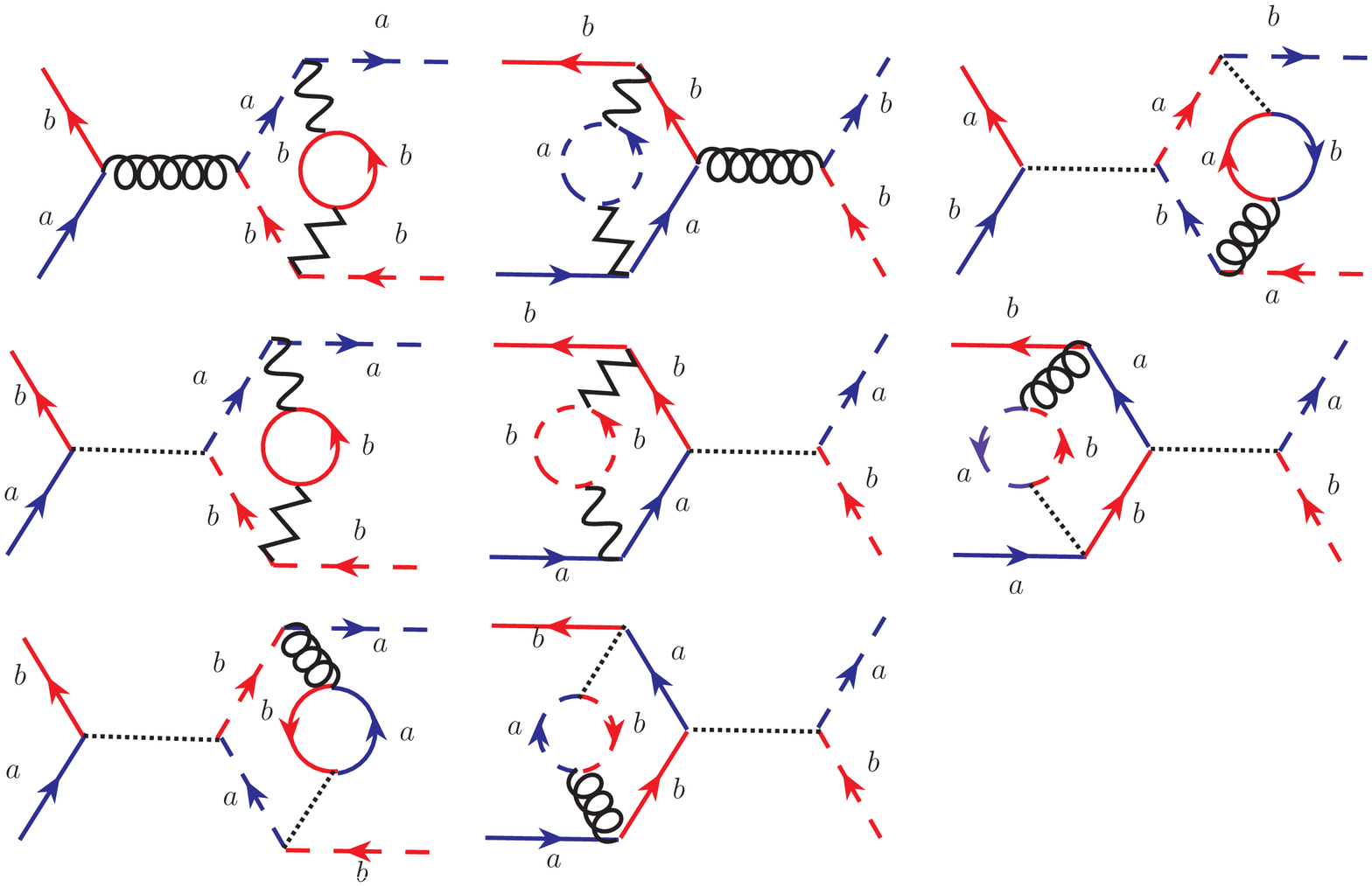}
\includegraphics[scale=0.3]{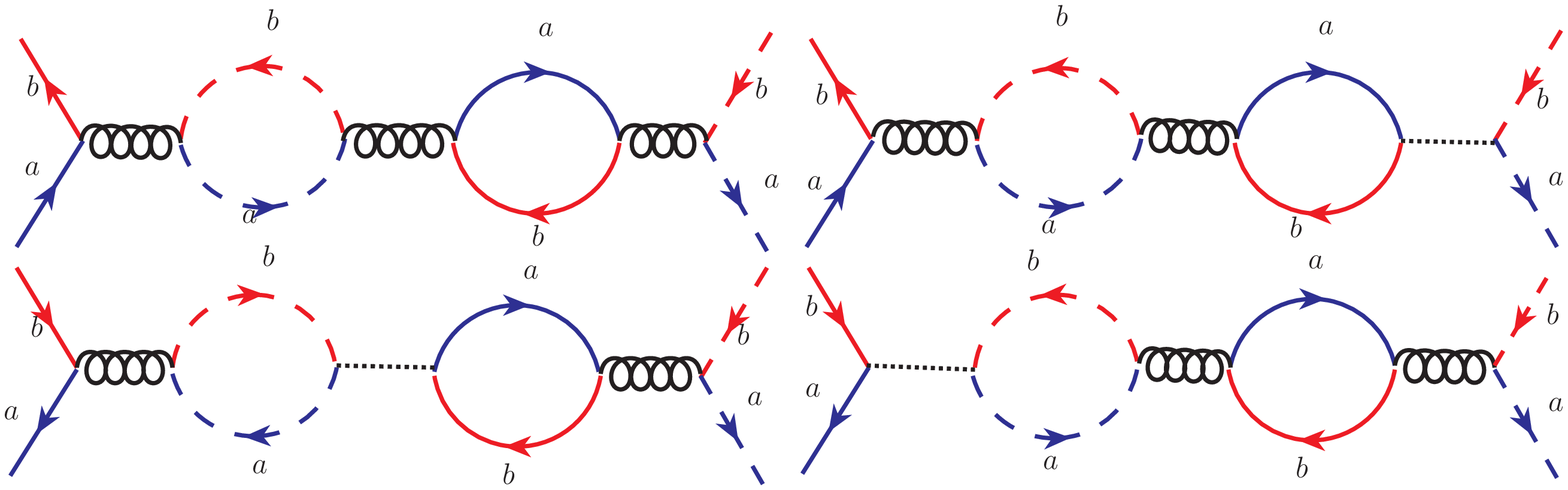}
\caption{(Color online) 2-loops non-parquet diagrams in the interbackscattering channel.}
\label{2bc2backscattering}
\end{figure}

Including the non-parquet contribution to the interbackscattering channel (figure \ref{2bc2backscattering}) 
the two-particle irreducible function in 2-loops has the following contribution at the Fermi surface
\begin{eqnarray}
\bar{\Gamma}_{\mathcal{B},R}^{(4)} &=& -i\bar{g}_{\mathcal{B},R} + 
i\bar{g}_{\mathcal{B},R}\bar{g}_{\mathcal{F},R}\ln\left(\frac{\Omega}{V_{F,R}\Delta k_{F,R}}\right) 
\nonumber \\
&+& i\bar{g}_{\mathcal{B},R}\bar{g}_{\mathcal{F},R}\ln\left(\frac{\Omega}{\omega}\right) - i\bar{g}_{0,R}\bar{g}_{\mathcal{B},R}\ln\left(\frac{\Omega}{V_{F,R}\Delta k_{F,R}}\right)\nonumber \\ 
&-&i\bar{g}_{0,R}\bar{g}_{\mathcal{B},R}\ln\left(\frac{\Omega}{\omega}\right) 
- i\bar{g}_{\mathcal{U},R}^{2}\bar{g}_{\mathcal{B},R}\ln\left(\frac{\Omega}{V_{F,R}\Delta k_{F,R}}\right) \nonumber \\
&-& 2 i\bar{g}_{0,R}\bar{g}_{\mathcal{F},R}\bar{g}_{\mathcal{B},R}\ln\left(\frac{\Omega}{V_{F,R}\Delta k_{F,R}}\right) + i\bar{g}_{\mathcal{U},R}^{2}\bar{g}_{\mathcal{B},R}\ln\left(\frac{\Omega}{\omega}\right) -i\delta \bar{g}_{\mathcal{B},R}, 
\label{Bch1}
\end{eqnarray}
where $-i\delta \bar{g}_{\mathcal{B},R}=-i\delta g_{\mathcal{B},R}/\pi V_{F,R}$ is the corresponding counterterm for the interbackscattering channel until 2-loops, where
$\bar{\Gamma}_{\mathcal{B},R}^{(4)}= \Gamma_{\mathcal{B},R}^{(4)}/\pi V_{F,R}$.

\begin{figure}[!htb]
\centering
\includegraphics[scale=0.3]{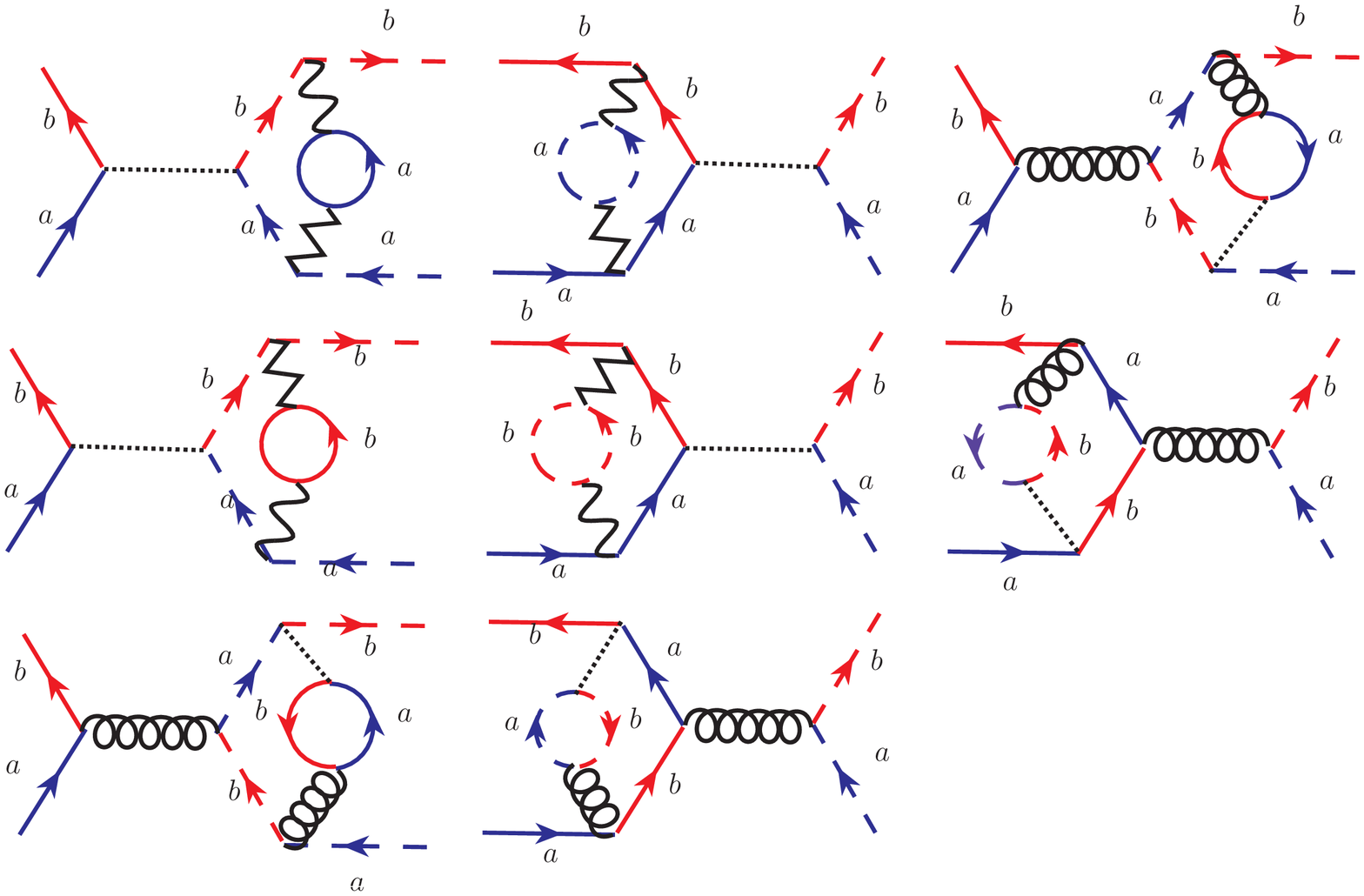}
\includegraphics[scale=0.3]{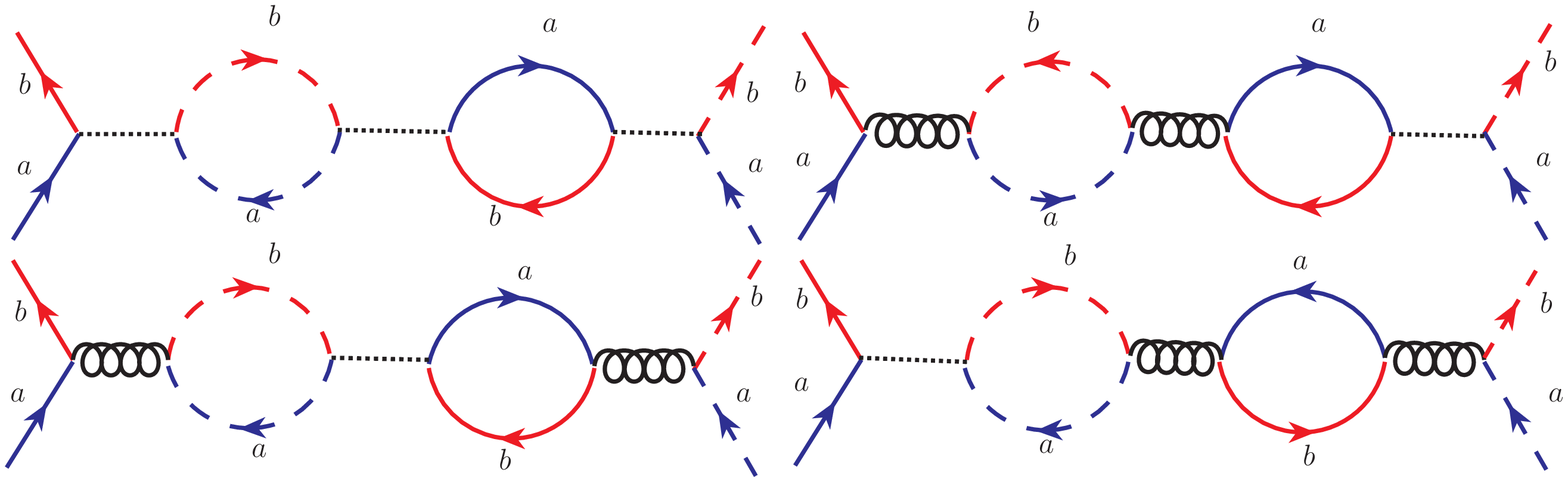}
\caption{(Color online) 2-loop non-parquet diagrams in the interumklapp scattering channel}
\label{2uc2umklap}
\end{figure}

Including the non-parquet contribution to the interumklapp channel (figure \ref{2uc2umklap}), we have the following 
contribution at the Fermi surface
\begin{eqnarray}
\bar{\Gamma}_{\mathcal{U},R}^{(4)} &=& -i\bar{g}_{\mathcal{U},R} + 2i\bar{g}_{0,R}\bar{g}_{\mathcal{U},R}\bar{g}_{\mathcal{B},R}\ln\left(\frac{\Omega}{\omega}\right) \nonumber \\
&-& 2i\bar{g}_{\mathcal{F},R}\bar{g}_{\mathcal{U},R}\ln\left(\frac{\Omega}{\omega}\right) \nonumber \\
&-& 2i\bar{g}_{0,R}\bar{g}_{\mathcal{F},R}\bar{g}_{\mathcal{U},R}\ln\left(\frac{\Omega}{\omega}\right) \nonumber \\
&-& 2i\bar{g}_{\mathcal{U},R}\bar{g}_{\mathcal{B},R}^{2}\ln\left(\frac{\Omega}{V_{F,R}\Delta k_{F,R}}\right) -i\delta \bar{g}_{\mathcal{U},R}, \nonumber \\
\label{uch1}
\end{eqnarray}
where $-i\delta \bar{g}_{\mathcal{U},R}=-i\delta g_{\mathcal{U},R}/\pi V_{F,R}$ is the corresponding 
counterterm for the interumklapp channel until 2-loops, where
$\bar{\Gamma}_{\mathcal{U},R}^{(4)}= \Gamma_{\mathcal{U},R}^{(4)}/\pi V_{F,R}$.

Taking into account the prescriptions for the scattering channels at the Fermi surface
\begin{eqnarray}
\bar{\Gamma}_{0,R}^{(4)} &=& -i\bar{g}_{0,R}, \label{g0}\\
\bar{\Gamma}_{\mathcal{F},R}^{(4)} &=& -i\bar{g}_{\mathcal{F},R}, \label{gf}\\
\bar{\Gamma}_{\mathcal{B},R}^{(4)} &=& -i\bar{g}_{\mathcal{B},R}, \label{gB}\\
\bar{\Gamma}_{\mathcal{U},R}^{(4)} &=& -i\bar{g}_{\mathcal{U},R}. \label{gu} 
\end{eqnarray}
We arrive at the explicit 2-loops contributions of the counterterms, including the non-parquet contributions. For the 
intraforward channel until 2-loops, we have
\begin{eqnarray}
\delta \bar{g}_{0,R}&=& \bar{g}_{\mathcal{U},R}^{2}\ln\left(\frac{\Omega}{\omega}\right) 
-\bar{g}_{\mathcal{B},R}^{2}\ln\left(\frac{\Omega}{V_{F,R}\Delta k_{F,R}}\right) \nonumber \\
&-& \bar{g}_{0,R}^{3}\ln\left(\frac{\Omega}{\omega}\right) 
-\bar{g}_{\mathcal{F},R}^{2}\bar{g}_{0,R}\ln\left(\frac{\Omega}{\omega}\right) \nonumber \\
&-& \bar{g}_{\mathcal{U},R}^{2}\bar{g}_{\mathcal{F},R}\ln\left(\frac{\Omega}{\omega}\right) - \bar{g}_{\mathcal{B},R}^{2}\bar{g}_{\mathcal{F},R}\ln\left(\frac{\Omega}{V_{F,R}\Delta k_{F,R}}\right),
\label{counteO}
\end{eqnarray}
the interforward counterterm until 2-loops 
\begin{eqnarray}
\delta \bar{g}_{\mathcal{F},R} &=& -\bar{g}_{\mathcal{U},R}^{2}\ln\left(\frac{\Omega}{\omega}\right) 
+ \bar{g}_{\mathcal{B},R}^{2}\ln\left(\frac{\Omega}{V_{F,R}\Delta k_{F,R}}\right) \nonumber \\
&-& \bar{g}_{0,R}^{2}\bar{g}_{\mathcal{F},R}\ln\left(\frac{\Omega}{\omega}\right) 
-\bar{g}_{\mathcal{F},R}^{3}\ln\left(\frac{\Omega}{\omega}\right) \nonumber \\
&-& \bar{g}_{\mathcal{U},R}^{2}\bar{g}_{0,R}\ln\left(\frac{\Omega}{\omega}\right) - \bar{g}_{\mathcal{B},R}^{2}\bar{g}_{0,R}\ln\left(\frac{\Omega}{V_{F,R}\Delta k_{F,R}}\right),
\label{counteF}
\end{eqnarray}
interbackscattering counterterm until 2-loops order
\begin{eqnarray}
\delta \bar{g}_{\mathcal{B},R} &=& \bar{g}_{\mathcal{B},R}\bar{g}_{\mathcal{F},R}\ln\left(\frac{\Omega}{V_{F,R}\Delta k_{F,R}}\right) 
+ \bar{g}_{\mathcal{B},R}\bar{g}_{\mathcal{F},R}\ln\left(\frac{\Omega}{\omega}\right) \nonumber \\
&-& \bar{g}_{0,R}\bar{g}_{\mathcal{B},R}\ln\left(\frac{\Omega}{V_{F,R}\Delta k_{F,R}}\right) 
-\bar{g}_{0,R}\bar{g}_{\mathcal{B},R}\ln\left(\frac{\Omega}{\omega}\right) - 2 \bar{g}_{0,R}\bar{g}_{\mathcal{F},R}\bar{g}_{\mathcal{B},R}\ln\left(\frac{\Omega}{V_{F,R}\Delta k_{F,R}}\right) \nonumber \\
&-& \bar{g}_{\mathcal{U},R}^{2}\bar{g}_{\mathcal{B},R}\ln\left(\frac{\Omega}{V_{F,R}\Delta k_{F,R}}\right) + \bar{g}_{\mathcal{U},R}^{2}\bar{g}_{\mathcal{B},R}\ln\left(\frac{\Omega}{\omega}\right), 
\label{counteB}
\end{eqnarray}
interumklapp counterterm in 2-loops order
\begin{eqnarray}
\delta \bar{g}_{\mathcal{U},R} &=& 2\bar{g}_{0,R}\bar{g}_{\mathcal{U},R}\bar{g}_{\mathcal{B},R}\ln\left(\frac{\Omega}{\omega}\right) \nonumber \\
&-& 2\bar{g}_{\mathcal{F},R}\bar{g}_{\mathcal{U},R}\ln\left(\frac{\Omega}{\omega}\right) \nonumber \\
&-& 2\bar{g}_{0,R}\bar{g}_{\mathcal{F},R}\bar{g}_{\mathcal{U},R}\ln\left(\frac{\Omega}{\omega}\right) - 2\bar{g}_{\mathcal{U},R}\bar{g}_{\mathcal{B},R}^{2}\ln\left(\frac{\Omega}{V_{F,R}\Delta k_{F,R}}\right).
\label{counteU}
\end{eqnarray}

\section{RG flow equations for the renormalized scattering couplings terms and confinement}

By considering the relation between the $\texttt{bare}$ coupling terms and the renormalized ones
\begin{eqnarray}
g_{\texttt{bare}}=Z^{-2}\left(g_{R} + \delta g_{R}\right),
\end{eqnarray}
we can calculate the flow of the renormalization group in the frequency scale $\omega$, described by means of the
equation
\begin{eqnarray}
\omega\frac{dg_{\texttt{bare}}}{d\omega}&=& \omega\frac{dZ^{-2}}{d\omega}\left(g_{R} + \delta g_{R}\right) \nonumber \\
&+& Z^{-2}\left(\omega\frac{dg_{R}}{d\omega} + \omega\frac{d\delta g_{R}}{d\omega}\right)=0, 
\end{eqnarray}
where we take into account the fact that the $\texttt{bare}$ quantities do not flow.

\begin{figure}[!htb]
\centering
\includegraphics[scale=0.5]{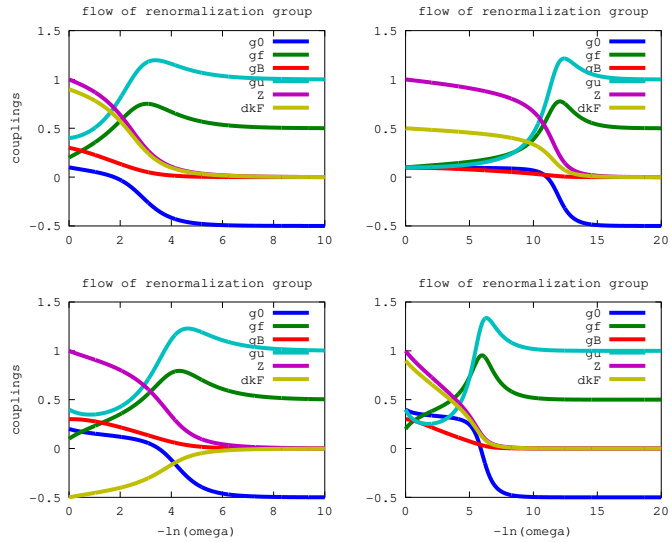}
\caption{(Color online) RG flow equations for initial positive couplings.}
\label{conf2pos}
\end{figure}

\begin{figure}[!htb]
\centering
\includegraphics[scale=0.5]{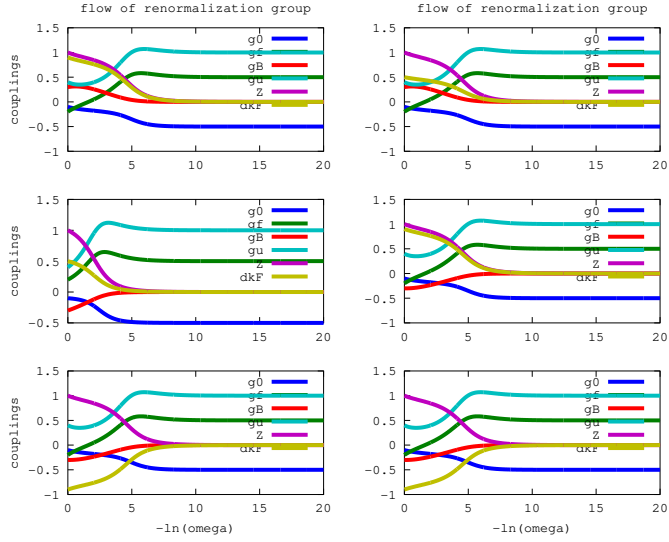}
\caption{(Color online) RG flow equations including initial negative couplings.}
\label{conf2neg}
\end{figure}

Taking into account the anomalous dimension
\begin{eqnarray}
\gamma =\frac{\omega}{Z}\frac{dZ}{d\omega}, \nonumber
\end{eqnarray}
we can write 
\begin{eqnarray}
\omega\frac{dZ^{-2}}{d\omega}&=& -2\omega Z^{-3}\frac{dZ}{d\omega} = -2\omega Z^{-2}Z^{-1}\frac{dZ}{d\omega} \nonumber \\
&=& -2 Z^{-2}\gamma,
\end{eqnarray}
and obtain the following result
\begin{eqnarray}
-2\gamma\left(g_{R} + \delta g_{R}\right) + 
\left(\omega\frac{dg_{R}}{d\omega} + \omega\frac{d\delta g_{R}}{d\omega}\right)=0.
\end{eqnarray}
Thus, we arrive at the RG flow equation
\begin{eqnarray} 
\omega\frac{dg_{R}}{d\omega}=2\gamma\left(g_{R} + \delta g_{R}\right) - \omega\frac{d\delta g_{R}}{d\omega}.
\end{eqnarray}
In the specific case of the scattering channels, we can write
\begin{eqnarray} 
\omega\frac{dg_{0R}}{d\omega} &=& 2\gamma\left(g_{0R} + \delta g_{0,R}\right) - \omega\frac{d\delta g_{0R}}{d\omega}, \\
\omega\frac{dg_{\mathcal{F},R}}{d\omega} &=& 2\gamma\left(g_{\mathcal{F},R} + \delta g_{\mathcal{F},R}\right) - \omega\frac{d\delta g_{\mathcal{F},R}}{d\omega}, \\
\omega\frac{dg_{\mathcal{B},R}}{d\omega} &=& 2\gamma\left(g_{\mathcal{B},R} + \delta g_{\mathcal{B},R}\right) - \omega\frac{d\delta g_{\mathcal{B},R}}{d\omega}, \\
\omega\frac{dg_{\mathcal{U},R}}{d\omega} &=& 2\gamma\left(g_{\mathcal{U},R} + \delta g_{\mathcal{U},R}\right) - \omega\frac{d\delta g_{\mathcal{U},R}}{d\omega}. 
\end{eqnarray}
Dividing both sides by $\pi V_{F,R}$, we have finally
\begin{eqnarray} 
\omega\frac{d\bar{g}_{0R}}{d\omega} &=& 2\gamma\left(\bar{g}_{0R} + \delta \bar{g}_{0,R}\right) - \omega\frac{d\delta \bar{g}_{0R}}{d\omega}, \\
\omega\frac{d\bar{g}_{\mathcal{F},R}}{d\omega} &=& 2\gamma\left(\bar{g}_{\mathcal{F},R} + \delta \bar{g}_{\mathcal{F},R}\right) - \omega\frac{d\delta \bar{g}_{\mathcal{F},R}}{d\omega}, \\
\omega\frac{d\bar{g}_{\mathcal{B},R}}{d\omega} &=& 2\gamma\left(\bar{g}_{\mathcal{B},R} + \delta \bar{g}_{\mathcal{B},R}\right) - \omega\frac{d\delta \bar{g}_{\mathcal{B},R}}{d\omega}, \\
\omega\frac{d\bar{g}_{\mathcal{U},R}}{d\omega} &=& 2\gamma\left(\bar{g}_{\mathcal{U},R} + \delta \bar{g}_{\mathcal{U},R}\right) - \omega\frac{d\delta \bar{g}_{\mathcal{U},R}}{d\omega}. 
\end{eqnarray}
Using the 2-loops counterterms, eq. (\ref{counteO}), (\ref{counteF}), (\ref{counteB}) and (\ref{counteU}), we can write the 
RG flow equations for the renormalized scattering coupling terms
\begin{eqnarray}
\omega\frac{d\bar{g}_{0,R}}{d\omega} &=& 2\gamma\bar{g}_{0,R} + \bar{g}^{2}_{\mathcal{U},R} -\bar{g}^{2}_{\mathcal{B},R}\frac{\omega}{\Delta k_{F,R}}\frac{d\Delta k_{F,R}}{d\omega}-\bar{g}_{0,R}^{3} \nonumber \\
&-& \bar{g}_{\mathcal{F},R}^{2}\bar{g}_{0,R} -\bar{g}_{\mathcal{U},R}^{2}\bar{g}_{\mathcal{F},R}  
\nonumber \\
&-&\bar{g}_{\mathcal{B},R}^{2}\bar{g}_{\mathcal{F},R}\frac{\omega}{\Delta k_{F,R}}\frac{d\Delta k_{F,R}}{d\omega}, 
\end{eqnarray}
\begin{eqnarray}
\omega\frac{d\bar{g}_{\mathcal{F},R}}{d\omega} &=& 2\gamma\bar{g}_{\mathcal{F},R} - \bar{g}^{2}_{\mathcal{U},R} 
+ \bar{g}^{2}_{\mathcal{B},R}\frac{\omega}{\Delta k_{F,R}}\frac{d\Delta k_{F,R}}{d\omega}-\bar{g}_{\mathcal{F},R}^{3} \nonumber \\
&-& \bar{g}_{\mathcal{0},R}^{2}\bar{g}_{\mathcal{F},R} -\bar{g}_{\mathcal{U},R}^{2}\bar{g}_{0,R} \nonumber \\
&-& \bar{g}_{\mathcal{B},R}^{2}\bar{g}_{0,R}\frac{\omega}{\Delta k_{F,R}}\frac{d\Delta k_{F,R}}{d\omega}, 
\end{eqnarray}
\begin{eqnarray}
\omega\frac{d\bar{g}_{\mathcal{B},R}}{d\omega} &=& 2\gamma\bar{g}_{\mathcal{B},R} 
+ \bar{g}_{\mathcal{B},R}\bar{g}_{\mathcal{F},R}\frac{\omega}{\Delta k_{F,R}}\frac{d\Delta k_{F,R}}{d\omega} 
+ \bar{g}_{\mathcal{B},R}\bar{g}_{\mathcal{F},R} \nonumber \\
&-&\bar{g}_{0,R}\bar{g}_{B,R}\frac{\omega}{\Delta k_{F,R}}\frac{d\Delta k_{F,R}}{d\omega}
- \bar{g}_{0,R}\bar{g}_{\mathcal{B},R} \nonumber \\
&-& 2\bar{g}_{0,R}\bar{g}_{\mathcal{F},R}\bar{g}_{\mathcal{B},R}\frac{\omega}{\Delta k_{F,R}}\frac{d\Delta k_{F,R}}{d\omega} \nonumber \\
&-&\bar{g}_{\mathcal{U},R}^{2}\bar{g}_{\mathcal{B},R}\frac{\omega}{\Delta k_{F,R}}\frac{d\Delta k_{F,R}}{d\omega} 
\nonumber \\
&-& \bar{g}_{\mathcal{U},R}^{2}\bar{g}_{\mathcal{B},R}, 
\end{eqnarray}
\begin{eqnarray}
\omega\frac{d\bar{g}_{\mathcal{U},R}}{d\omega} &=& 2\gamma\bar{g}_{\mathcal{U},R} 
+ 2\bar{g}_{0,R}\bar{g}_{\mathcal{U},R} 
- 2\bar{g}_{\mathcal{F},R}\bar{g}_{\mathcal{U},R} \nonumber \\
&-& 2\bar{g}_{0,R}\bar{g}_{\mathcal{F},R}\bar{g}_{U,R}, 
\nonumber \\
&-& 2\bar{g}_{\mathcal{U},R}\bar{g}_{\mathcal{B},R}^{2}\frac{\omega}{\Delta k_{F,R}}\frac{d\Delta k_{F,R}}{d\omega}. 
\end{eqnarray}
Where the RG flow equation for $\Delta k_{F,R}$ , eq.(\ref{deltakfrg}), at 2-loops can be written, for large cutoff $\Omega$, as
\begin{eqnarray}
\omega\frac{d \ln\Delta k_{F,R}}{d\omega}
&=& \gamma.
\end{eqnarray}
Now we can calculate the flow to $Z$
\begin{eqnarray}
Z &=& 1+\delta Z \nonumber \\
&=& 1 -\left[\frac{\bar{g}_{fR}^{2}}{2}
+\frac{\bar{g}_{0R}^{2}}{2} + \frac{\bar{g}_{uR}^{2}}{2}\right]\ln\left(\frac{\Omega}{\omega}\right) \nonumber \\
&-& \frac{\bar{g}_{bR}^{2}}{2}\ln\left(\frac{\Omega}{V_{F}\Delta k_{F}}\right)
\end{eqnarray}
and the anomalous dimension
\begin{eqnarray}
\gamma = \frac{\bar{g}_{fR}^{2}}{2}
+\frac{\bar{g}_{0R}^{2}}{2} + \frac{\bar{g}_{uR}^{2}}{2} 
- \frac{\bar{g}_{bR}^{2}}{2}\frac{\omega}{\Delta k_{F}}\frac{d\Delta k_{F}}{d\omega},
\end{eqnarray}
that reduces, at 2-loops, to the following result
\begin{eqnarray}
2\gamma = \bar{g}_{fR}^{2}
+ \bar{g}_{0R}^{2} + \bar{g}_{uR}^{2}.
\end{eqnarray}
Taking into account only terms up to 2-loops order in the renormalized coupling terms, we can finally write 
the RG flow equations for the intraforward
\begin{eqnarray}
\omega\frac{d\bar{g}_{0,R}}{d\omega} &=& 2\gamma\bar{g}_{0,R} + 
\bar{g}^{2}_{\mathcal{U},R} -\bar{g}_{0,R}^{3} - \bar{g}_{\mathcal{F},R}^{2}\bar{g}_{0,R} \nonumber \\
&-& \bar{g}_{\mathcal{U},R}^{2}\bar{g}_{\mathcal{F},R}, 
\end{eqnarray}
interforward
\begin{eqnarray}
\omega\frac{d\bar{g}_{\mathcal{F},R}}{d\omega} &=& 2\gamma\bar{g}_{\mathcal{F},R} - \bar{g}^{2}_{\mathcal{U},R} 
-\bar{g}_{\mathcal{F},R}^{3} - \bar{g}_{0,R}^{2}\bar{g}_{\mathcal{F},R} \nonumber \\ &-&\bar{g}_{\mathcal{U},R}^{2}\bar{g}_{0,R}, 
\end{eqnarray}
interbackscattering 
\begin{eqnarray}
\omega\frac{d\bar{g}_{\mathcal{B},R}}{d\omega} &=& 2\gamma\bar{g}_{\mathcal{B},R} 
+ \bar{g}_{\mathcal{B},R}\bar{g}_{\mathcal{F},R} -\bar{g}_{0,R}\bar{g}_{\mathcal{B},R} \nonumber \\
&-& \bar{g}_{\mathcal{U},R}^{2}\bar{g}_{\mathcal{B},R}, 
\end{eqnarray}
interumklapp 
\begin{eqnarray}
\omega\frac{d\bar{g}_{\mathcal{U},R}}{d\omega} &=& 2\gamma\bar{g}_{\mathcal{U},R} 
+ 2\bar{g}_{0,R}\bar{g}_{\mathcal{U},R} 
- 2\bar{g}_{\mathcal{F},R}\bar{g}_{\mathcal{U},R} \nonumber \\
&-& 2\bar{g}_{0,R}\bar{g}_{\mathcal{F},R}\bar{g}_{\mathcal{U},R}, 
\end{eqnarray}
and the difference of the renormalized interband Fermi points 
\begin{eqnarray}
\omega\frac{d\Delta k_{F,R}}{d\omega}
&=& \gamma \Delta k_{F,R}.
\end{eqnarray}
By solving these RG flow equations and the corresponding RG flow equation 
for the quasi-particle weight $Z$, we found that in the RG flows to the
confinement, where the renormalized $\Delta k_{F,R}$ vanishes, that the corresponding 
quasi-particle weight $Z$ flows to zero and the interaction coupling terms flow to Luttinger liquid fixed points, with
intraforward, interforward and interumklapp interactions RG flowing to non-vanishing
 fixed points and the interbackscattering interactions RG flowing to zero as a consequence of its relation to the confinement 
term $\Delta k_{F,R}$. This result in fact is applied for many different cases, positive initial renormalized couplings
 (figure \ref{conf2pos}), including negative initial renormalized coupling 
terms (except interumklapp) (figure \ref{conf2neg}). 
For initial negative values, except for the interumklapp interaction, the negative initial 
values still move towards the fixed points. As the intraforward and interforward interactions flow to opposite value fixed points, their
sum RG flows to zero and the corresponding Luttinger theorem is satisfied in the fixed points.

As a consequence, in the confinement of renormalized chiral interband Fermi points, the bands do not stop comunicating but they still 
interact under intraforward, interforward and interumklapp in a dinamical equilibrium characterized by the Luttinger liquid fixed 
point regimes. 


\section{Chiral tranformations, Ward-Takahashi Identity and Quantum Anomaly}

\begin{figure}[H]
\centering
\includegraphics[scale=0.4]{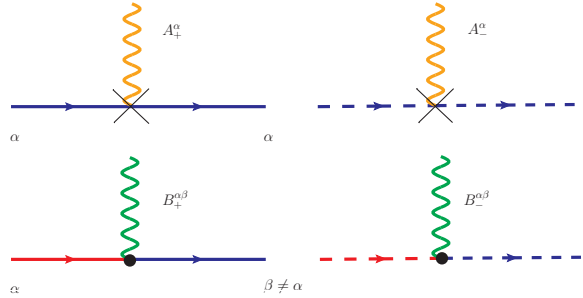}
\caption{(Color online) Interactions that change color and that do not change color.}
\label{bandint}
\end{figure}

In $1+1$ spacetime coordinates, we can also represent the action of the quasi-1D system as $1+1$ Dirac field under external 
field interactions that 
do not change color $A_{s}^{\alpha}$ and interactions that change color $B_{s}^{\alpha\beta}$ (figure \ref{bandint})
\begin{eqnarray}
\mathcal{S}
&=& \int d^{2}x\sum_{\alpha}\bar{\psi}^{\alpha}\slashed\partial\psi^{\alpha} + \sum_{\alpha, s} \int d^{2}x\psi_{s}^{\alpha\dagger}A_{s}^{\alpha}\psi_{s}^{\alpha} \nonumber \\
&+& \sum_{\alpha\neq\beta,s}\int d^{2}x \psi_{s}^{\alpha\dagger}B_{s}^{\alpha\beta}\psi_{s}^{\beta},
\end{eqnarray}
where $\alpha=a,b$; $\beta=a,b$; $s=\pm$.
From the previous sections, we showed that only interumklapp and interbackscattering interactions change color, such that these 
interactions represent color exchange interaction $B^{\alpha\beta}_{s}$. Note that $B^{\alpha\beta}_{s}=0$, if $\alpha=\beta$. 

As a whole, we can represent the external fields as a single external field interaction 
\begin{eqnarray}
\mathcal{A}^{\alpha\beta}_{s} = A^{\alpha}_{s} + B^{\alpha\beta}_{s}
\end{eqnarray}
that couples to the $1+1$ free Dirac fields as follows
\begin{eqnarray}
\mathcal{L}&=&\sum_{\alpha}\bar{\psi}^{\alpha}\slashed\partial \psi^{\alpha} 
+ \sum_{\alpha,\beta,s}\psi_{s}^{\alpha\dagger}\mathcal{A}^{\alpha\beta}_{s}\psi_{s}^{\beta}, 
\end{eqnarray}
where $\alpha=a,b$, $\beta=a,b$, $s=\pm$.

As a procedure in $1+1$ \cite{bala}, let us consider a chiral transformation, involving the $+$ side,
\begin{eqnarray}
\psi_{+} &\rightarrow& e^{-i\theta(x)}\psi_{+} = (1-i\theta(x))\psi_{+} \\
\psi_{+}^{\dagger} &\rightarrow& e^{i\theta(x)}\psi^{\dagger}_{+} = (1+i\theta(x))\psi_{+}^{\dagger}
\end{eqnarray}
and the action will change as
\begin{eqnarray}
\mathcal{S} &\rightarrow& \mathcal{S} + \int d^{2}x J_{+}^{\alpha}(-i\partial_{+}\theta(x)), 
\end{eqnarray}
where $J_{+}^{\alpha}=\psi_{+}^{\alpha\dagger}\psi_{+}^{\alpha}$ is the corresponding current in the $+$ side, 
and $\partial_{+}= \partial_{t} -iv_{F}\partial_{x}$.

The functional generator will change as follows \cite{ryder}
\small
\begin{eqnarray}
\mathcal{Z'}[J,\eta, \eta^{\dagger}]&=& \int \mathcal{D}\left(\mathcal{A}',\psi',\psi'^{\dagger}\right) \nonumber \\
&\times&e^{\left[\mathcal{S} + \int d^{2}x \left( \sum_{\alpha,\beta;s} J_{s}^{\alpha\beta}\mathcal{A}_{s}^{\alpha\beta} + \sum_{\alpha;s} 
\eta_{s}^{\alpha\dagger}\psi_{s}^{\alpha} 
- \sum_{\alpha;s} \psi_{s}^{\alpha\dagger}\eta_{s}^{\alpha} \right) \right]}\nonumber \\
&\times& e^{
\int d^{2}x \left( \sum_{\alpha} [-i\partial_{+}\theta(x)] J_{+}^{\alpha} + \sum_{\alpha} 
i\theta(x)\eta_{+}^{\alpha\dagger}\psi_{+}^{\alpha} 
- \sum_{\alpha} i\theta(x)\psi_{+}^{\alpha\dagger}\eta_{+}^{\alpha} \right)},\nonumber \\ 
\end{eqnarray}
\normalsize
leading to the following Ward-Takahashi identity
\begin{eqnarray}
\theta(x)\left( \sum_{\alpha} i\partial_{+}J_{+}^{\alpha} + \sum_{\alpha} 
i\eta_{+}^{\alpha\dagger}\psi_{+}^{\alpha} 
- \sum_{\alpha} i\psi_{+}^{\alpha\dagger}\eta_{+}^{\alpha} \right)Z=0. \nonumber \\
\end{eqnarray}
As usual we will make the Ward-Takahashi associations\cite{ryder}
\begin{eqnarray}
A^{\alpha}_{+} &\rightarrow&  \frac{1}{i}\frac{\delta}{\delta J_{+}^{\alpha}} \\
\psi^{\alpha}_{+} &\rightarrow&  \frac{1}{i}\frac{\delta}{\delta \eta_{+}^{\alpha\dagger}} \\
\psi^{\alpha\dagger}_{+} &\rightarrow&  \frac{1}{i}\frac{\delta}{\delta \eta_{+}^{\alpha}} 
\end{eqnarray}
Such that 
\begin{eqnarray}
\theta(x)\left( \sum_{\alpha} i\partial_{+}J_{+}^{\alpha\beta} + \sum_{\alpha} 
i\eta_{+}^{\alpha\dagger}\frac{1}{i}\frac{\delta}{\delta \eta_{+}^{\alpha\dagger}}
+ \sum_{\alpha} i\frac{1}{i}\eta_{+}^{\alpha}\frac{\delta}{\delta \eta_{+}^{\alpha}} \right)Z=0 \nonumber \\
\end{eqnarray}
Making the substitution $Z=e^{iW}$
\begin{eqnarray}
\theta(x)\left( \sum_{\alpha} i\partial_{+}J_{+}^{\alpha\beta} + \sum_{\alpha} 
i\eta_{+}^{\alpha\dagger}\frac{\delta W}{\delta \eta_{+}^{\alpha\dagger}}
+ \sum_{\alpha} i\eta_{+}^{\alpha}\frac{\delta W}{\delta \eta_{+}^{\alpha}} \right)=0 \nonumber \\
\end{eqnarray}
As a consequence we have the Ward-Takahashi relation
\begin{eqnarray}
\sum_{\alpha} i\partial_{+}J_{+}^{\alpha} &=& -\left(\sum_{\alpha} 
i\eta_{+}^{\alpha\dagger}\frac{\delta W}{\delta \eta_{+}^{\alpha\dagger}} 
+ \sum_{\alpha} i\eta_{+}^{\alpha}\frac{\delta W}{\delta \eta_{+}^{\alpha}} \right). \nonumber \\
\label{wit}
\end{eqnarray}
Note that the current contribution is due to one-color interactions.
 
\begin{figure}[ht]
\begin{minipage}[b]{0.4\linewidth}
\centering
\includegraphics[scale=0.3]{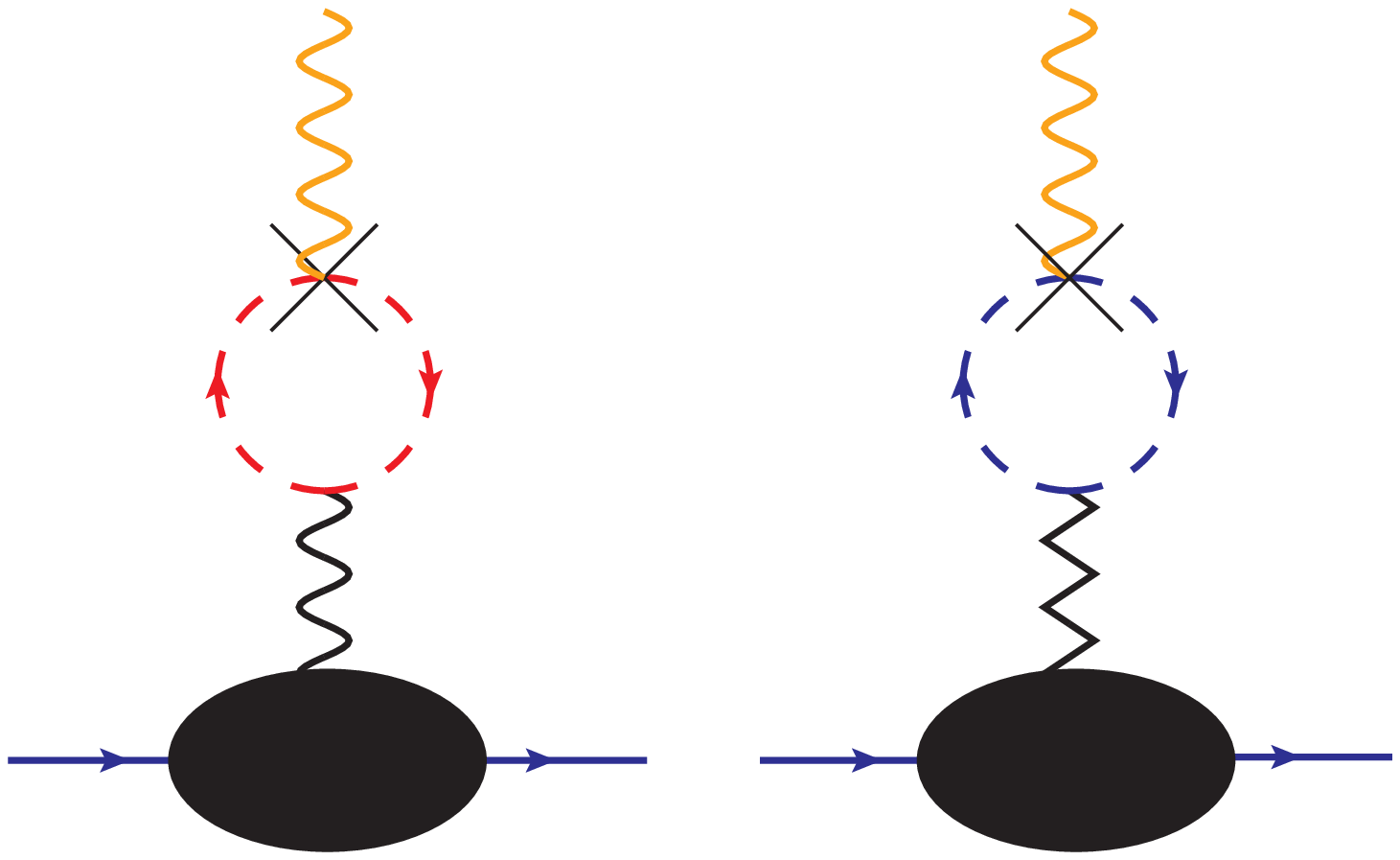}
\includegraphics[scale=0.3]{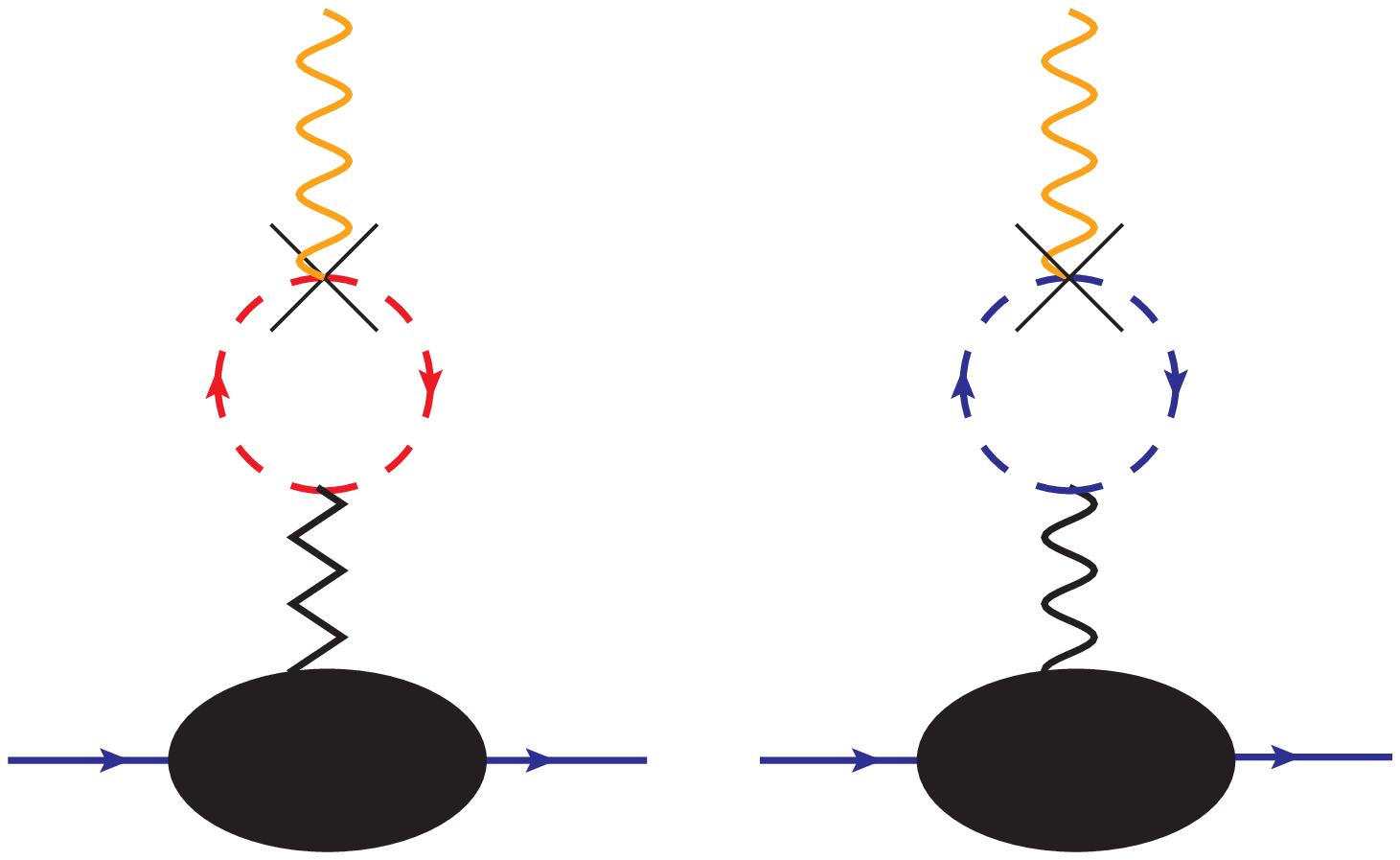}
\label{fig:figure1}
\end{minipage}
\hspace{0.2cm}
\begin{minipage}[b]{0.4\linewidth}
\centering
\includegraphics[scale=0.3]{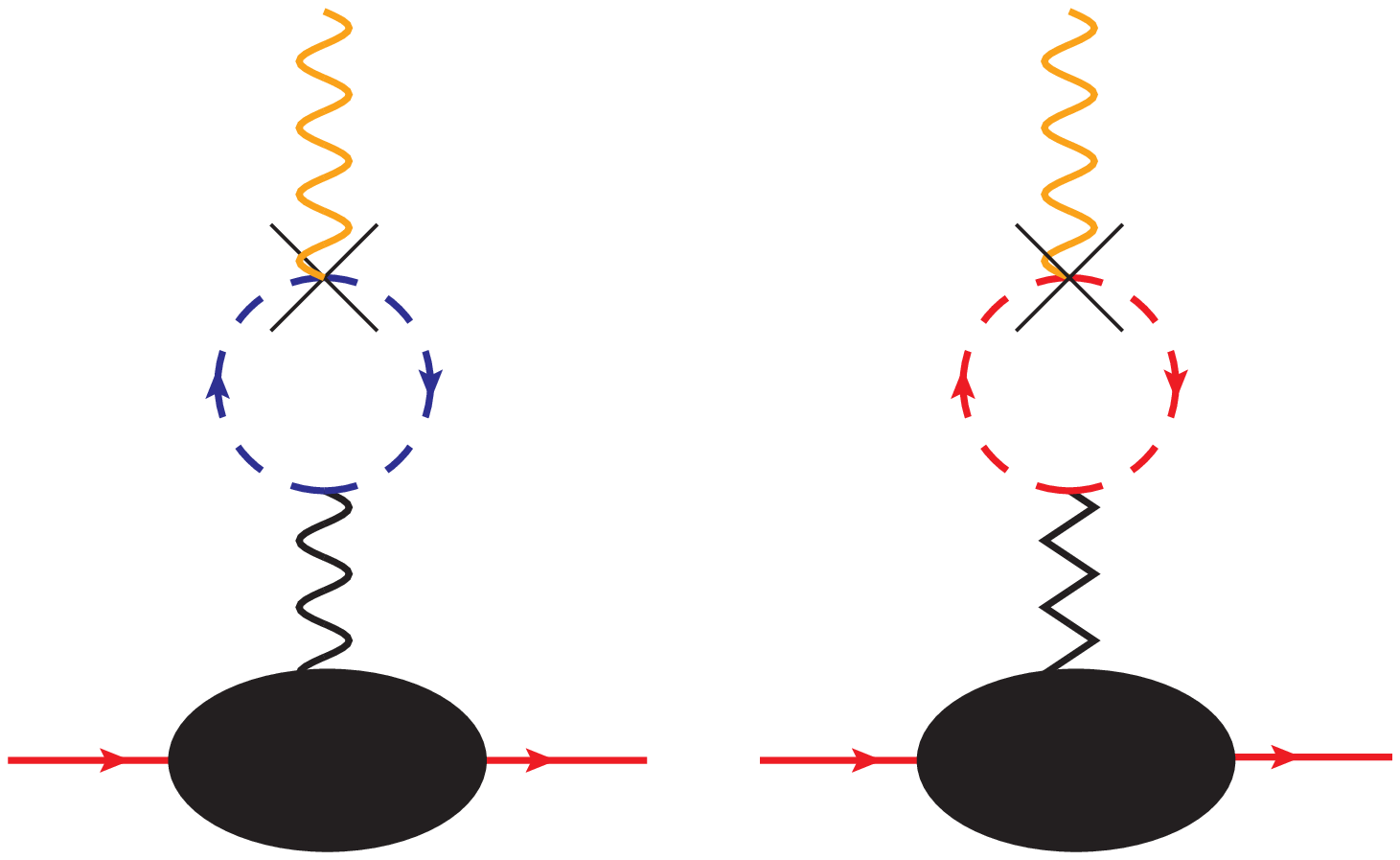}
\includegraphics[scale=0.3]{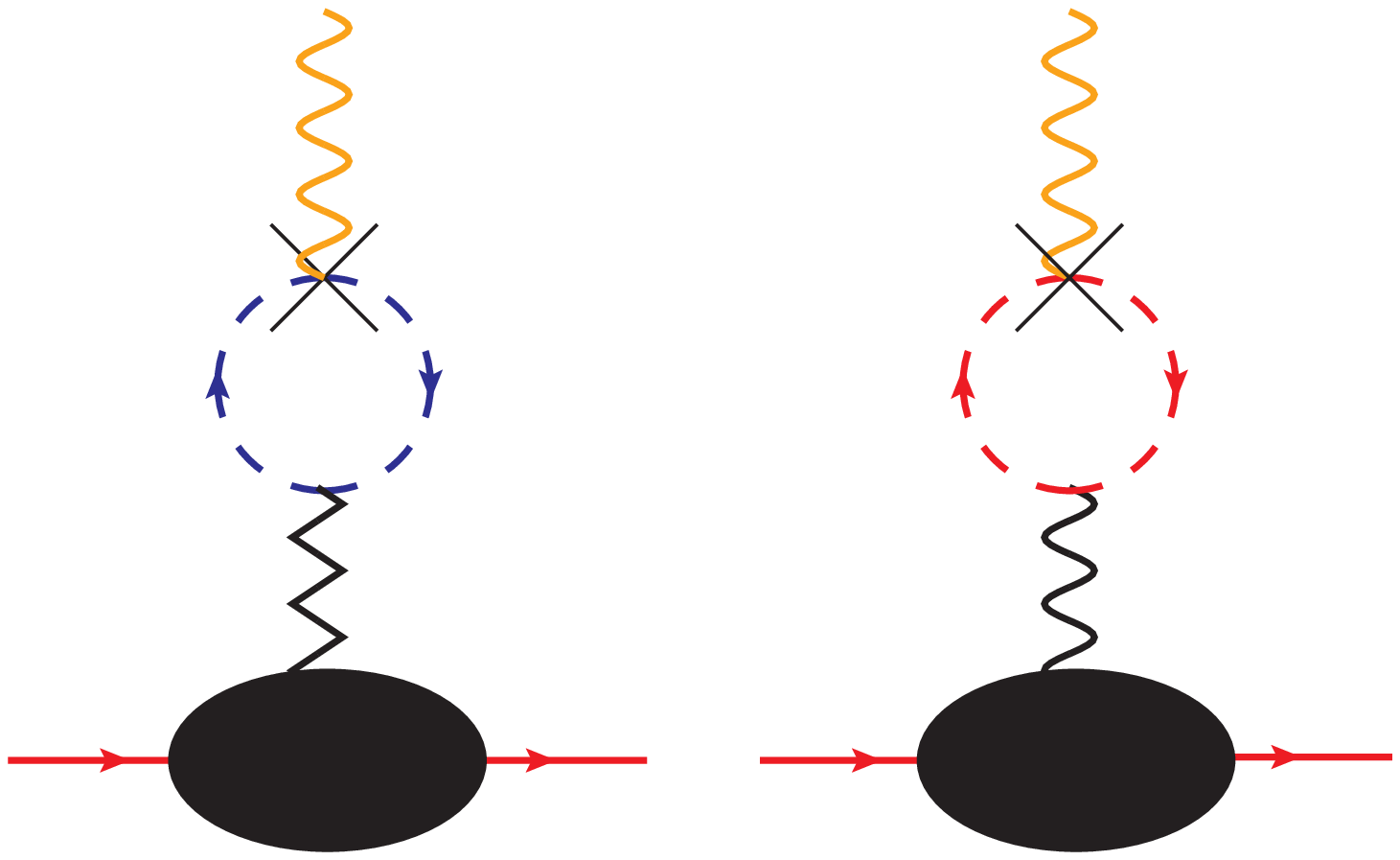}
\end{minipage}
\begin{minipage}[b]{0.4\linewidth}
\centering
\includegraphics[scale=0.3]{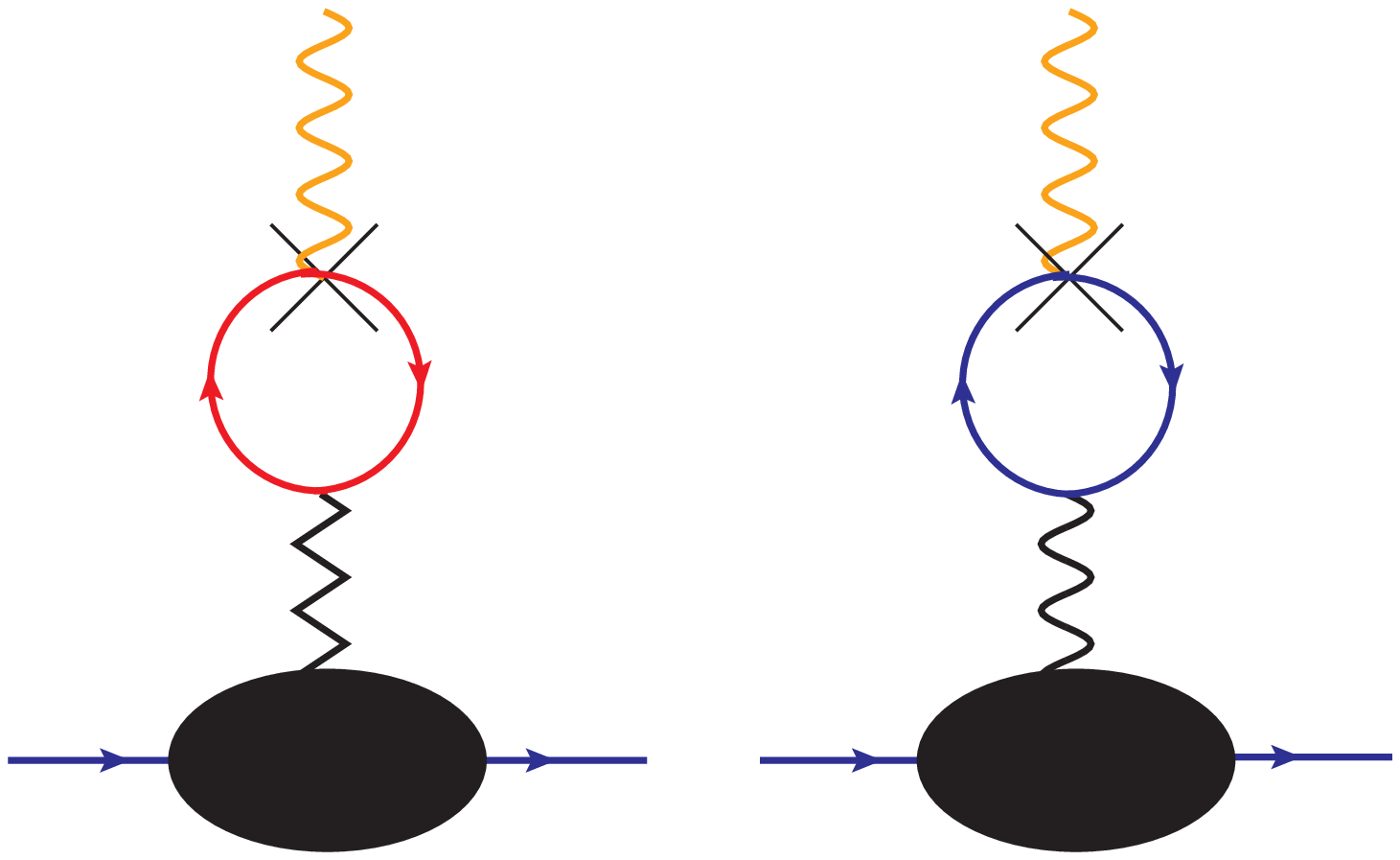}
\includegraphics[scale=0.3]{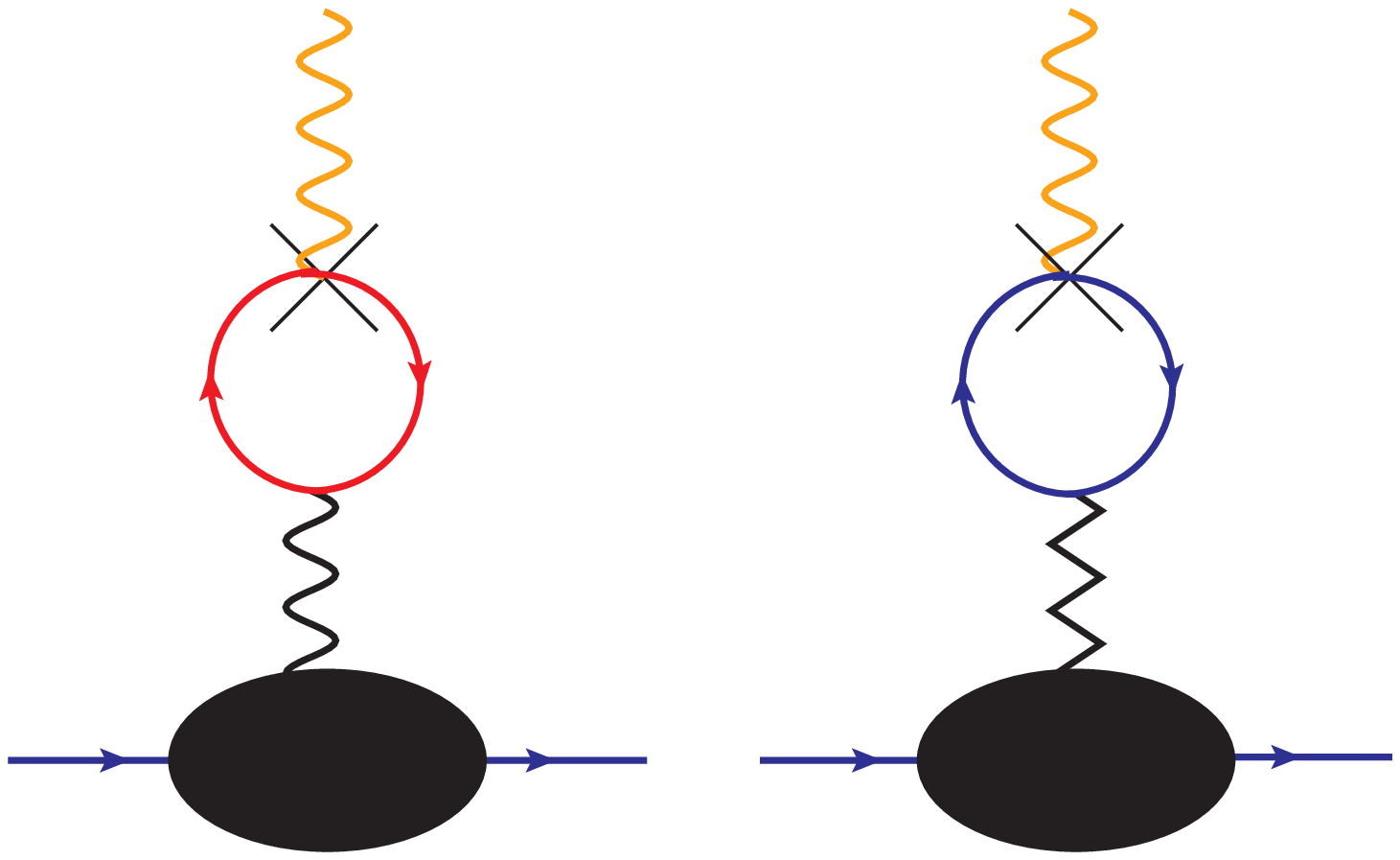}
\label{fig:figure1}
\end{minipage}
\hspace{0.2cm}
\begin{minipage}[b]{0.4\linewidth}
\centering
\includegraphics[scale=0.3]{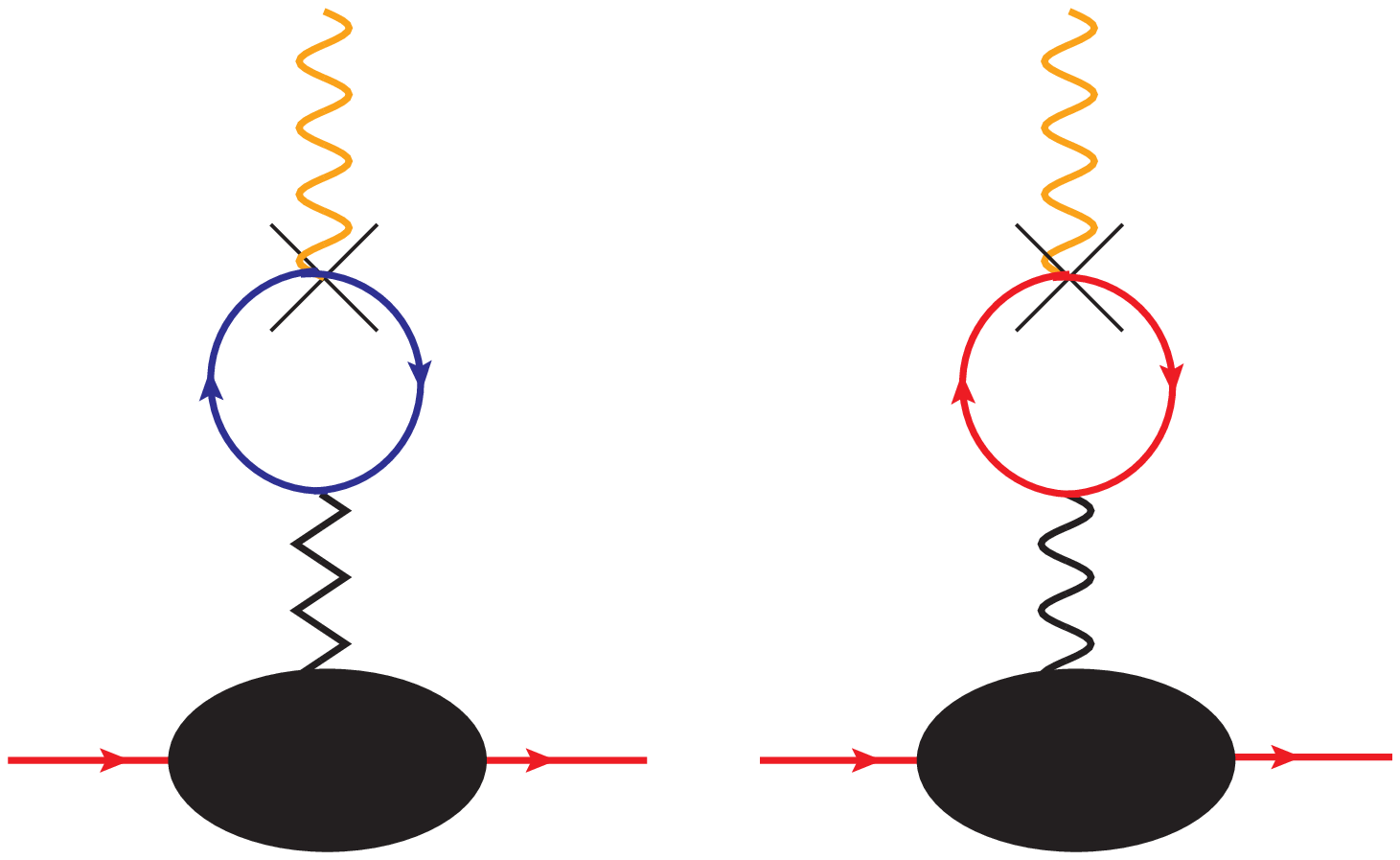}
\includegraphics[scale=0.3]{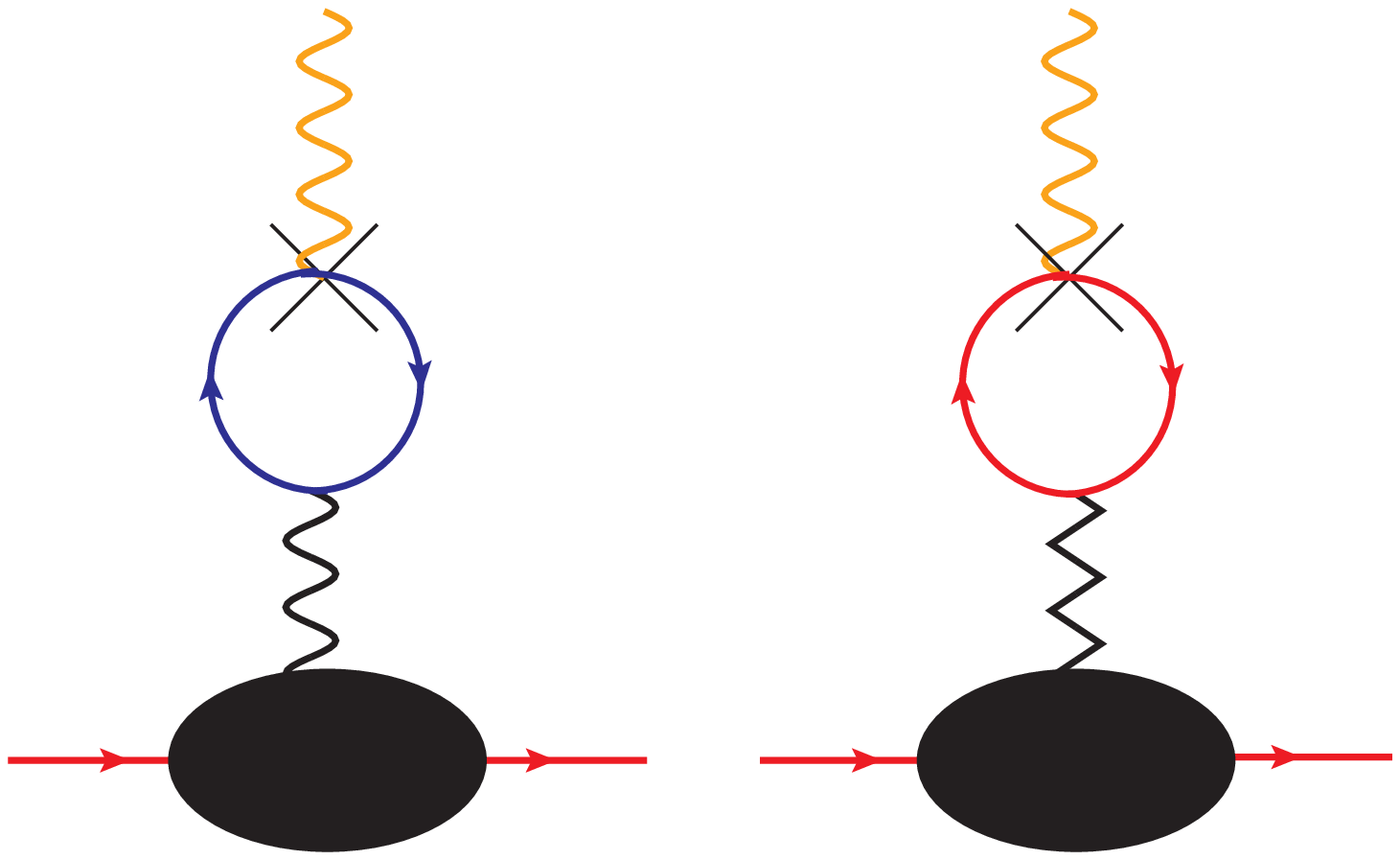}
\end{minipage}
\caption{(Color online) Generic anomalous terms in the Vertex functions for right-hand side.}
\label{totalan}
\end{figure}

Now we will make an usual Legendre transform of quantum field theory\cite{ryder}
\begin{eqnarray}
\Gamma[\psi,\psi^{\dagger},A]= W[\eta,\eta^{\dagger},J] - \int d^{2}x [\psi^{\dagger}\eta + \eta^{\dagger}\psi + A J] \nonumber \\
\end{eqnarray}
Where we have
\begin{eqnarray}
\frac{\delta \Gamma}{\delta A^{\alpha\beta}_{+}} &=& -J^{\alpha\beta}_{+}, \\
\frac{\delta \Gamma}{\delta \psi^{\alpha}_{+}} &=& -\eta^{\alpha\dagger}_{+}, \\
\frac{\delta \Gamma}{\delta \psi^{\alpha\dagger}_{+}} &=& -\eta^{\alpha}_{+}, 
\end{eqnarray}
and
\begin{eqnarray}
\frac{\delta W}{\delta J^{\alpha\beta}_{+}} &=& A^{\alpha\beta}_{+}, \\
\frac{\delta W}{\delta \eta^{\alpha\dagger}_{+}} &=& \psi^{\alpha}_{+}, \\
\frac{\delta W}{\delta \eta^{\alpha}_{+}} &=& \psi^{\alpha\dagger}_{+}. 
\end{eqnarray}
Then, from (\ref{wit}), we have the following 
\begin{eqnarray}
-\sum_{\alpha} i\partial_{+}^{\vec{x}}\frac{\delta \Gamma}{\delta A^{\alpha}_{+}(\vec{x})} &=& \sum_{\alpha} 
i\frac{\delta \Gamma}{\delta \psi^{\alpha}_{+}(\vec{x})}\psi^{\alpha}_{+}(\vec{x}) \nonumber \\
&+& \sum_{\alpha} i\frac{\delta \Gamma}{\delta \psi^{\alpha\dagger}_{+}(\vec{x})}\psi^{\alpha\dagger}_{+}(\vec{x})  \nonumber \\
\end{eqnarray}
Taking the functional derivatives with respect to $\delta \psi_{s'}^{\alpha'\dagger}(\vec{z})\delta \psi_{s'}^{\alpha'}(\vec{y})$, we arrive 
at the Ward-Takahashi identity
\begin{eqnarray}
\sum_{\alpha} \partial_{+}^{\vec{x}}\frac{\delta^{3} \Gamma}{\delta \psi_{s'}^{\alpha'\dagger}(\vec{z})\delta \psi_{s'}^{\alpha'}(\vec{y})\delta A^{\alpha}_{+}(\vec{x})} &=& \nonumber \\  
\frac{\delta^{2} \Gamma}{\delta \psi_{s'}^{\alpha'\dagger}(\vec{z})\delta \psi^{\alpha'}_{+}(\vec{y})}\delta_{+,s'}\delta(\vec{x}-\vec{y}) 
&-& \frac{\delta^{2} \Gamma}{\delta \psi^{\alpha'\dagger}_{+}(\vec{z})\delta \psi_{s'}^{\alpha'}(\vec{y})}\delta_{+,s'}\delta(\vec{x}-\vec{z}). \nonumber 
\end{eqnarray}
It follows the two relations
\begin{eqnarray}
\sum_{\alpha} \partial_{+}^{\vec{x}}\frac{\delta^{3} \Gamma}{\delta \psi_{-}^{\alpha'\dagger}(\vec{z})\delta \psi_{-}^{\alpha'}(\vec{y})\delta A^{\alpha}_{+}(\vec{x})} &=&  
0, \nonumber
\end{eqnarray}
\begin{eqnarray}
\sum_{\alpha} \partial_{+}^{\vec{x}}\frac{\delta^{3} \Gamma}{\delta \psi_{+}^{\alpha'\dagger}(\vec{z})\delta \psi_{+}^{\alpha'}(\vec{y})\delta A^{\alpha}_{+}(\vec{x})} &=&  
\frac{\delta^{2} \Gamma}{\delta \psi_{+}^{\alpha'\dagger}(\vec{z})\delta \psi^{\alpha'}_{+}(\vec{y})}\delta(\vec{x}-\vec{y}) \nonumber
\nonumber \\
&-& \frac{\delta^{2} \Gamma}{\delta \psi^{\alpha'\dagger}_{+}(\vec{z})\delta \psi_{+}^{\alpha'}(\vec{y})}\delta(\vec{x}-\vec{z}). \nonumber
\end{eqnarray}
Taking into account the corresponding Fourier transforms
\begin{eqnarray}
\int dzdye^{i(p'z-py)}\Gamma^{(2)}(z,y)&=& \nonumber \\
(2\pi)^{2}\delta(\vec{p'}-\vec{p})\Gamma^{(2)}(\vec{p}), &&\nonumber \\
\int dxdzdye^{i(p'z-py-qx)}\Gamma^{(2,1)}(z,y,x)&=& \nonumber \\
(2\pi)^{2}\delta(\vec{p'}-\vec{p}-\vec{q})\Gamma^{(2,1)}(\vec{p},\vec{q},\vec{p'}) && \nonumber 
\end{eqnarray}
\begin{eqnarray}
\partial_{x}\Gamma^{(2,1)}(z,y;x)=\delta(x-z)\Gamma^{(2)}(z,y) -\delta(x-y)\Gamma^{(2)}(z,y),  \nonumber
\end{eqnarray}
Integrating in $dxdydz$ and $e^{i(p'z-py)}e^{-iqx}$,
\begin{eqnarray}
\int dxdydze^{i(p'z-py)}e^{-iqx}\partial_{x}\Gamma^{(2,1)}&=& \nonumber \\
-iq(2\pi)^{2}\delta(p'-q-p)\Gamma^{(2,1)}(p,q,p+q) && 
\end{eqnarray}
\begin{eqnarray}
\int dx dy_{1} dx_{1} e^{i(p'x_{1} -py_{1})}e^{-iqx}\delta(x-y_{1})\Gamma^{(2)}(x_{1},y_{1})&=& \nonumber\\
\int dy_{1}dx_{1}e^{ip'x_{1}}e^{-i(p+q)y_{1}}\Gamma^{(2)}(x_{1},y_{1})&=& \nonumber \\
 i(2\pi)\delta(p'-q-p)\Gamma^{(2)}(p+q),&&  
\end{eqnarray}
\begin{eqnarray}
\int dx dy_{1} dx_{1} e^{i(p'x_{1} -py_{1})}e^{-iqx}\delta(x-x_{1})\Gamma^{(2)}(x_{1},y_{1})&=& \nonumber\\
\int dy_{1}dx_{1}e^{i(p'-q)x_{1}}e^{-ipy_{1}}\Gamma^{(2)}(x_{1},y_{1}) &=& \nonumber \\
i(2\pi)\delta(p'-q-p)\Gamma^{(2)}(p).&& 
\end{eqnarray}
we arrive at the Ward-Takahashi identity 
\begin{eqnarray}
q_{+}\sum_{\alpha}\Gamma_{\alpha\alpha',++}^{(2,1)}(p,p+q;q)=\Gamma_{\alpha',+}^{(2)}(p+q)
-\Gamma_{\alpha',+}^{(2)}(p), \nonumber\\
\end{eqnarray}
where $q_{s}=q_{0}-sv_{F}q$, $s=\pm$. Note that this is in fact a generalization 
of the Ward-Takahashi identity for the Luttinger liquid \cite{castro}. 
\begin{figure}[H]
\centering
\includegraphics[scale=0.3]{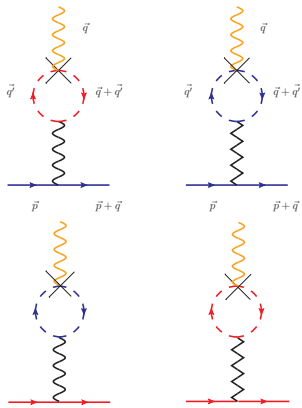}
\caption{(Color online) Vertex functions in 1-loop.}
\label{vertex1loop}
\end{figure}  
\begin{figure}[H]
\centering
\includegraphics[scale=0.2]{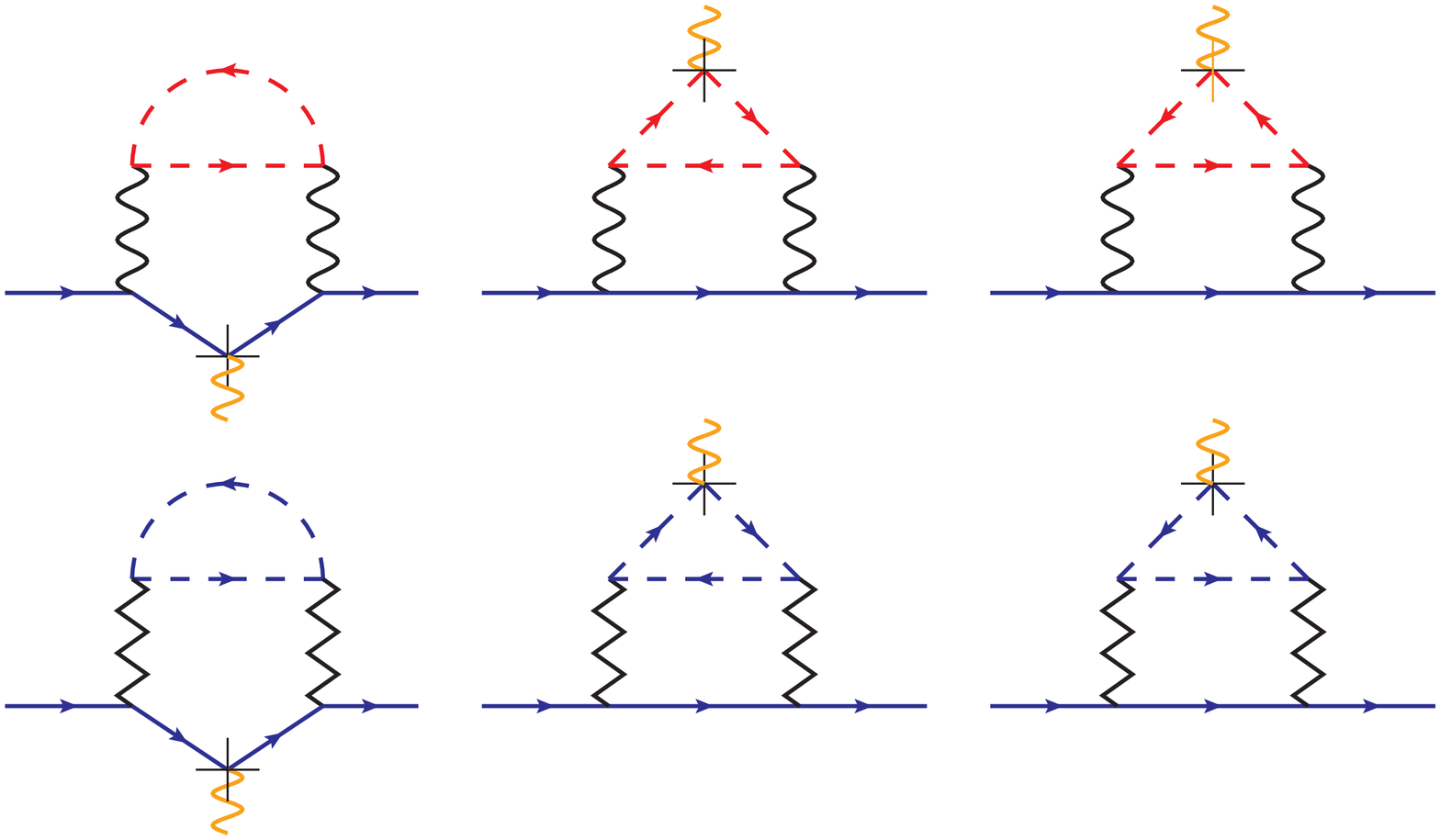}
\includegraphics[scale=0.2]{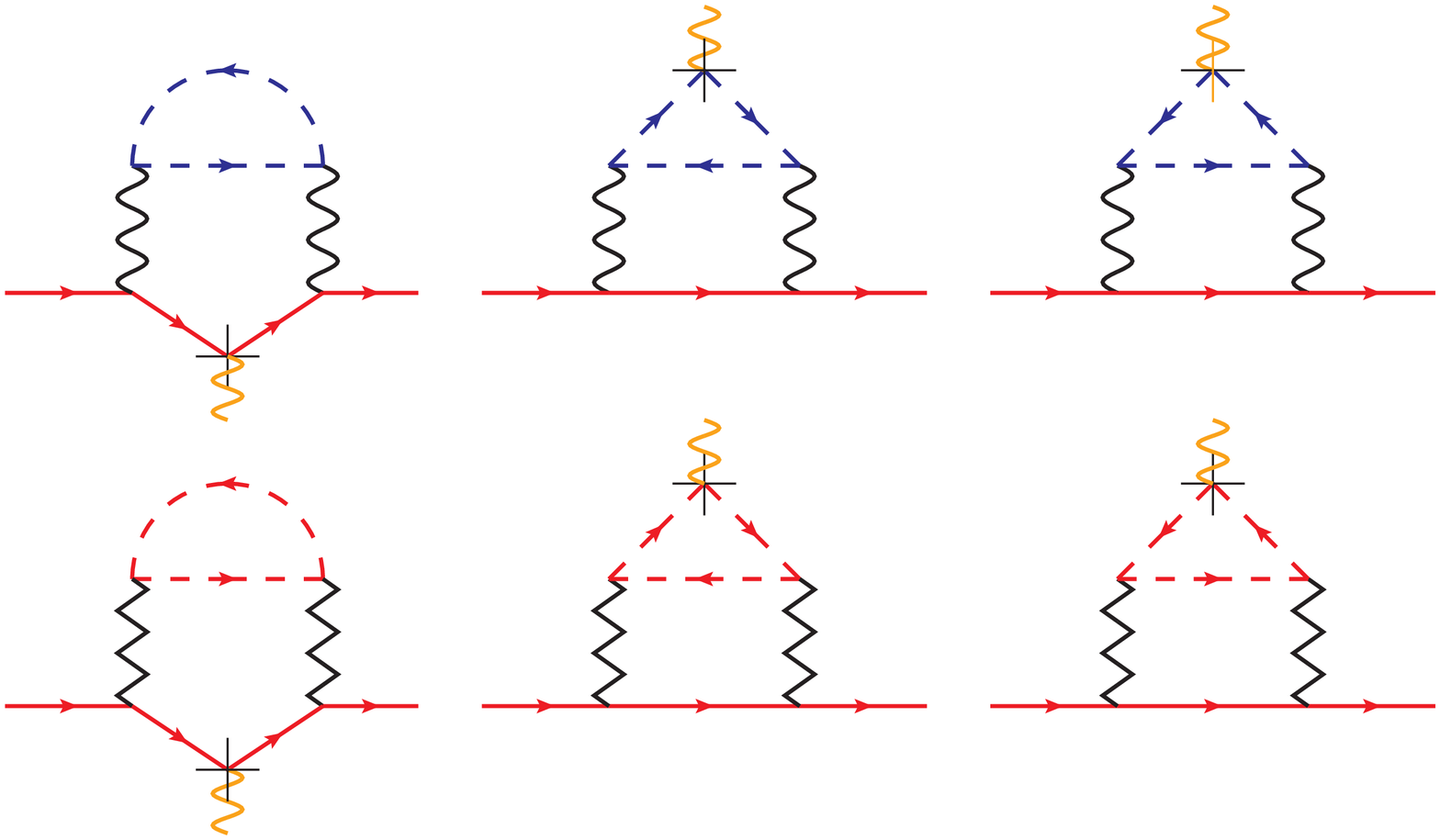}
\caption{(Color online) 2-loops vertex function diagrams with interforward and intraforward interactions.}
\label{vertexFig4isso}
\end{figure}
\begin{figure}[H]
\centering
\includegraphics[scale=0.2]{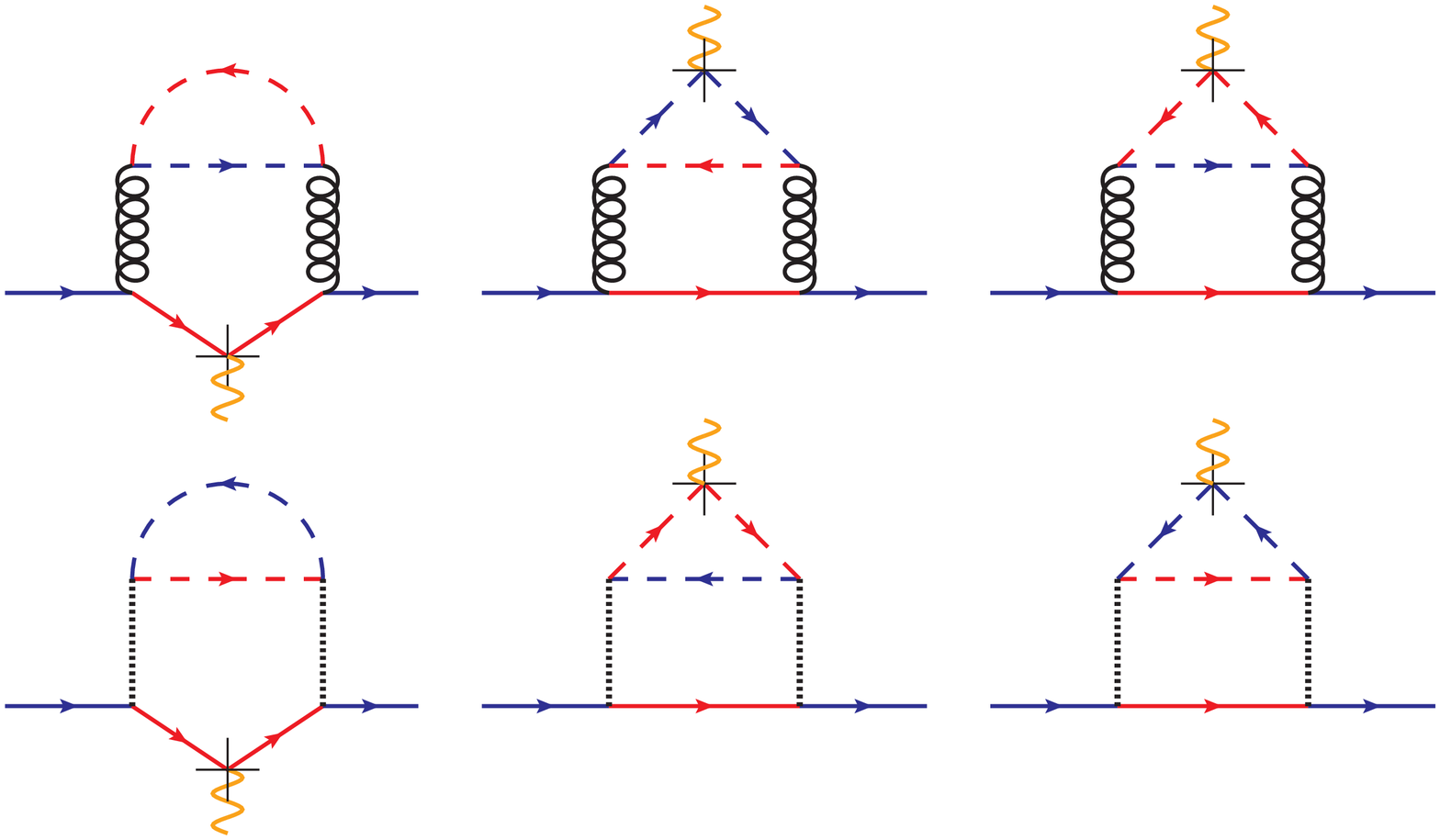}
\includegraphics[scale=0.2]{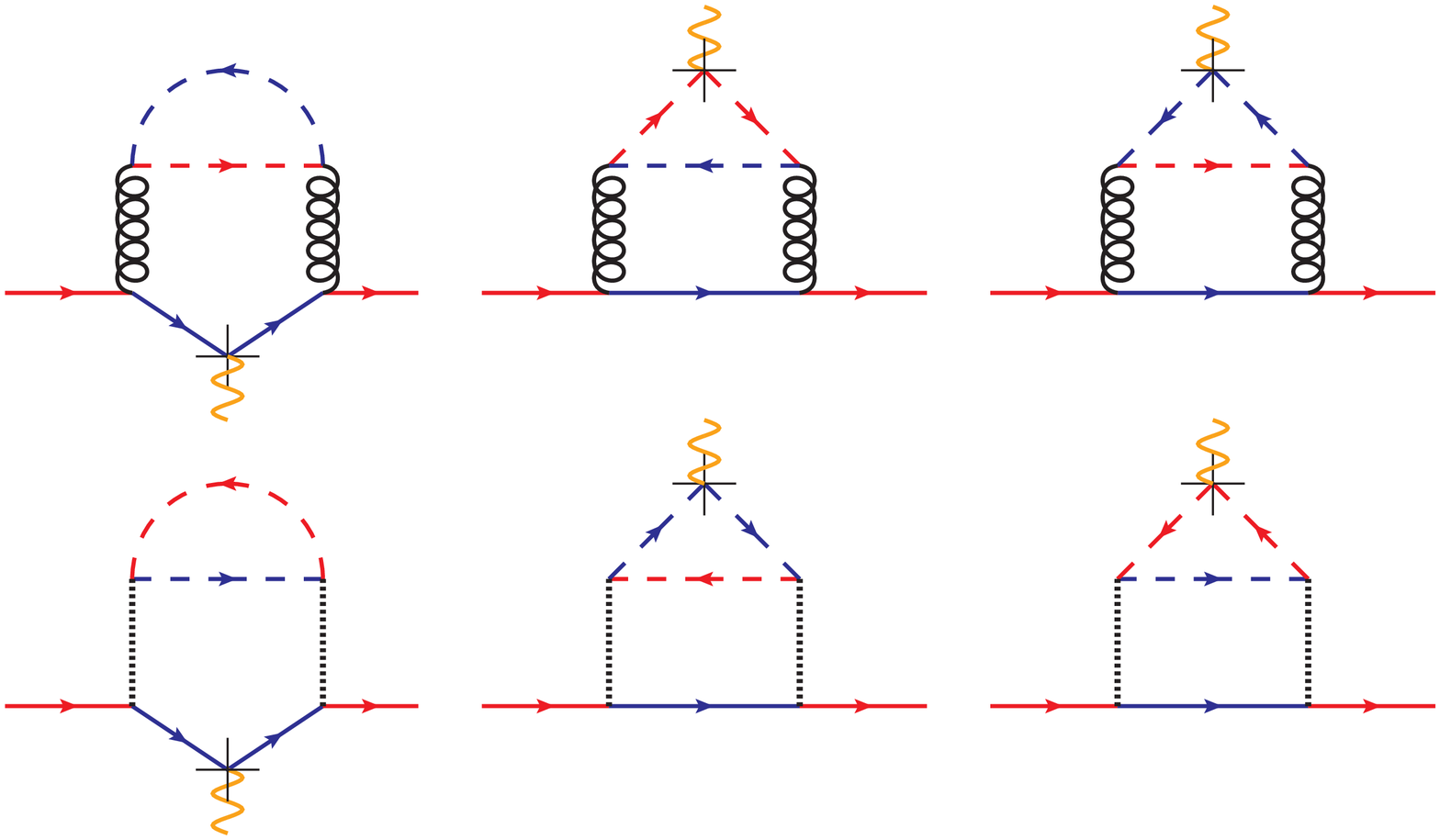}
\caption{(Color online) 2-loops vertex function with interbackscattering and interumklapp interactions.}
\label{vertexFig5isso}
\end{figure}
This WTI contains anomalous terms, due to the one color polarization bubbles, the $1+1$ corresponding 
triangle anomalies terms, 
\begin{eqnarray}
\Gamma_{ab,++}^{(2,1)(\texttt{1-loop})}(\vec{q})&=& -ig_{\mathcal{F}}\chi_{-}^{b}(\vec{q}) \nonumber \\
&=& -g_{\mathcal{F}}\frac{q}{2\pi}G_{-}^{b}(q_{0},q-k_{F}^{b}),  \\
\Gamma_{aa,++}^{(2,1)(\texttt{1-loop})}(\vec{q})&=&-ig_{0,R}\chi_{-}^{a}(\vec{q})\nonumber\\
&=&-g_{0}\frac{q}{2\pi}G_{-}^{a}(q_{0},q-k_{F}^{a}),\\
\Gamma_{ba,++}^{(2,1)(\texttt{1-loop})}(\vec{q})&=& -ig_{\mathcal{F}}\chi_{-}^{a}(\vec{q}) \nonumber \\
&=& -g_{\mathcal{F}}\frac{q}{2\pi}G_{-}^{a}(q_{0},q-k_{F}^{a}),  \\
\Gamma_{bb,++}^{(2,1)(\texttt{1-loop})}(\vec{q})&=&-ig_{0}\chi_{-}^{b}(\vec{q})\nonumber\\
&=&-g_{0,R}\frac{q}{2\pi}G_{-}^{b}(q_{0},q-k_{F}^{b}),
\end{eqnarray}
that are the 1-loop expansion of 
the $(2,1)$-vertex function $\Gamma_{\alpha\alpha',++}^{(2,1)}(p,p+q;q)$ (figures \ref{vertex1loop}), corresponding to 
the coupling of the one color interaction $A_{+}^{\alpha}$. 
\begin{figure}[H]
\centering
\includegraphics[scale=0.2]{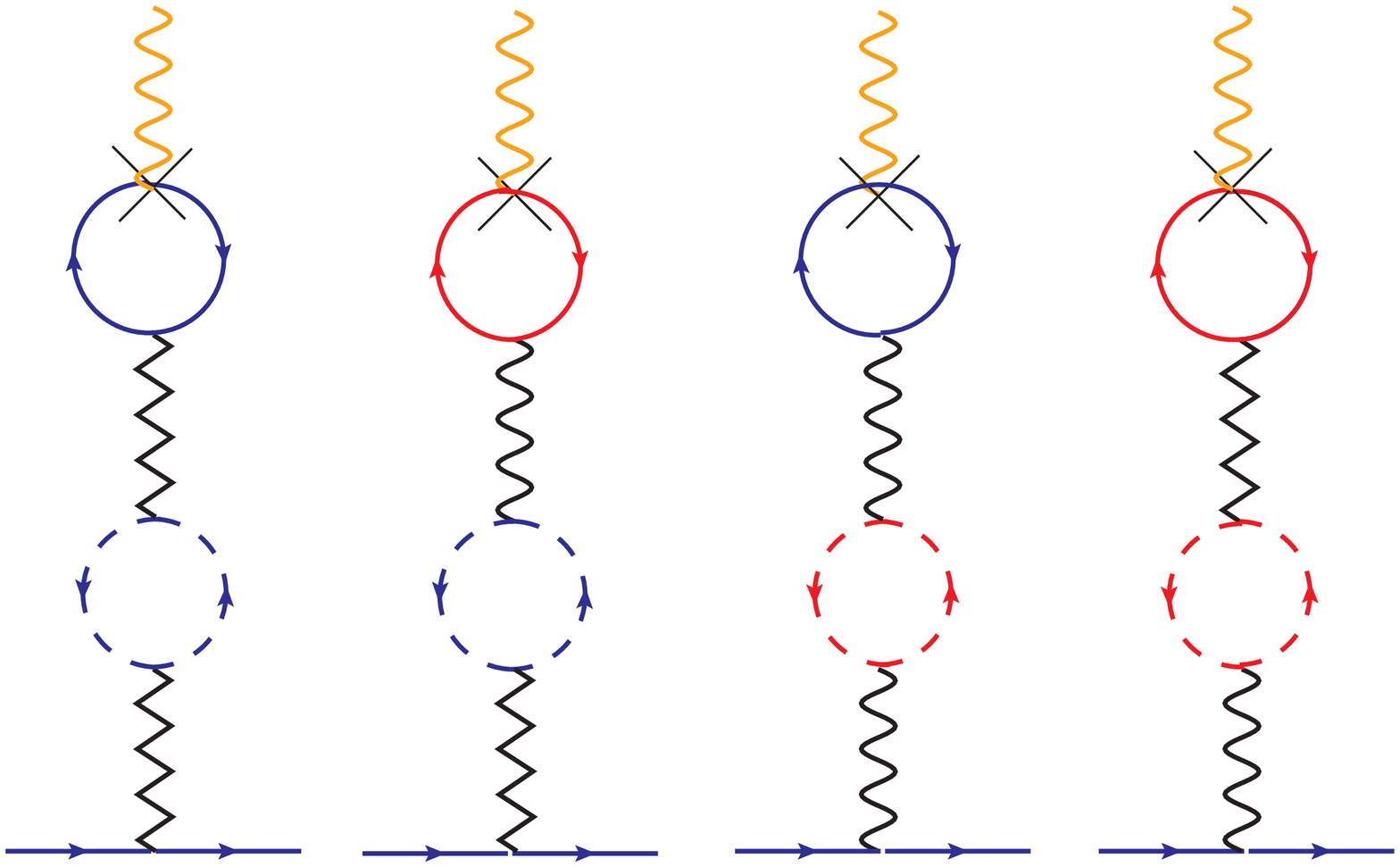}
\includegraphics[scale=0.2]{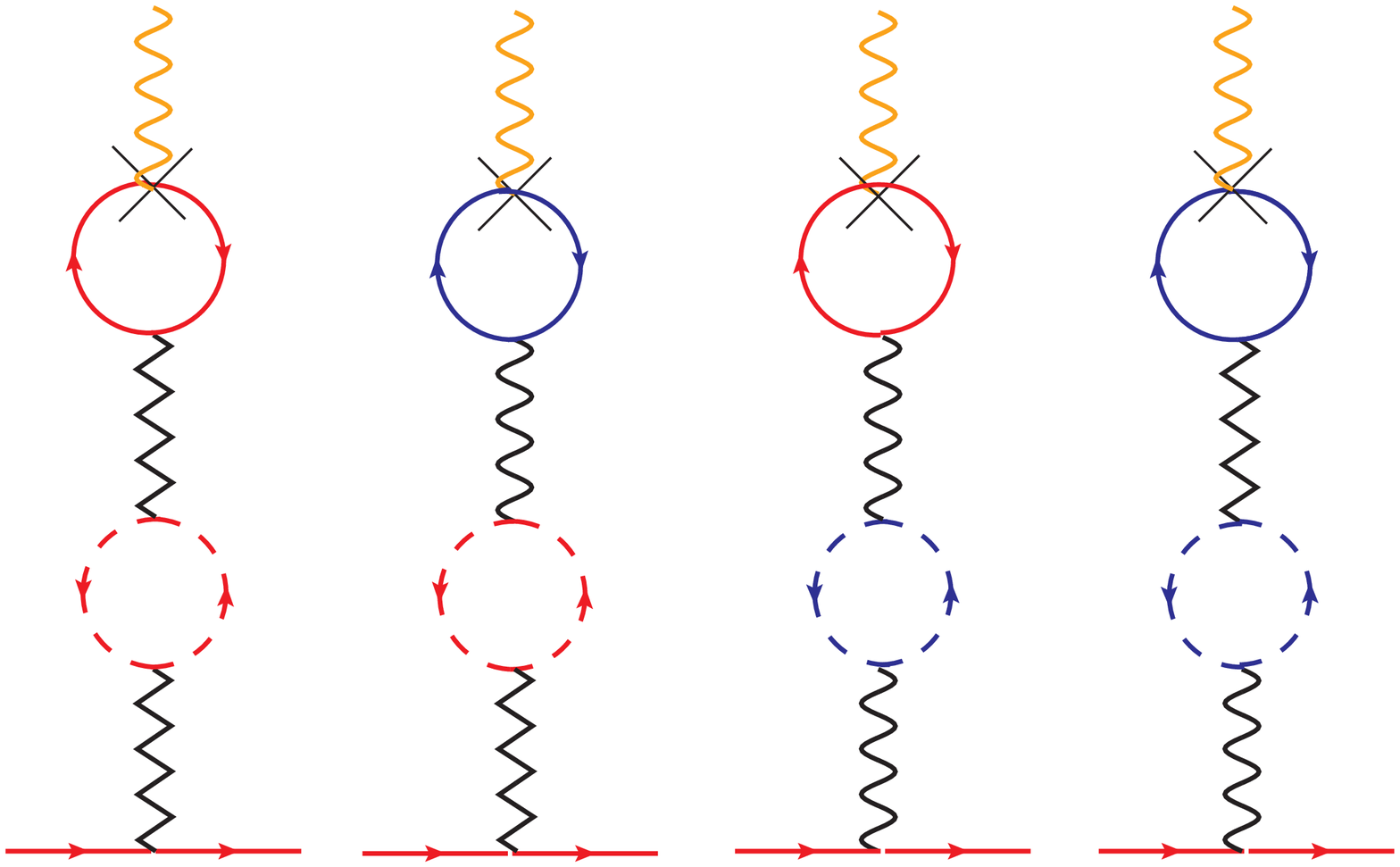}
\caption{(Color online) Anomalous terms in the Vertex functions at 2-loops.}
\label{v}
\end{figure}
On the other hand, the 2-loops order (2,1)-vertex function contributions not involving the 1-loop one-color polarization bubbles 
(figures \ref{vertexFig4isso} and \ref{vertexFig5isso}) are identically cancelled, 
as predicted in the Adler-Bardeen theorem \cite{adler}, while the 2-loops anomalous terms are parquet diagrams involving 
the 1-loop one-color polarization bubbles (figure \ref{v}). At higher order anomalous terms 
are given by the generic terms with the coupled 1-loop one-color polarization bubbles (figure \ref{totalan}). 

\begin{figure}[ht]
\begin{minipage}[b]{0.4\linewidth}
\centering
\includegraphics[scale=0.3]{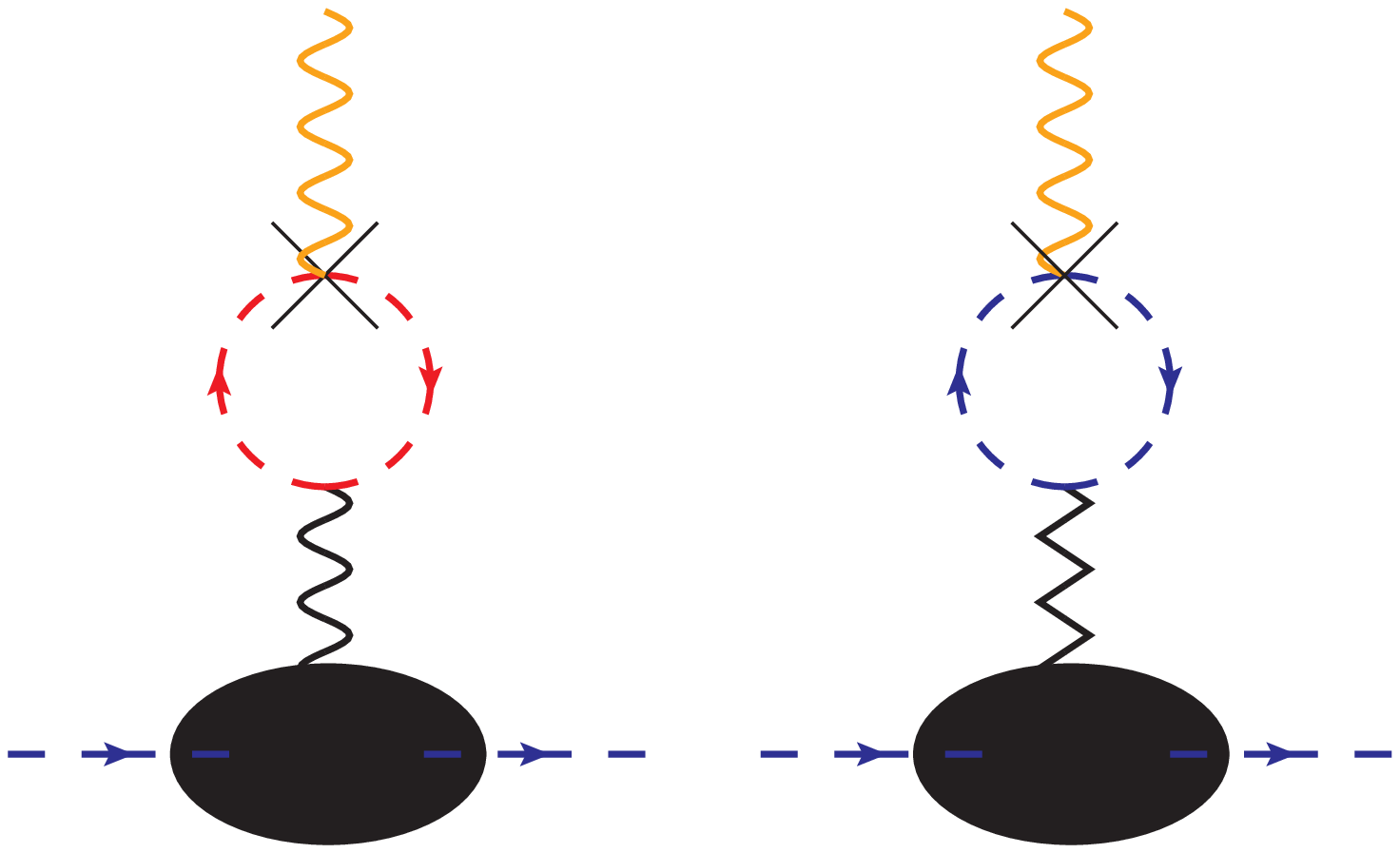}
\includegraphics[scale=0.3]{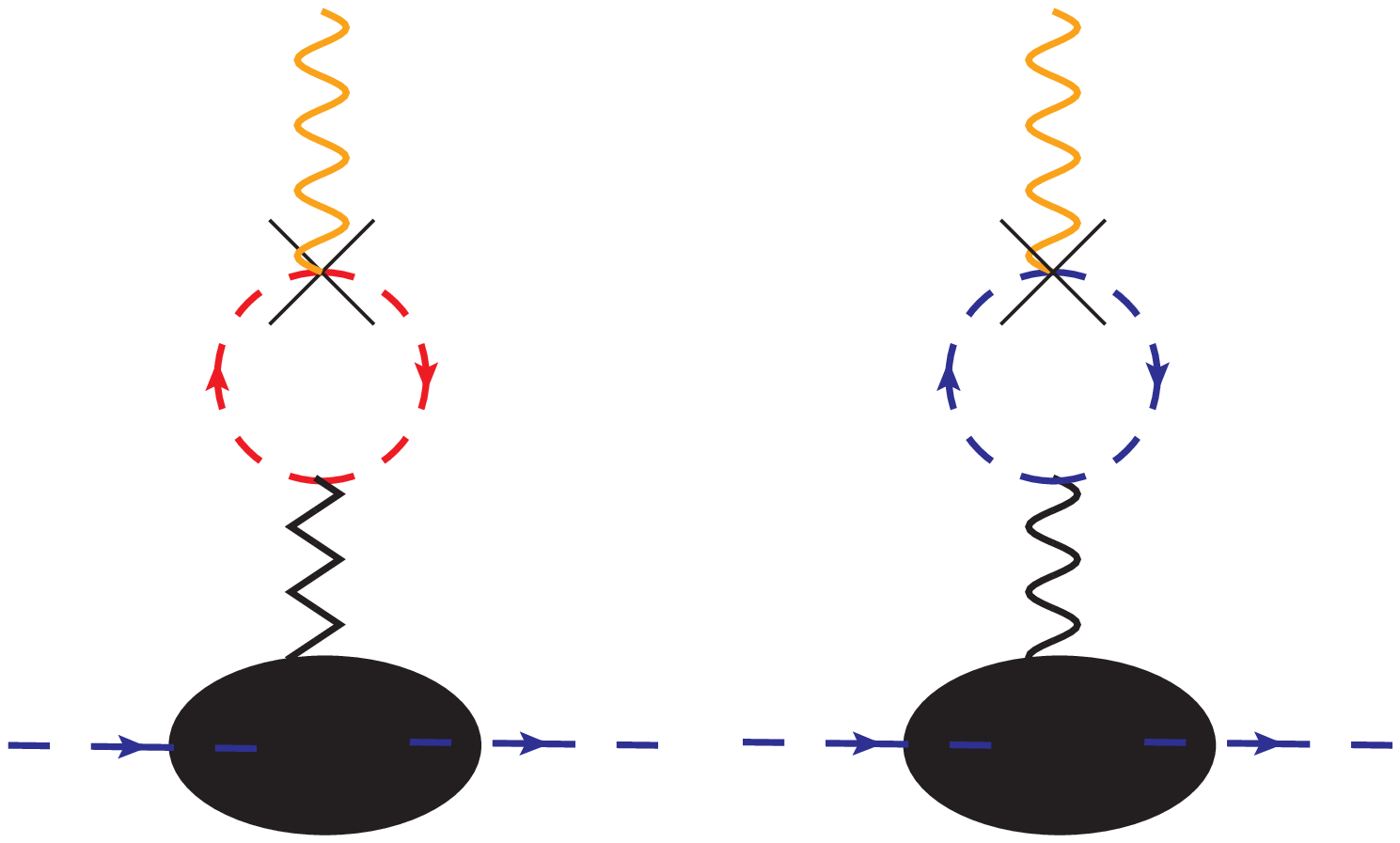}
\label{fig:figure1}
\end{minipage}
\hspace{0.2cm}
\begin{minipage}[b]{0.4\linewidth}
\centering
\includegraphics[scale=0.3]{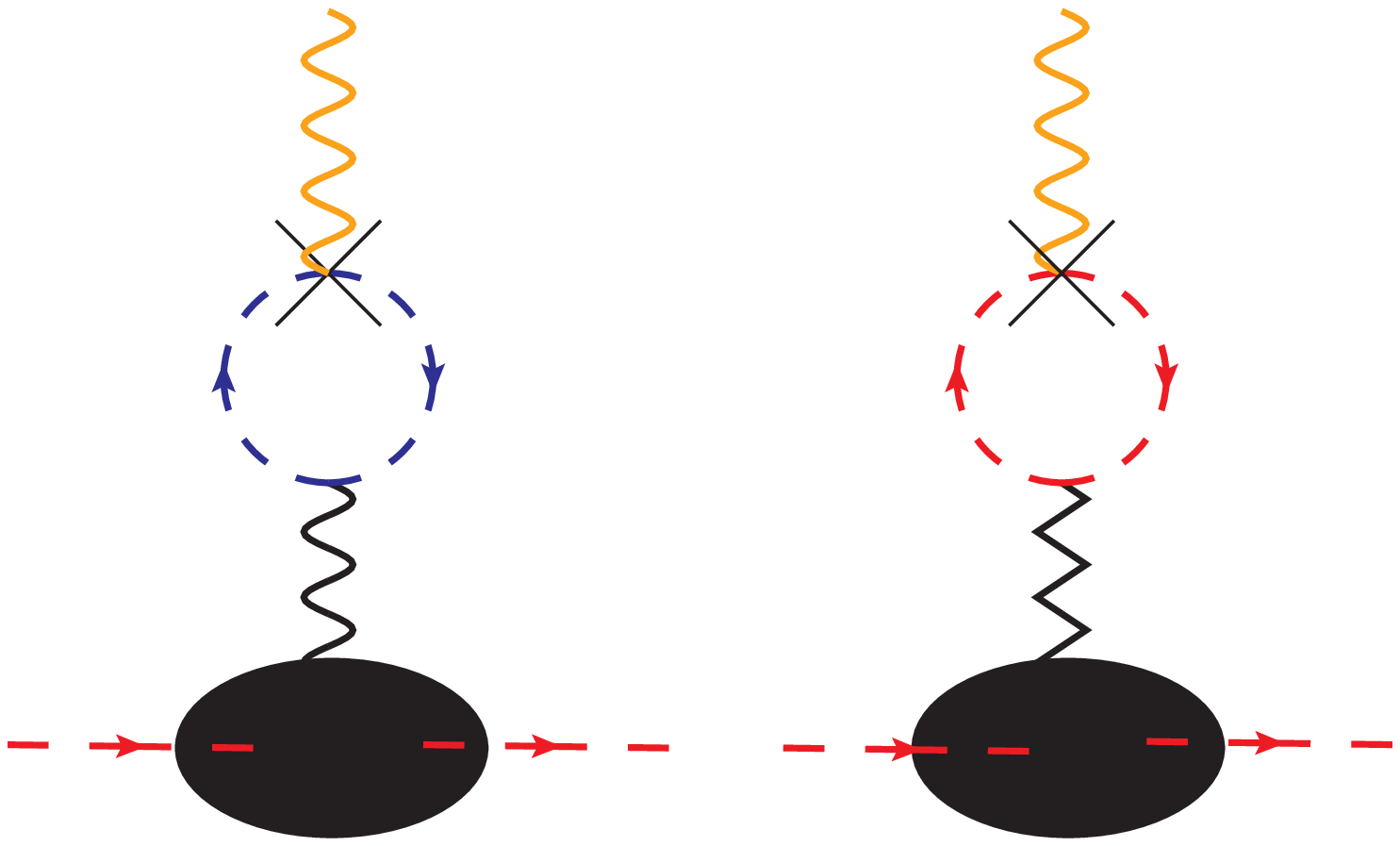}
\includegraphics[scale=0.3]{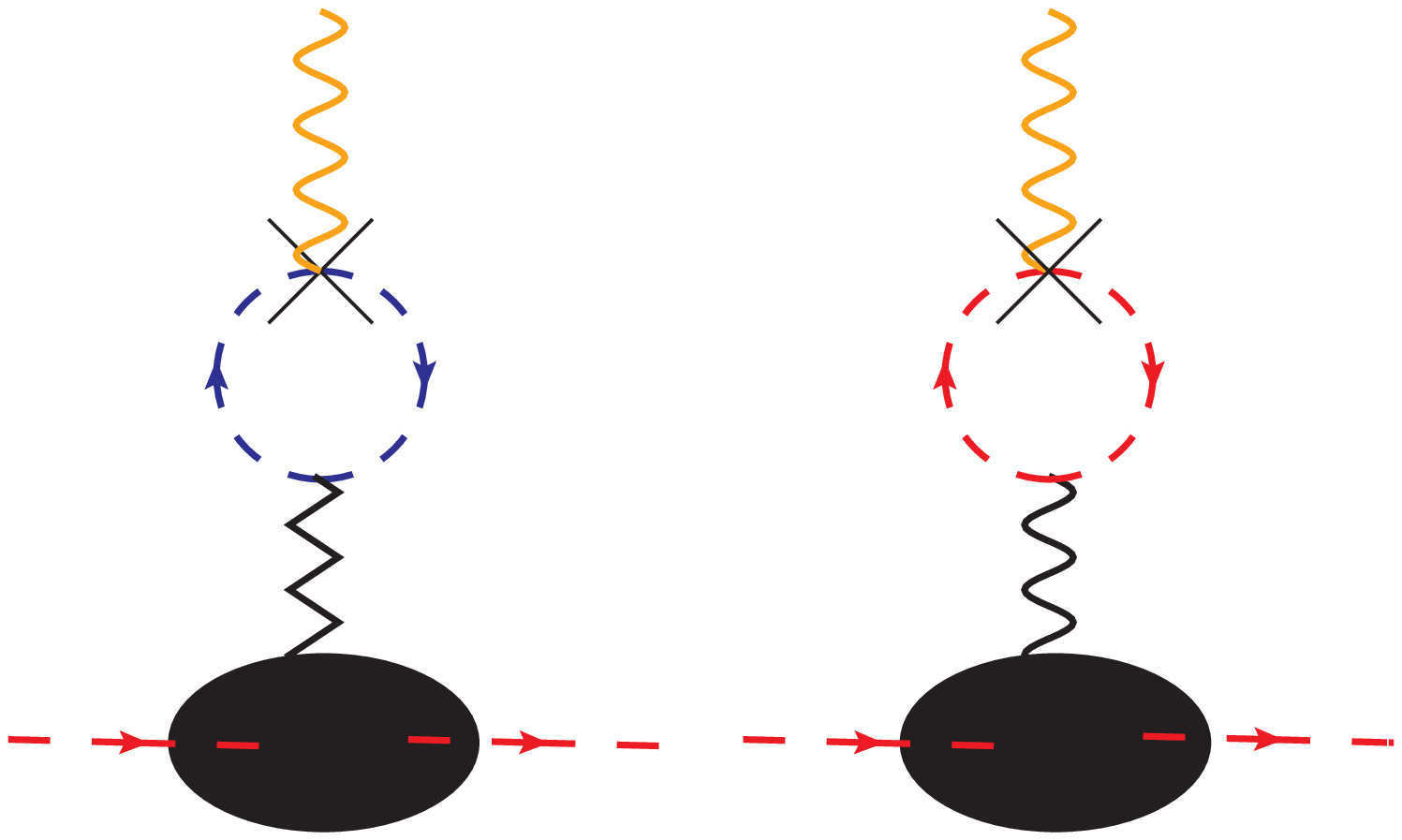}
\end{minipage}
\begin{minipage}[b]{0.4\linewidth}
\centering
\includegraphics[scale=0.3]{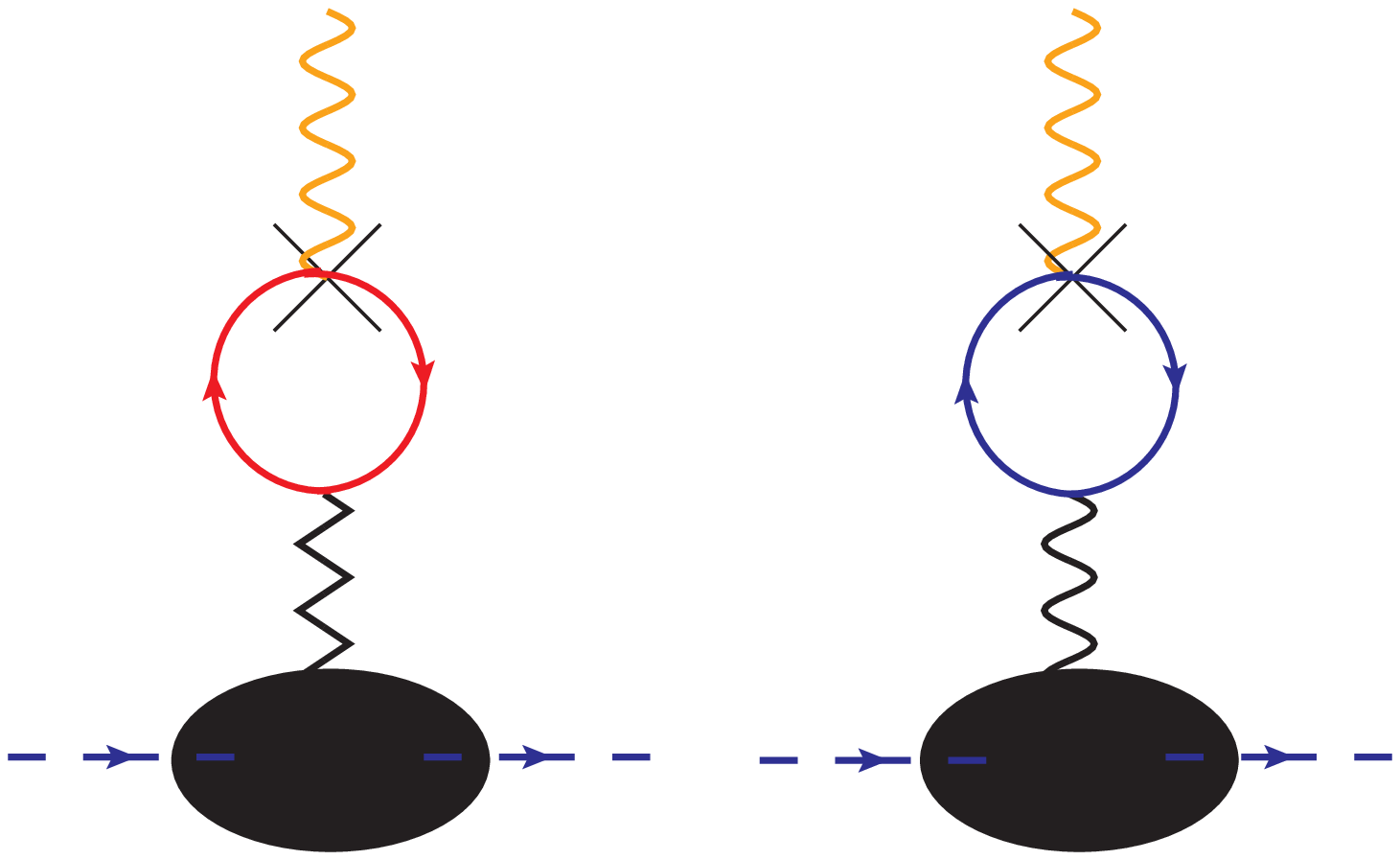}
\includegraphics[scale=0.3]{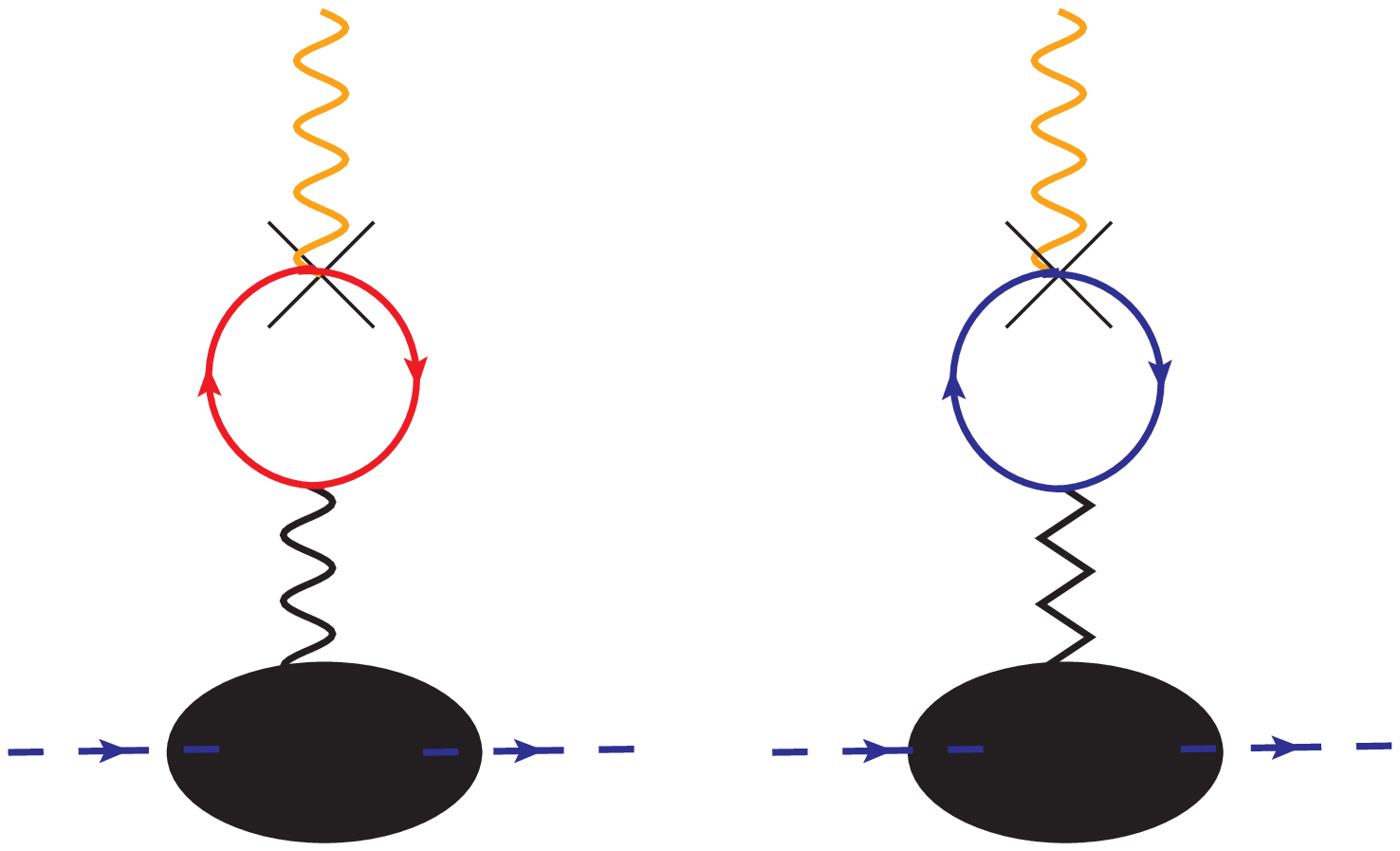}
\label{fig:figure1}
\end{minipage}
\hspace{0.2cm}
\begin{minipage}[b]{0.4\linewidth}
\centering
\includegraphics[scale=0.3]{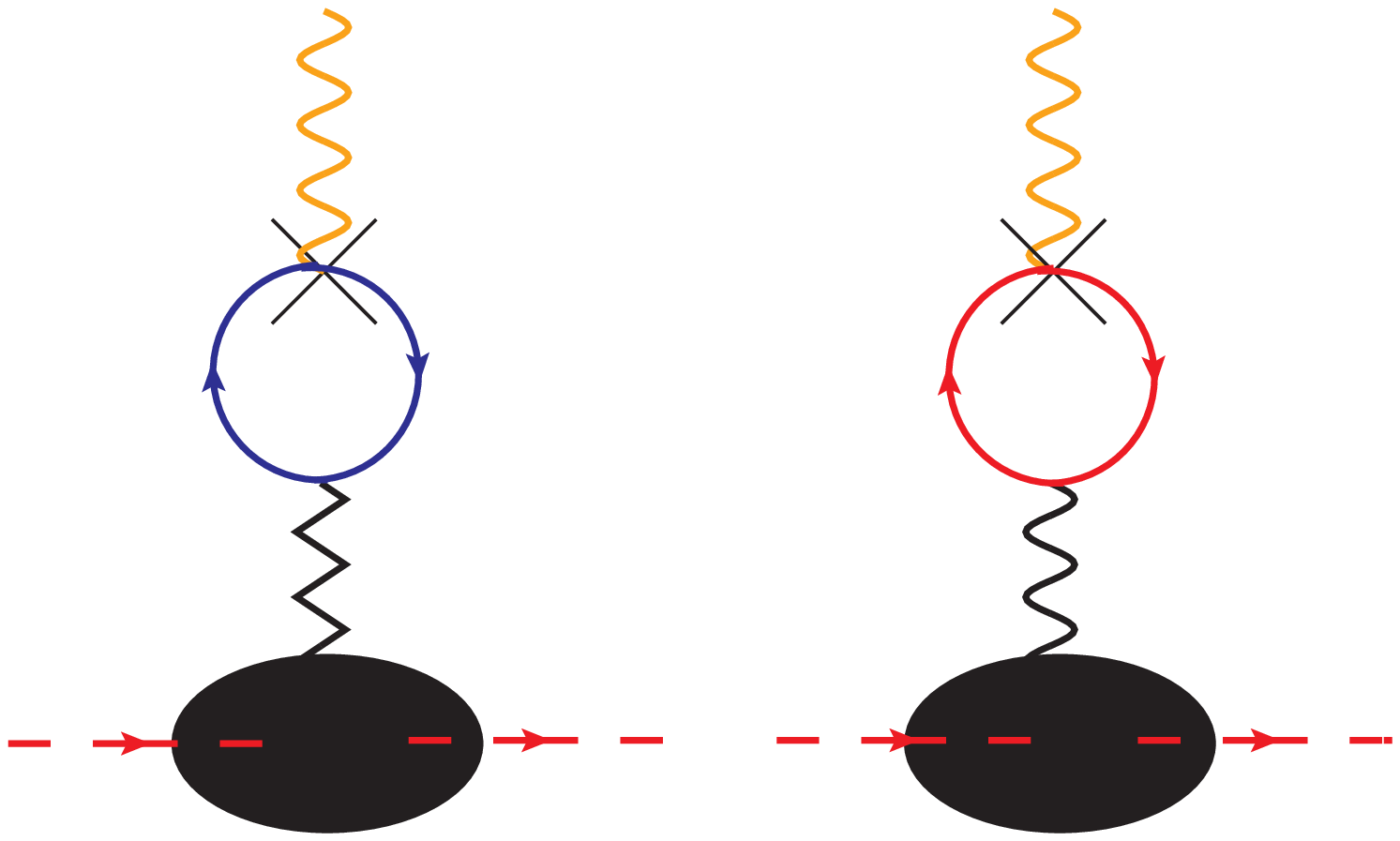}
\includegraphics[scale=0.3]{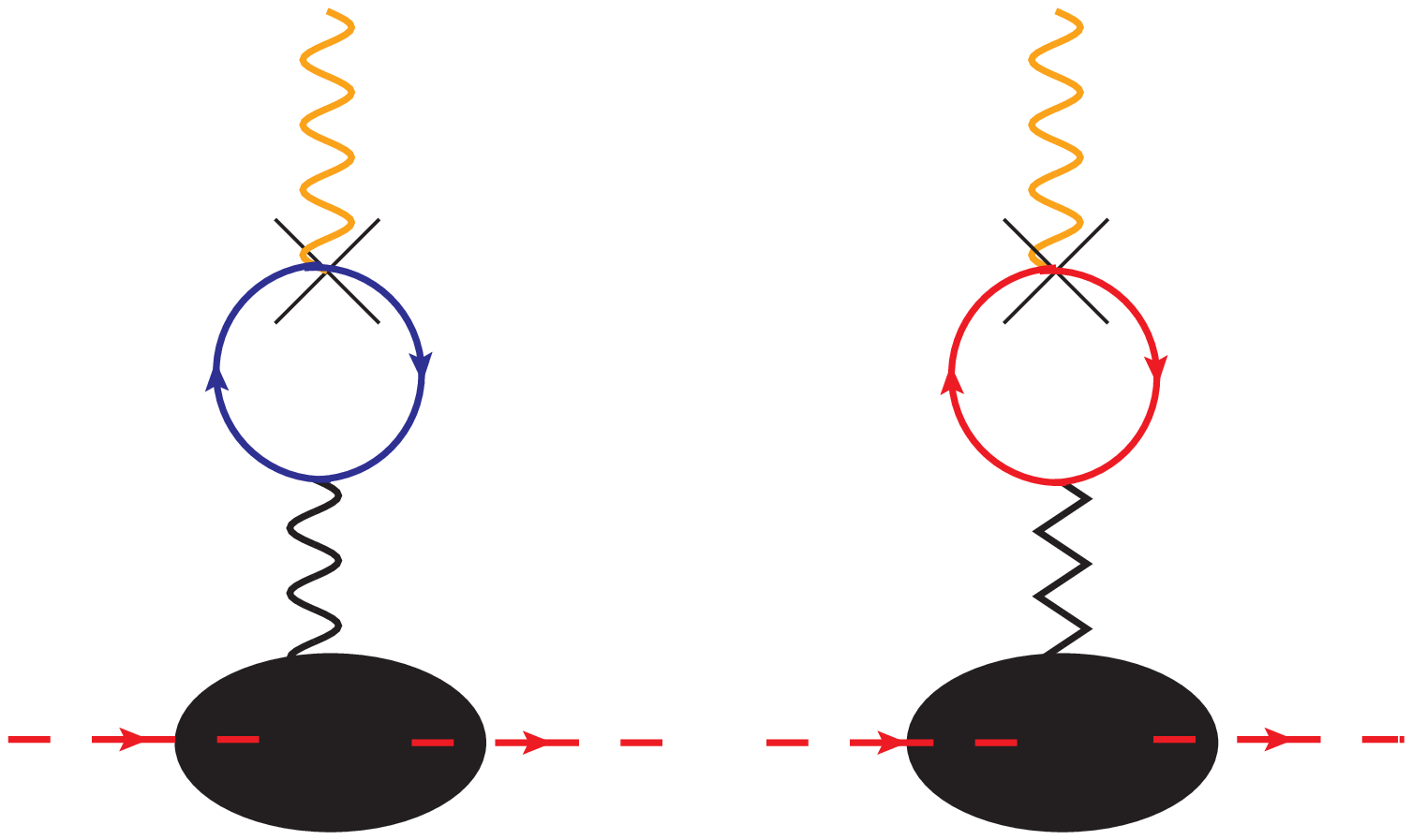}
\end{minipage}
\caption{(Color online) Generic anomalous terms in the Vertex functions for left-hand side.}
\label{totalanm}
\end{figure}

The anomalous terms that leads to the quantum anomaly (figure \ref{totalan}), in the right-hand side, associated to the one-color 
interaction $A^{\alpha}_{+}$, can be written as
\begin{eqnarray}
F_{\alpha\alpha',++}(p,p+q;q)&=& (g_{0}\chi_{-}^{a} + g_{f}\chi_{-}^{b})\Gamma^{(2,1)}_{\alpha\alpha',++}(p,p+q;q) \nonumber \\
&+& (g_{0}\chi_{+}^{a} + g_{f}\chi_{+}^{b})\Gamma^{(2,1)}_{\alpha\alpha',++}(p,p+q;q) \nonumber \\
\end{eqnarray}
and for the left-hand side (figure \ref{totalanm}), associated to the one-color interaction $A^{\alpha}_{-}$,
\begin{eqnarray}
F_{\alpha\alpha',--}(p,p+q;q)&=& (g_{0}\chi_{-}^{a} + g_{f}\chi_{-}^{b})\Gamma^{(2,1)}_{\alpha\alpha',--}(p,p+q;q) \nonumber \\
&+& (g_{0}\chi_{+}^{a} + g_{f}\chi_{+}^{b})\Gamma^{(2,1)}_{\alpha\alpha',--}(p,p+q;q) \nonumber \\
\end{eqnarray}
As a consequence, the anomalous Ward-Takahashi identity \cite{das} for the right-hand side is given by
\begin{eqnarray}
q_{+}\sum_{\alpha}\left(\Gamma_{\alpha\alpha',++}^{(2,1)}(p,p+q;q)-F_{\alpha\alpha',++}(p,p+q;q)\right)&=& \nonumber \\
\Gamma_{\alpha',+}^{(2)}(p+q)-\Gamma_{\alpha',+}^{(2)}(p)
\end{eqnarray}
and for the left-hand side,
\begin{eqnarray}
q_{-}\sum_{\alpha}\left(\Gamma_{\alpha\alpha',--}^{(2,1)}(p,p+q;q)-F_{\alpha\alpha',--}(p,p+q;q)\right)&=& \nonumber \\
\Gamma_{\alpha',-}^{(2)}(p+q)-\Gamma_{\alpha',-}^{(2)}(p). 
\end{eqnarray}

As in the Schwinger model $1+1$ \cite{manton}, the quantum anomaly $1+1$ here is a consequence of the coupling interaction in the free 
Dirac fields involving axial vectors. In the case, the polarization bubble of one-color is associated to quantum anomaly is not 
coincidence. As we will
 show, the interactions that do not change color are associated to axial vectors, while the interactions that 
change color (interbackscattering and interumklapp) are not. Indeed, we can write the intra 
and interband interactions explicitly in terms of 
current components
\begin{eqnarray}
g_{0}\psi_{+}^{a\dagger}\psi_{-}^{b\dagger}\psi_{-}^{b}\psi_{+}^{a} &=& g_{0}J_{+}^{a}J_{-}^{a} \\
g_{\mathcal{F}}\psi_{+}^{a\dagger}\psi_{-}^{b\dagger}\psi_{-}^{b}\psi_{+}^{a} &=& g_{\mathcal{F}}J_{+}^{a}J_{-}^{b} \\
g_{\mathcal{B}}\psi_{+}^{b\dagger}\psi_{-}^{a\dagger}\psi_{-}^{b}\psi_{+}^{a}&=& g_{\mathcal{B}}J_{+}^{ba}J_{-}^{ab} \\
g_{\mathcal{U}}\psi_{+}^{b\dagger}\psi_{-}^{b\dagger}\psi_{-}^{a}\psi_{+}^{a} &=& g_{\mathcal{U}}J_{+}^{ba}J_{-}^{ba}
\end{eqnarray}
In $1+1$ spacetime coordinates, the axial and vector currents 
\begin{eqnarray}
J^{\mu}_{A} &=& \bar{\psi}\gamma^{\mu}\gamma^{5}\psi,\\
J^{\mu}_{V} &=& \bar{\psi}\gamma^{\mu}\psi,
\end{eqnarray}
can be written, respectivelly, as
\begin{eqnarray}
J_{A}&=&\left( 
\begin{array}{c}
\psi_{+}^{\dagger}\psi_{+} - \psi_{-}^{\dagger}\psi_{-}  \\ 
\psi_{+}^{\dagger}\psi_{+} + \psi_{-}^{\dagger}\psi_{-} 
\end{array}%
\right)= \left( 
\begin{array}{c}
J_{A}^{0}  \\ 
J_{A}^{1} 
\end{array}%
\right), \\
J_{V}&=&\left( 
\begin{array}{c}
\psi_{+}^{\dagger}\psi_{+} + \psi_{-}^{\dagger}\psi_{-}  \\ 
\psi_{+}^{\dagger}\psi_{+} - \psi_{-}^{\dagger}\psi_{-} 
\end{array}%
\right)= \left( 
\begin{array}{c}
J_{V}^{0}  \\ 
J_{V}^{1} 
\end{array}%
\right). 
\end{eqnarray}
The following $1+1$ relation is satisfied $J_{A}=\gamma^{0}J_{V}$ and we can also write
\begin{eqnarray}
J_{A}^{0} + J_{A}^{1} &=& 2J_{+}, \\
J_{A}^{0} - J_{A}^{1} &=& -2J_{-}.  
\end{eqnarray}
As a consequence the intraforward and interforward interactions can be written in terms of axial current components  
\begin{eqnarray}
g_{0}J_{+}^{a}J_{-}^{a} &=& -\frac{g_{0}}{4}\left(J_{A}^{0a} + J_{A}^{1a}\right)\left(J_{A}^{0a} - J_{A}^{1a}\right),\\
g_{\mathcal{F}}J_{+}^{a}J_{-}^{b} &=& -\frac{g_{\mathcal{F}}}{4}\left(J_{A}^{0a} + J_{A}^{1a}\right)\left(J_{A}^{0b} 
- J_{A}^{1b}\right).
\end{eqnarray}
On the other hand, the interbackscattering and the interumklapp interactions are well defined in terms of one-side spinors 
defined as
\begin{eqnarray}
\psi_{C+}= \left( 
\begin{array}{c}
\psi_{+}  \\ 
\psi_{+}^{\dagger} 
\end{array}%
\right), \psi_{C-}= \left( 
\begin{array}{c}
\psi_{-}  \\ 
\psi_{-}^{\dagger} 
\end{array}%
\right)
\end{eqnarray}
\begin{eqnarray}
\psi_{C+}^{\dagger}= \left( 
\begin{array}{cc}
\psi_{+} ^{\dagger} & \psi_{+} 
\end{array}%
\right), \psi_{C-}^{\dagger}= \left( 
\begin{array}{cc}
\psi_{-}^{\dagger} &  \psi_{-}  
\end{array}%
\right)
\end{eqnarray}
with the corresponding one-side currents
\begin{eqnarray}
J^{\mu}_{C+}=\bar{\psi}_{C+}\gamma^{\mu}\gamma^{5}\psi'_{C+}, \\
J^{\mu}_{C-}=\bar{\psi}_{C-}\gamma^{\mu}\gamma^{5}\psi'_{C-}. 
\end{eqnarray}
We can write
\begin{eqnarray}
J_{C+}=  \left( 
\begin{array}{c}
\psi_{+}^{\dagger}\psi'_{+} + \psi_{+}^{'\dagger}\psi_{+}  \\ 
\psi_{+}^{\dagger}\psi_{+} - \psi_{+}^{'\dagger}\psi_{+} 
\end{array}%
\right)= \left( 
\begin{array}{c}
J_{C+}^{0}  \\ 
J_{C+}^{1} 
\end{array}%
\right),
\end{eqnarray}
\begin{eqnarray}
J_{C-}=  \left( 
\begin{array}{c}
\psi_{-}^{\dagger}\psi'_{-} + \psi_{-}^{'\dagger}\psi_{-}  \\ 
\psi_{-}^{\dagger}\psi'_{-} - \psi_{-}^{'\dagger}\psi_{-} 
\end{array}%
\right)= \left( 
\begin{array}{c}
J_{C-}^{0}  \\ 
J_{C-}^{1} 
\end{array}%
\right),
\end{eqnarray}
where $\psi'=\psi(x')$ and $\psi=\psi(x)$. Note that, if $x'=x$, we have that one component of the one-side current is zero.
\begin{eqnarray}
J_{C+}^{ab}=  \left( 
\begin{array}{c}
\psi_{+}^{a\dagger}\psi^{b}_{+} + \psi_{+}^{b\dagger}\psi_{+}^{a}  \\ 
\psi_{+}^{a\dagger}\psi^{b}_{+} - \psi_{+}^{b\dagger}\psi_{+}^{a} 
\end{array}%
\right)= \left( 
\begin{array}{c}
J_{C+}^{0ab}  \\ 
J_{C+}^{1ab} 
\end{array}%
\right),\nonumber \\
\end{eqnarray}
\begin{eqnarray}
J_{C-}^{ab}=  \left( 
\begin{array}{c}
\psi_{-}^{a\dagger}\psi^{b}_{-} + \psi_{-}^{b\dagger}\psi_{-}^{a}  \\ 
\psi_{-}^{a\dagger}\psi^{b}_{-} - \psi_{-}^{b\dagger}\psi_{-}^{a} 
\end{array}%
\right)= \left( 
\begin{array}{c}
J_{C-}^{0ab}  \\ 
J_{C-}^{1ab} 
\end{array}%
\right).\nonumber \\
\end{eqnarray}
Consequently
\begin{eqnarray}
J_{C-}^{0ab} &=& J^{ab}_{-} + J^{ba}_{-}, \\
J_{C-}^{1ab} &=& J^{ab}_{-} - J^{ba}_{-}, \\
J_{C+}^{0ab} &=& J^{ab}_{+} + J^{ba}_{+}, \\
J_{C+}^{1ab} &=& J^{ab}_{+} - J^{ba}_{+} 
\end{eqnarray}
and we can write the interbackscattering and interumklapp in the one-side currents
\begin{eqnarray}
g_{\mathcal{B}}J_{+}^{ba}J_{-}^{ab} &=& \frac{g_{\mathcal{B}}}{4}\left(J_{C+}^{0ab} - J_{C+}^{1ab}\right)\left(J_{C-}^{0ab} + J_{C-}^{1ab}\right), \\
g_{\mathcal{U}}J_{+}^{ba}J_{-}^{ba} &=& \frac{g_{\mathcal{U}}}{4}\left(J_{C+}^{0ab} - J_{C+}^{1ab}\right)\left(J_{C-}^{0ab} - J_{C-}^{1ab}\right).
\end{eqnarray}
Alternativelly, we can write 
\begin{eqnarray}
J_{C-}^{0ba} &=& J^{ba}_{-} + J^{ab}_{-}, \\
J_{C-}^{1ba} &=& J^{ba}_{-} - J^{ab}_{-}, \\
J_{C+}^{0ba} &=& J^{ba}_{+} + J^{ab}_{+}, \\
J_{C+}^{1ba} &=& J^{ba}_{+} - J^{ab}_{+}, 
\end{eqnarray}
and then
\begin{eqnarray}
g_{\mathcal{B}}J_{+}^{ba}J_{-}^{ab} &=& \frac{g_{\mathcal{B}}}{4}\left(J_{C+}^{0ba} + J_{C+}^{1ba}\right)\left(J_{C-}^{0ba} - J_{C-}^{1ba}\right), \\
g_{\mathcal{U}}J_{+}^{ba}J_{-}^{ba} &=& \frac{g_{\mathcal{U}}}{4}\left(J_{C+}^{0ba} + J_{C+}^{1ba}\right)\left(J_{C-}^{0ba} + J_{C-}^{1ba}\right).
\end{eqnarray}
Now we can also consider the term involving $v_{F}k_{F}^{\alpha}$, 
\begin{eqnarray}
\sum_{\alpha,s} v_{F}k_{F}^{\alpha}\psi_{s}^{\alpha\dagger}\psi_{s}^{\alpha}= \sum_{\alpha}v_{F}k_{F}^{\alpha}\left(J^{\alpha}_{+} + J^{\alpha}_{-}\right).
\end{eqnarray}
In terms of the axial vector this will be written as
\begin{eqnarray}
\sum_{\alpha,s} v_{F}k_{F}^{\alpha}\psi_{s}^{\alpha\dagger}\psi_{s}^{\alpha}= \sum_{\alpha}v_{F}k_{F}^{\alpha}J^{1\alpha}_{A},
\end{eqnarray}
where we have used $J^{1}_{A}= J_{+} + J_{-}$.

The consequence of quantum anomaly in $1+1$ is more appropriatelly related to the component $J^{0}_{A}$ of the axial vector, 
that leads to the fermion number anomaly \cite{fujikawa}, appearing in the intraforward and interforward interactions, i.e., 
while $J_{A}^{0}=\psi_{+}^{\dagger}\psi_{+} - \psi_{-}^{\dagger}\psi_{-}$ is related to the chiral symmetry, the component 
$J_{A}^{1}=\psi_{+}^{\dagger}\psi_{+} + \psi_{-}^{\dagger}\psi_{-}$ is related to the charge conservation \cite{gaume}.

From the point of view of the regularization, the renormalized theory is given in terms of the 
anomalous Ward-Takahashi identity, if we want require the invariance under chiral symmetry. Alternativelly, a different axial 
vector could be formulated that would satisfy the normal Ward-Takahashi identity, but violating gauge invariance \cite{itizikson}.

\section{Conclusion}

We have considered a quasi-1D system of two-coupled spinless fermions chains under 
intraforward, interforward, interbackscattering and interumklapp interactons. We have showed that this system 
can be equivalently described in terms the one-color and two-color interactions, i.e., interactions that change color 
and interactions that do not change color. 

By means of field renormalization group, we derived appropriatelly the bare and renormalized quantities that 
lead to the set of RG flow equations for this system. We have showed that the renormalized interband Fermi points, $\Delta k_{F,R}$, flow to zero, 
leading consequently to a confinement of the bonding and antibonding bands. The interbackscattering interaction flows to zero, as a 
consequence its relation to $\Delta k_{F,R}$, a result also verified in previous works \cite{ledowski}. The intraforward, interforward 
and interumklapp flows to Luttinger liquid fixed points, leading to interband communications in the confinement by 
constant interforward and interumklapp interactions. 

Considering the quantum anomaly, inherent to the system, as a consequence of coupling to $1+1$ Dirac fields to one-color and two-color 
interactions, we have arrived at the Ward-Takahashi identities that contains the anomalous terms and the corresponding, gauge invariant, 
anomalous Ward-Takahashi identities, derived first by \cite{alvaro2}, using other methods. 

We have showed that the quantum anomaly is associated to one-color interactions, as a consequence of the fact such interactions are 
related to axial vectors, while two-color interactions are associated to one-side vectors. More appropriatelly in $1+1$, 
the zero-component of axial vector is related to the chiral symmetry and the one-component to the charge conservation.

\section{Acknowledgements}

T.P. thanks the CAPES (Brazil) and IIP-UFRN-Natal (Brazil), where part of this
work were realized.


\begin{thebibliography}{99}
\bibitem{alvaro1} A. Ferraz, J. Phys. A: Math. Gen. {\bf 39} (2006) 7963.
\bibitem{alvaro4} A. Ferraz, Phys. Rev. B, \textbf{75} (2007) 233103.
\bibitem{alvaro2} L. Costa, A. Ferraz, V. Mastropietro, arXiv:1103.2744.
\bibitem{haldane} F. D. M. Haldane, J. Phys. C: Solid State Phys., \textbf{14} (1981) 2585.
\bibitem{ledowski2} S. Ledowski, P. Kopietz, Phys. Rev. B \textbf{76} (2007) 121403R.
\bibitem{kondo} T. Kondo et al., Phys. Rev. Lett. \textbf{105} (2010) 267003.
\bibitem{correa} E. Correa, H. Freire, A. Ferraz, Phys. Rev. B, {\bf 78} (2008) 195108.
\bibitem{ledowski} S. Ledowski, P. Kopietz, A. Ferraz, Phys. Rev. B, {\bf 71} (2005) 235106. 
\bibitem{ledowski3} S. Ledowski, P. Kopietz, Phys. Rev. B \textbf{75} (2007) 045134.
\bibitem{mastropietro1} G. Benfatto, V. Mastropietro, Commun. Math. Phys., \textbf{258} (2005) 609.
\bibitem{mastropietro2} V. Mastropietro, Phys. Rev. B {\bf 84} (2011) 035109.
\bibitem{kopietz1} P. Kopietz, L. Bartosch, L. Costa, A. Isidori, A. Ferraz, J. Phys. A: Math. Theor. \textbf{43} (2010) 385004.
\bibitem{schwinger1} J. Schwinger, Phys. Rev. \textbf{82} (1951) 664.
\bibitem{schwinger2} J. Schwinger, Phys. Rev. \textbf{125} (1962) 2542.
\bibitem{ward} J. C. Ward, Phys. Rev. \textbf{78} (1950) 182.
\bibitem{takahashi} Y. Takahashi, Nuovo Cimento \textbf{6} (1957) 371.
\bibitem{das} A. Das, \textit{Field Theory A Path Integral Approach} (World Scientific, Singapore ,2006).
\bibitem{kochetov} A. Ferraz, E. A. Kochetov, Nuc. Phys. B \textbf{853} (2011) 710.
\bibitem{freire3} H. Freire, E. Correa, A. Ferraz, Phys. Rev. B, \textbf{78} (2008) 125114.
\bibitem{freire1} H. Freire, E. Correa, A. Ferraz, J. Phys. A: Math. Gen. \textbf{39} (2006) 7977.
\bibitem{freire2} H. Freire, E. Correa, A. Ferraz, Phys. Rev. B, \textbf{71} (2005) 165113. 
\bibitem{alvaro5} A. Ferraz, Phys. Rev. B, \textbf{68} (2003) 075115.
\bibitem{alvaro6} A. Ferraz, Modern Physics Letters B, \textbf{17} (2003) 167.
\bibitem{alvaro3} S. Ledowski, P. Kopietz, A. Ferraz, Phys. Rev. B, {\bf 71} (2005) 235106.
\bibitem{fabrizio} M. Fabrizio, Phys. Rev. B {\bf 48} (1993) 15838.
\bibitem{ledermann} Ledermann, Phys. Rev. B {\bf 61} (2000) 2497.
\bibitem{collins} J. C. Collins, \textit{Renormalization} (Cambridge University Press,
Cambridge, 1984).
\bibitem{castro} C. Di Castro, W. Metzner, Phys. Rev. Lett. \textbf{67} (1991) 3857.
\bibitem{adler} S. L. Adler, W. A. Bardeen, Phys. Rev. \textbf{182} (1969) 1517.
\bibitem{solyom} J. S\'olyom, Advances in Physics, \text{28} (1979) 201.
\bibitem{bala} A. P. Balachandram, G. Marmo, B. S. Skargestam, A. Stern, \textit{Classical Topology and Quantum States}, 
(World Scientific, Singapore, 1991).
\bibitem{ryder} L. H. Ryder,  \textit{Quantum Field Theory} (Cambridge University Press, Cambridge, 1996).
\bibitem{manton} N. S. Manton, Annals of Physics, \textbf{159} (1985) 220. 
\bibitem{fujikawa} K. Fujikawa, S. Suzuki, \textit{Path Integrals and Quantum Anomalies} (Claredon Press, Oxford, 2004).
\bibitem{gaume} L. Alvarez-Gaum\'e, M. A. Vazquez-Mozo, \textit{An invitation to quantum field theory} (Springer Verlag, Berlin, 2012).
\bibitem{itizikson} C. Itzykson, B. Zuber, \textit{Quantum Field Theory} (Mc-GrawHill Inc., New York, 1980).  
\end{thebibliography}
\end{document}